\documentclass[fleqn,usenatbib]{mnras}
\usepackage{newtxtext,newtxmath}
\usepackage[T1]{fontenc}
\usepackage{ae,aecompl}
\usepackage{graphicx}   
\usepackage{epsfig}
\usepackage{psfig}
\usepackage{color}

\def\gs{\mathrel{\raise0.35ex\hbox{$\scriptstyle >$}\kern-0.5em
\lower0.40ex\hbox{{$\scriptstyle \sim$}}}}
\def\ls{\mathrel{\raise0.35ex\hbox{$\scriptstyle <$}\kern-0.5em
\lower0.40ex\hbox{{$\scriptstyle \sim$}}}}

\makeatother

\date{Accepted ---. Received ---; in original form 2024 January 14}

\pubyear{2024}

\title[A SCUBA-2 survey of $z$\,$=$\,1.6--2.0 clusters]{Obscured star formation in clusters at \emph{z}\,$=$\,1.6--2.0: massive  galaxy formation and the reversal of the star formation--density relation}

\author[Smail]{Ian Smail$^{1}$\thanks{E-mail: ian.smail@durham.ac.uk}\\
  $^{1}$ Centre for Extragalactic Astronomy, Department of Physics, Durham University, South Road, Durham DH1 3LE, UK\\
  }

\begin{document}

\label{firstpage}
\pagerange{\pageref{firstpage}--\pageref{lastpage}}
\maketitle

\begin{abstract}
Clusters of galaxies at $z$\,$\gs$\,1 are expected to be increasingly active sites of star formation.  To test this, an  850\,$\mu$m survey was undertaken of eight clusters at $z$\,$=$\,1.6--2.0 using  SCUBA-2 on the James Clerk Maxwell Telescope. Mid-infrared properties were used to identify 53 probable counterparts to 45 SCUBA-2 sources with colours that suggested they were cluster members. This  uncovered a modest  overdensity of 850-$\mu$m sources, with far-infrared luminosities of $L_{\rm IR}$\,$\geq$\,10$^{12}$\,L$_\odot$ (SFR\,$\gs$\,100\,M$_\odot$\,yr$^{-1}$) and  colours consistent with being cluster members, of a factor of 4\,$\pm$\,1  within the central 1\,Mpc radius of the clusters.  The submillimetre photometry of these galaxies was used to estimate the total cluster star formation rates. These showed that the mass-normalised rates in the clusters are  two orders of magnitude higher than in local systems, evolving as $(1+z)^{5.5\pm 0.6}$.  This rapid evolution means that the mass-normalised star formation rates in these clusters  matched that of average halos in the field at $z$\,$\sim$\,1.8\,$\pm$\,0.2 marking the epoch where the local star formation--density relation reverses in massive halos. The estimated stellar masses of the cluster submillimetre galaxies suggests that their descendants will be amongst the most massive galaxies in $z$\,$\sim$\,0 clusters. This reinforces the suggestion that the majority of the massive early-type galaxy population in $z$\,$\sim$\,0 clusters were likely to have formed at $z$\,$\gs$\,1.5--2 through very active, but dust-obscured, starburst events.
\end{abstract}

\begin{keywords}
cosmology: observations --- galaxies: evolution --- galaxies: formation  --- submillimetre: galaxies
\end{keywords}

\section{Introduction}

Surveys of galaxy clusters at $z$\,$\sim$\,0.5--2 suggest that star formation was increasingly prevalent in these dense environments at higher redshifts (e.g.,  \citealt{Webb05,Geach06,Popesso12,Wagner17,Smith19},  see \citealt{Alberts22} for an extensive review).   This is expected in a hierarchical galaxy formation model, where the most massive halos (which represent the progenitors of today's massive clusters of galaxies) and the galactic sub-halos within them, collapsed at earlier times \citep{Cole89},  a trend that is supported by observations of local clusters \citep[e.g.,][]{Bower90}.  Indeed observations suggest that  intense star formation activity in dense environments  extends out to proto-clusters at the highest redshifts, $z$\,$\gs$\,2--5  \citep[e.g.,][]{Stevens03,Casey15,Umehata15,Kato16,Casey16,MacKenzie17,Martinache18,Zeballos18,Rotermund21}.  However, the interpretation of this evolution is complicated by the selection of these stuctures:  they were frequently discovered as over-densities of star-forming galaxies (which   biases them to atypically active systems) or using sign-post  active sources (e.g., radio galaxies or QSOs), the evolution of which results in complex selection functions \citep[e.g.,][]{Rigby14,Greenslade18,Cheng19,Nowotka22,Polletta22,Zhang22}.  To obtain a less biased view of the redshift evolution of cluster activity  the target clusters  need to be identified using more robust tracers of their  total mass, such as their X-ray luminosity, Sunyaev-Zel'dovich (SZ) decrements or the integrated stellar mass of their galaxy populations \citep[e.g.,][]{Webb13,Ma15,Noble17,Wu18,Smith19}.

%
%
\begin{table*}
\caption{Cluster sample.   The  integrated star formation rates within the central 1\,Mpc radius of the clusters, $\Sigma_{\rm SFR}$, are derived in \S4.3. }
\begin{tabular}{llcccccl}
\hline \noalign {\smallskip}
Cluster & Long Name & R.A.\ & Dec.\ & $z$ &  $M_{200}$ & $\Sigma_{\rm SFR}$ &  References \\
  & & \multicolumn{2}{c}{(J2000)} & & (10$^{14}$\,M$_\odot$) & (M$_\odot$\,yr$^{-1}$) & \\
\hline \noalign {\smallskip}
XLSSC122&  ACT-CL\,J0217.7$-$0345 & 02\,17\,44.1 & $-$03\,46\,10  & 1.98 &  2.3\,$\pm$\,0.3 &  740\,$\pm$\,40 & \citet{vanMarrewijk23} \\
SpARCSJ0224 & SpARCS\,J022426$-$032330 & 02\,24\,26.3 & $-$03\,23\,30 & 1.63 &  2.0\,$\pm$\,0.3  & 530\,$\pm$\,50 & \citet{Babyk14} \\
SpARCSJ0225 & SpARCS\,J022545$-$035517 & 02\,25\,45.6 & $-$03\,55\,17 & 1.60 & $\sim$\,1 &  180\,$\pm$\,30 & \citet{Noble17}  \\
  JKCS041 & ... & 02\,26\,44.0 &  $-$04\,41\,36  & 1.80 & $\sim$\,2 & 530\,$\pm$\,50 & \citet{Mei15} \\
LH146 & XMMU\,J105324.7$+$572348 & 10\,53\,21.6 & $+$57\,24\,00 & 1.71 & 1.4\,$\pm$\,0.2 & 700\,$\pm$\,60 & \citet{Henry14} \\
IDCSJ1426 & IDCS\,J1426.5$+$3508 & 14\,26\,32.7  & +35\,08\,29  & 1.75 & 4.1\,$\pm$\,1.1 & 170\,$\pm$\,30 & \citet{Brodwin12} \\
IDCSJ1433 & IDCS\,J1433.2$+$3306 &  14\,33\,11.5 & +33\,06\,39    & 1.89 & $\sim$\,1 & $\ls$\,160 & \citet{Zeimann12} \\    
ClJ1449  & Cl\,J1449+0856  & 14\,49\,14.0  & +08\,56\,21  & 1.99 & 0.53\,$\pm$\,0.09 & 620\,$\pm$\,70 &  \citet{Gobat13} \\
\hline \noalign {\smallskip}
\end{tabular}
\end{table*}

The stellar populations of  passive, massive early-type  galaxy population that dominate  clusters at $z$\,$\sim$\,0 are metal rich \citep[e.g.,][]{Poggianti01,Nelan05} and so it was expected that  the star formation activity associated with their formation was   obscured by dust.   Indeed,  early ISO and Spitzer mid-infrared surveys of $z$\,$\ls$\,1 clusters suggested that  they hosted previously unappreciated populations of dusty star-forming galaxies \citep[e.g.,][]{Coia05,Geach06, Marcillac07,Bai09}.  The launch of Herschel and the extension of such surveys into the far-infrared (which is a more robust tracer of the obscured star formation than the restframe mid-infrared at $z$\,$\gs$\,1) provided compelling evidence for strong evolution of the far-infrared luminosity function of cluster galaxies  and suggested obscured star formation rates (SFR) far in excess of those measured using tracers in the optical or UV wavebands \citep[e.g.,][]{Popesso12,Santos15,Alberts16,Alberts21}.    At longer wavelengths, which provide sensitive probes of the most massive dusty (and gas rich) galaxies, SCUBA, SCUBA-2 and now ALMA have strengthened the evidence for significant populations of obscured, active galaxies in well-defined cluster samples out to $z$\,$\sim$\,1--1.5 \citep[e.g.,][]{Best02,Webb05,Stach17,Cooke19}, and a single example of a well-studied X-ray-detected cluster at $z$\,$=$\,2.0 \citep{Coogan18,Smith19}. The accelerated evolution in these systems  means that the cores of $z$\,$\gs $\,1 clusters hosted significant (but variable) numbers of  dusty star-forming galaxies \citep[e.g.,][]{Tran10,Tadaki12,Cooke19}.

The most reliable method to select dusty star-forming galaxies in clusters at $z$\,$>$\,1 uses far-infrared or submillimetre observations that select the sources in the restframe far-infrared.  One of the most efficient facilities for undertaking such studies is the SCUBA-2 submillimetre camera \citep{Holland13}  on the James Clerk Maxwell Telescope (JCMT), due to its  8$'$\,$\times$\,8$'$ field of view, corresponding to $\sim$\,4\,Mpc at $z$\,$>$\,1, and hence sufficient to map a massive cluster in one pointing.  \cite{Cooke19} therefore undertook a SCUBA-2 study of the submillimetre population in eight virialised, mass-selected clusters at $z$\,$= $\,0.8--1.6 (see also \citealt{Smail14,Ma15,Stach17}).   The clusters all showed significant  over-densities of submillimetre galaxies, with the integrated star formation rates, normalized by the corresponding cluster mass, showing an increase out to $z$\,$\sim$\,1.5 that was consistent with evolution of the form $(1 + z)^\gamma$ with $\gamma$\,$\sim$\,6, potentially more rapid than the $\gamma$\,$\sim$\,4 trend in the field \citep[see also][]{Kodama04,Finn05,Geach06,Bai09,Popesso12,Webb13,Alberts16,Smith20}. However, they also reported hints of a flattening in the evolution at $z$\,$\gs$\,1, and in addition the mass-normalized star formation rate in clusters at $z$\,$<$\,1.6 was still lower than the field by a factor of 1.5\,$\pm$\,0.3, suggesting no evidence  for a reversal of the local SFR--density relation \citep[e.g.,][]{Spitzer51,Dressler80} in   massive  clusters at $z$\,$\ls $\,1.5 \citep[c.f.,][]{Tran10,Alberts14,Santos15}.    \cite{Smith19} subsequently published a similar SCUBA-2 and Herschel survey of a single  $z$\,$=$\,2.0 cluster \citep{Gobat13}, showing a much higher mass normalised star formation rate, above the surrounding field and suggesting even more rapid evolution,  $\gamma$\,$\sim$\,7. Given the variation seen in the Cooke et al.\ sample, it was possible that the Smith et al.\ cluster was simply an outlier -- but to test this SCUBA-2  observations of a larger sample of clusters at $z$\,$\gs$\,1.5 was needed.

%
%
\begin{figure*}
\centerline{
\psfig{file=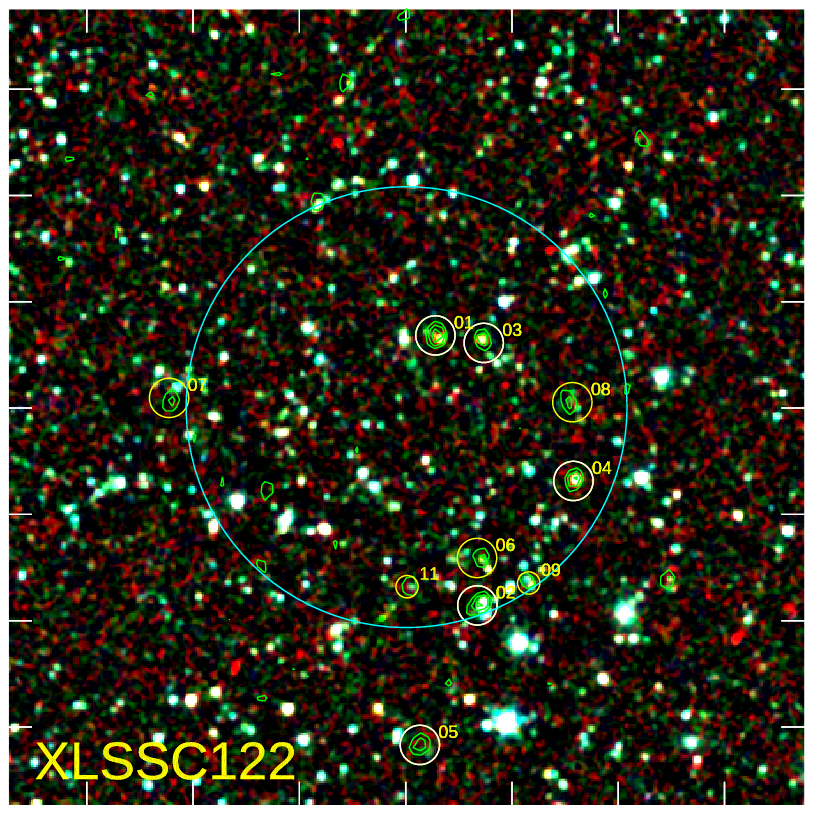, width=2.2in, angle=0}\hspace*{-0.5in}
\psfig{file=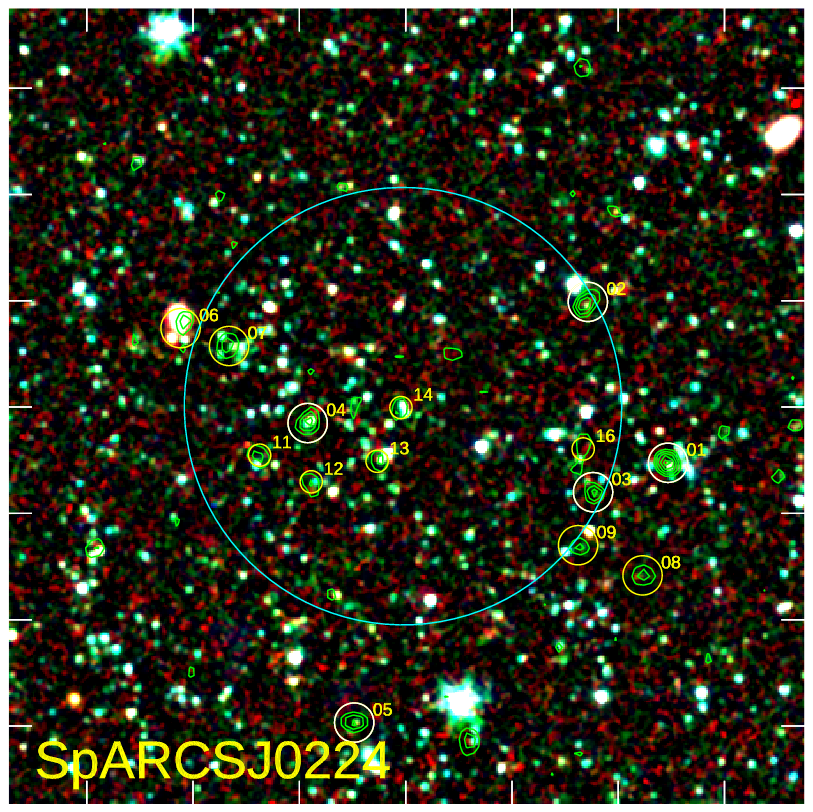, width=2.2in, angle=0}\hspace*{-0.5in}
\psfig{file=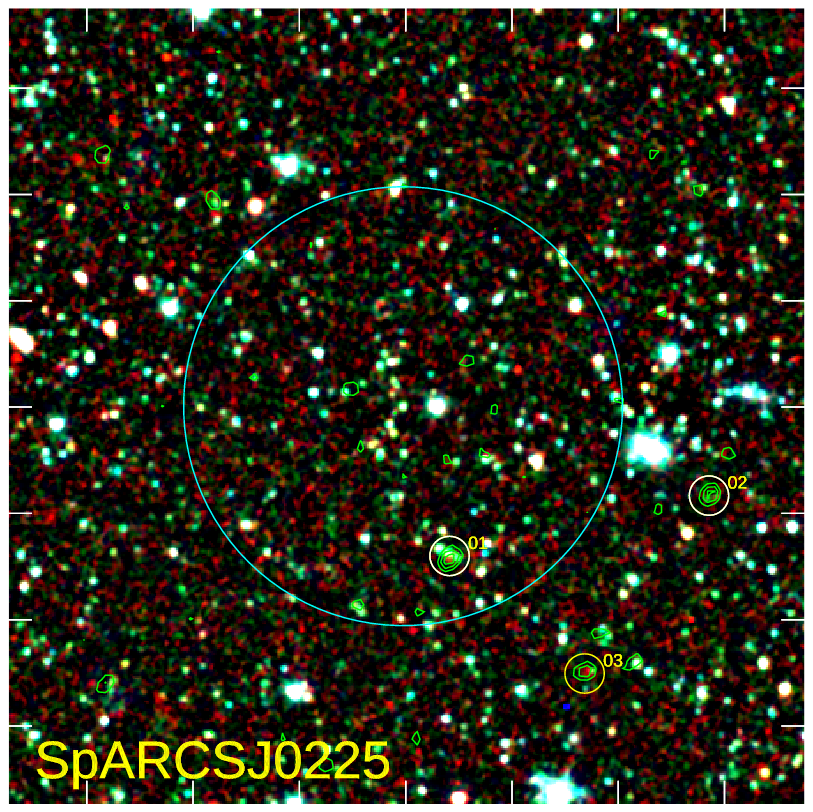, width=2.2in, angle=0}\hspace*{-0.5in}
\psfig{file=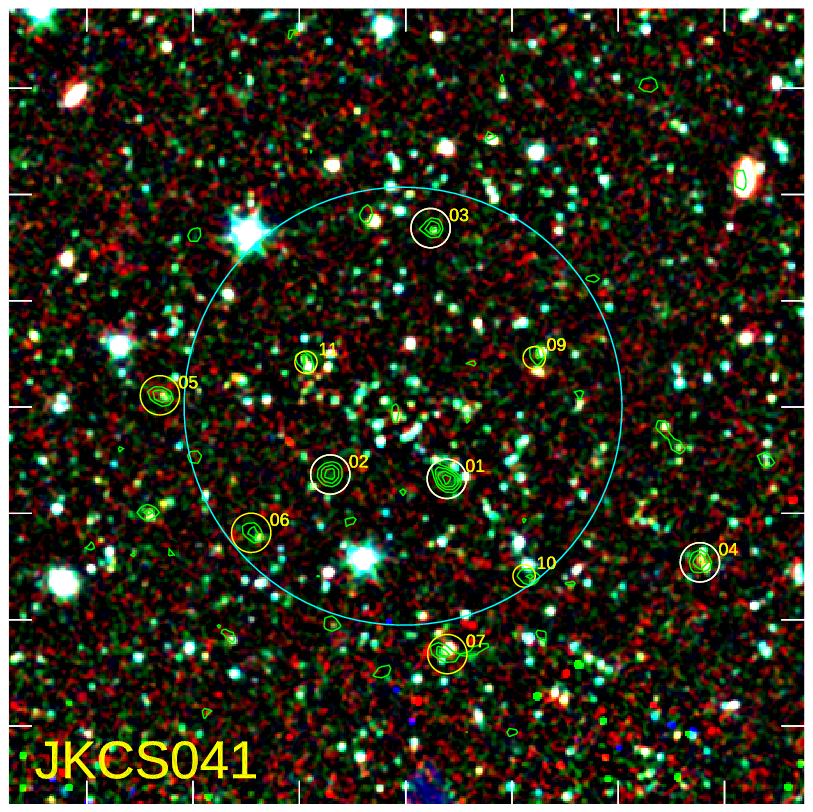, width=2.2in, angle=0}}\vspace*{-0.5in}
\centerline{
\psfig{file=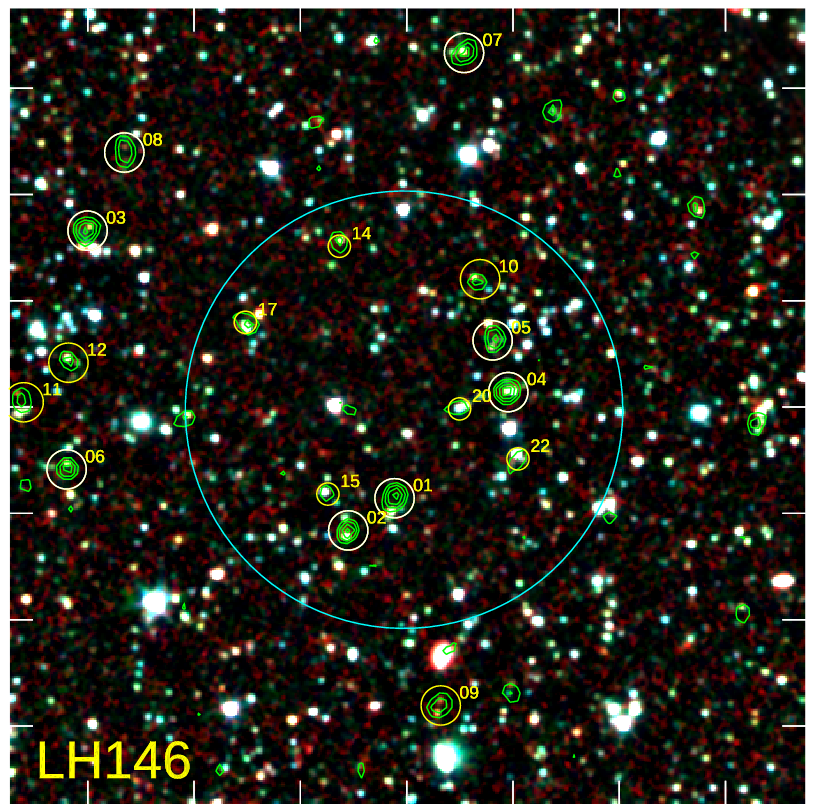, width=2.2in, angle=0}\hspace*{-0.5in}
\psfig{file=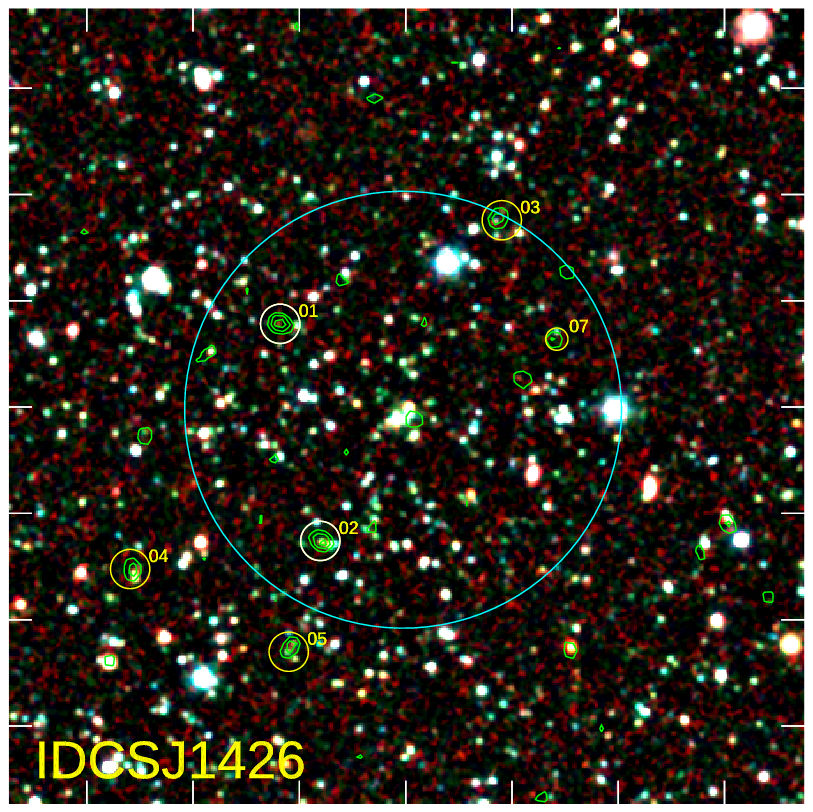, width=2.2in, angle=0}\hspace*{-0.5in}
\psfig{file=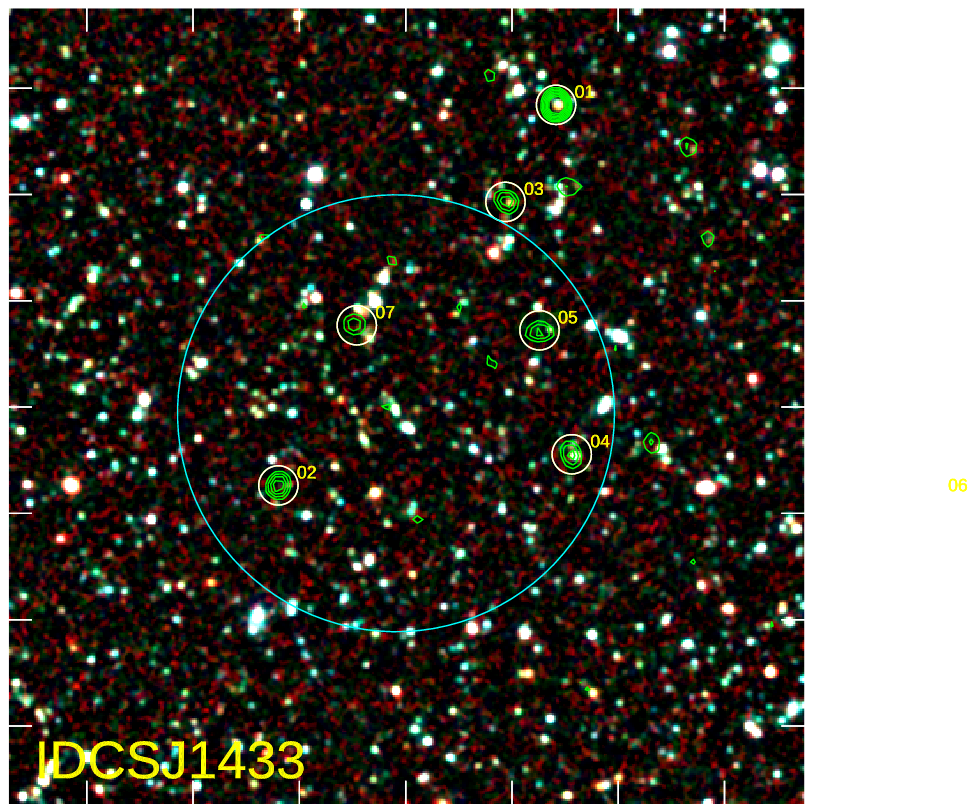, width=2.2in, angle=0}\hspace*{-0.5in}
\psfig{file=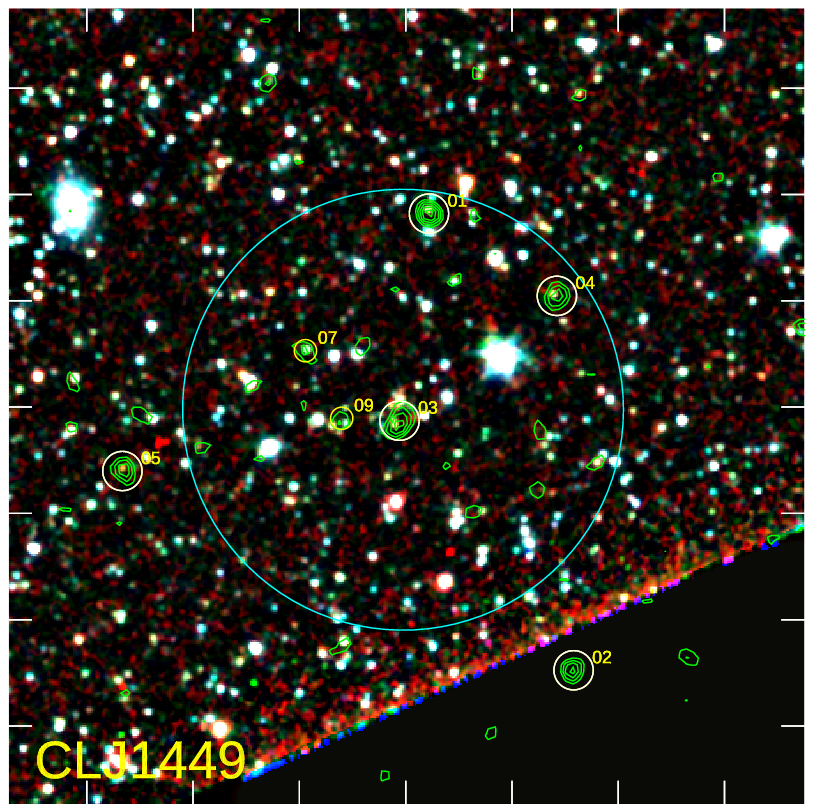, width=2.2in, angle=0}}
\caption{\small  $\sim$\,8$'$\,$\times$\,8$'$ Spitzer IRAC colour images of the eight  clusters in the sample, using 3.6\,$\mu$m as the blue channel, 4.5\,$\mu$m as green and 5.8+8.0\,$\mu$m as red.  The SCUBA-2   850-$\mu$m signal-to-noise maps are contoured over these (contours are in 1\,$\sigma$ increments starting at 2\,$\sigma$) and large circles mark the   {\sc Main} catalogue sources, yellow for those detected at 3.5--4.0\,$\sigma$, with those satisfying $\geq$\,4.0\,$\sigma$ significance shown in white.  The smaller yellow circles identify  {\sc Supplementary} catalogue sources with 3.0--3.5\, $\sigma$ significance and projected  radii from the cluster centres of $\theta$\,$<$\,2\,arcmin  ($\ls$\,1\,Mpc).  Sources are labelled by their catalogue numbers from Tables~3 and 4.  The cyan circle shows a 1.0\,Mpc radius at the cluster redshifts.   A significant variation is seen in the numbers of submillimetre sources in the central regions of the clusters.
}
\end{figure*}

The study presented here aims to assess the evolution in the star formation rate of submillimetre-selected galaxy populations within massive clusters at $z$\,$\sim$\,1.5--2.0.  By adopting a similar observational strategy and methodology to \cite{Cooke19}, the intention was to provide a homogeneous extension of that  analysis out to $z$\,$\sim$\,2 to better quantify the evolution of the star formation activity within massive clusters and potentially the processes that are driving it.      This study assumes a \cite{Chabrier03} IMF and a cosmology with $\Omega_{\rm M}$\,$=$\,0.3, $\Omega_\Lambda$\,$=$\,0.7 and $H_0$\,$=$\,70\,km\,s$^{-1}$\,Mpc$^{-1}$.   In this cosmology at the median redshift of the cluster sample, $z$\,$\sim$\,1.8 ($\sim$\,27 per cent of the current age of the Universe), 1 arcsec corresponds to 8.6\,kpc.   All quoted magnitudes are on the AB system and errors on median values are estimated using bootstrap resampling.

\section{Observations and Reduction}

The sample analysed here  comprises eight well-studied  clusters at $z$\,$\sim$\,1.6--2.0   (Table~1), which were chosen to extend the $z$\,$=$\,0.8--1.6 redshift range covered by \citet{Cooke19}.   These clusters were originally discovered either from spatially extended X-ray emission (XLSSC122, \citealt{Willis13}; LH146, \citealt{Henry14}) or as overdensities of near-infrared colour-selected galaxies (SpARCSJ0224 and SpARCSJ0225, \citealt{Nantais16}; JKCS041, \citealt{Andreon09}; IDCSJ1426 and IDCSJ1433, \citealt{Zeimann12, Brodwin16}; ClJ1449, \citealt{Gobat11}).  Several of the latter have subsequently been  confirmed as having extended X-ray emission and/or SZ detections confirming that they are  massive collapsed halos  (e.g., XLSSC122, \citealt[][]{Mantz14,vanMarrewijk23};
JKCS041, \citealt[][]{Andreon09,Andreon23}; IDCSJ1426, \citealt[][]{Brodwin12,Andreon21}; ClJ1449, \citealt[][]{Gobat11,Gobat19}).  Cluster mass estimates are given in Table~1 and were taken from the references cited in the table, as tabulated by \citet{Mei15}.  As none of the eight clusters are known to be strong lenses and this work focused on submillimetre sources that are members of the clusters,  the following analysis assumed that neither the clusters, nor the individual cluster galaxies, are  significantly gravitationally magnifying any of the  submillimetre sources.

The median redshift of the sample is $z$\,$=$\,1.77\,$\pm$\,0.08 (a cosmological age of $\sim$\,4\,Gyrs) and the median mass is $M_{200}$\,$=$\,(1.7\,$\pm$\,0.4)\,$\times$\,10$^{14}$\,M$_\odot$.   These compare to a median redshift  for the sample studied by \citet{Cooke19} of  $z$\,$=$\,1.25\,$\pm$\,0.09 (around 1\,Gyr later than the clusters in this work) and a median mass of $M_{200}$\,$=$\,(4.0\,$\pm$\,0.4)\,$\times$\,10$^{14}$\,M$_\odot$. 

\subsection{Observations} 

The eight clusters were observed with SCUBA-2  \citep{Holland13} on the 15-m JCMT simultaneously at 850\,$\mu$m and 450\,$\mu$m in typically good weather conditions suitable for sensitive 850\,$\mu$m observations ($\tau_{\rm 225GHz}$ values are reported in Table~2). Each cluster was observed for an average of $\sim$\,10\,h as a series of $\sim$\,0.5\,h integrations (Table~2) using a standard constant-velocity daisy mapping pattern.  Observations of four clusters were obtained through projects M21BP030 and M22AP039, while data for the remaining  four clusters (JKCS041, ClJ1449, IDCSJ1426 and IDCSJ1433) were taken from suitable archival SCUBA-2 programmes observed during 2012--2016 (Table~2).   It should be noted that the SCUBA-2 observations of ClJ1449 are a rereduction of those presented by \cite{Smith19} and the observations of JKCS041 are discussed in \cite{Smith20}, both of those studies also included archival Herschel observations.   The SCUBA-2 observations of the remaining six clusters have not been presented before.   

For two of the  fields  the cluster centres were revised in light of new evidence about their positions.   For XLSSC122 an updated SZ-based position was adopted from \cite{vanMarrewijk23}, which moved it south from the SCUBA-2 map centre by $\sim$\,0.5\,arcmin. For IDCSJ1433   the centroid of the spectroscopically confirmed members  from \cite{Zeimann12} was used to determine the centre, shifting this by $\sim$\,2\,arcmin east relative to the original archival SCUBA-2 pointing.  The adopted  centres for all clusters are listed in Table~1.

%
%
\begin{table*}
\caption{Log of the observations}
\begin{tabular}{lccccccl}
\hline \noalign {\smallskip}
  Cluster & $T^{\rm SCUBA-2}_{\rm exp}$ & $\sigma_{\rm 850\mu m}$ & $\sigma_{\rm 450\mu m}$ & $\tau_{\rm 225GHz}$& $S^{\rm 5\sigma}_{\rm 4.5\mu m}$& $S^{\rm 3\sigma}_{\rm 24\mu m}$& JCMT Project  \\
     & [h] & [mJy] & [mJy] & & [$\mu$Jy] & [$\mu$Jy] & \\
\hline \noalign {\smallskip}
XLSSC122         & 11.0 & 1.14 & 35 & 0.04--0.12 & 4.6 & 180 & M22AP039 \\
SpARCSJ0224   & 11.8 & 1.04 & 37 & 0.04--0.11 & 5.0 & 180 & M21BP030 \\
SpARCSJ0225   & 11.3 & 1.01 & 32 & 0.04--0.11 & 5.1 & 180 & M21BP030 \\
JKCS041           &  ~8.2 & 0.98 & 10 &  0.02--0.04 & 5.0 & 170 &  M15BI038 \\
LH146              &  13.8 & 0.98 & 33 & 0.04--0.12 &  0.7 & 20 & M22AP039 \\
IDCSJ1426       & 10.3 & 0.93 & ~5 & 0.04--0.08 & 2.0 & 70 & M12AI01, M15AI39, M15AI09 \\
IDCSJ1433       & 19.3 & 0.81 & 11 & 0.01--0.17 & 2.0 & 70 & M15AI39, M16AP087 \\
ClJ1449           &  ~7.8 & 0.99 & ~8 & 0.02--0.04 & 2.3 & 40 & M15AI51, M16AP047  \\
  \hline \noalign {\smallskip}
\end{tabular}
\end{table*}

\subsection{Data Reduction}

The new and archival SCUBA-2 observations were reduced using the Dynamical Iterative Map Maker ({\sc dimm}) within  {\sc smurf} (Submillimeter User Reduction Facility, \citealt{Chapin13}) from the 2018A EAO {\sc starlink} release, with additional software from the {\sc starlink} {\sc kappa} software package \citep{WarrenSmith93,Jenness09} to manipulate the images.     The faint point-source recipe in the {\sc orac-dr}   pipeline was used for the reduction.  A summary of the main reduction steps is given here and a detailed description of the  data reduction process with {\sc smurf} is provided by \citet{Chapin13}.
 
Firstly,  the  time-series data stored in each $\sim$ 30 minute observation were flat fielded and then  a number of cleaning steps were applied, including removing steps and spikes in the time-streams.  After  cleaning, an iterative map-making procedure fitted the data with a model comprising a common-mode signal, astronomical signal, and noise.  In this process, the pipeline estimated and removed the common-mode signal and derived the best solution to apply an extinction correction. Then several noise sources in the data were estimated and removed. These steps were repeated until the solution converged. Flux calibration was then applied to convert the reduced map into units of Janskys adopting  flux conversion factors (FCF) of  FCF$_{\rm 850\mu m}$\,$=$\,537\,$\pm$\,26\,Jy\,beam$^{-1}$\,pW$^{-1}$  and FCF$_{\rm 450\mu m}$\,$=$\,491\,$\pm$\,67\,Jy\,beam$^{-1}$\,pW$^{-1}$ (\citealt{Dempsey13}, see also \citealt{Mairs21}) and assuming $\sim$\,10 per cent systematic uncertainties.  The individual reduced observations were  combined using inverse-variance weighting to create a final map per cluster at each wavelength. To improve point source detection, the resulting  850\,$\mu$m and 450\,$\mu$m  maps were match-filtered with  15$''$  and 8$''$ FWHM Gaussian profiles, respectively. This match-filtering step introduces a small (13 per cent) loss of flux for point sources \citep[e.g.,][]{Geach17,Simpson19} and a  corresponding correction was applied to the measured fluxes (this correction is now included in the default SCUBA-2 pipeline).   Finally, combined maps were generated with  4.0$''$ pixel$^{-1}$ sampling (Nyquist sampling at both wavelengths) and cropped to a radius of 4\,arcmin (beyond which the noise increases).   This radius corresponded to approximately 2\,Mpc at $z$\,$\sim$\,1.8, the median redshift of the sample.   At 850\,$\mu$m the median noise in the centre of the maps was $\sigma_{\rm 850\mu m}$\,$=$\,1.0\,$\pm$\,0.1 mJy (Table~2).    IRAC images of the eight clusters are shown in Figure~1 with the corresponding 850-$\mu$m signal-to-noise maps contoured over each field. The clusters display a wide range of activity at 850\,$\mu$m within the central 1\,Mpc.

\subsection{Source detection}

Sources were identified in the SCUBA-2 maps using the approach described in  \citet{Simpson19} employing a simple top-down peak-finding algorithm.   This  involved detecting prominent peaks in the filtered 850-$\mu$m signal-to-noise ratio (SNR) maps  down to a minimum  threshold of 3.0\,$\sigma$  in a ``first-pass'' catalogue.  After this first detection pass,  the detected sources were subtracted from the map using an empirical PSF (if two sources lay within 40\,arcsec a double PSF model was used).   The detection step was then repeated on this  source-subtracted map  for a second pass. If additional sources were detected within 7.5\,arcsec of the first-pass sources, then these were assumed to be the same  as the first-pass sources and removed from the catalogue. Further details of the method can be found in \citet{Simpson19}.

To assess the robustness of the resulting source catalogue, the source detection was also run on ``jack-knife'' realisations of the data constructed by inverting the signal in half of the individual 30-minute observations used to construct the final cluster maps \citep[e.g.,][]{Hyun23}.  These jack-knife images had noise properties identical to the actual data, but had no flux from astrophysical sources.   This analysis indicated that the false detection rate for sources at a $>$\,3.5\,$\sigma$ significance limit was $\sim$\,8 per cent (this is  consistent with Gaussian statistics and the number of resolution elements across the eight maps) and this dropped to $\sim$\,1 per cent  for those with significance of $>$\,4.0\,$\sigma$.

An identical analysis was applied to the 450-$\mu$m maps, but due to the typically modest atmospheric transparency in the observed weather conditions  there were no significant 450-$\mu$m sources detected in the maps of XLSSC122, SpARCSJ0224, SpARCSJ0225 or LH146, reflecting the depth of the maps (Table~2).   As the goal of this study was a homogeneous analysis of the eight clusters, the 450-$\mu$m observations were therefore not considered further in this work (c.f., \citealt{Smith19,Smith20}).

As the detection significances of the  850-$\mu$m sources were modest, their measured flux densities suffer from flux boosting \citep{Coppin06}.   The boosting factor ($B$) was estimated from the ratio of output (observed) flux density and the input flux density of sources injected into the jack-knife maps. The  average boosting factor was found to be very close to the power-law form in signal-to-noise reported by \citet{Geach17}  and for consistency with that work (which was used to estimate the field source densities) and \citet{Cooke19}, who also used this relation, the following correction was applied to the measured flux densities:  $B = 1 + 0.2 \times    (\rm SNR/5)^{-2.3}$.

The  850-$\mu$m maps of the eight $z$\,$=$\,1.6--2.0 clusters yielded a total of 95 detections with  SNR\,$\geq$\,3 and projected separation from the cluster centres of $\theta$\,$\leq$\,4 arcmin.   Of these, 38 have SNR\,$\geq$\,4.0 and a further 18 have SNR\,$=$\,3.5--4.0, of which  1--2 are expected to be false positives.  A limit of SNR\,$\geq$\,3.5 was therefore adopted to construct a robust ``{\sc Main}'' sample comprising 56 sources with a false-positive rate of $\sim$\,3\,per cent.  In addition, to improve the completeness of the measurements of the star formation rate in the cluster centres,  a {\sc Supplementary} selection was also made for statistical purposes that included sources with  SNR\,$=$\,3.0--3.5, but only within a projected separation from the cluster centres of $\theta$\,$\leq$\,2\,arcmin ($\sim$\,1\,Mpc).  This included an additional 18 faint sources, of which 15 were subsequently found to be  coincident within 4\,arcsec with red IRAC counterparts (see \S3.2 below). This suggested that the majority of these sources were real as the expected false match rate was $\sim$\,1 source. This is slightly less that the false-positive rate estimated  from the jack-knife simulations which suggested $\sim$\,5 false-positive sources at the median signal-to-noise of the {\sc Supplementary} catalogue,  broadly consistent with the  3--4 expected from Gaussian statistics in this  SNR\,$=$\,3.0--3.5 subset.   The false positive rate for the 74 sources in the full {\sc Main}+{\sc Supplementary} sample is therefore expected to be $\sim$\,6\,$\pm$\,2 per cent.

%
%
\begin{figure*}
\centerline{\psfig{file=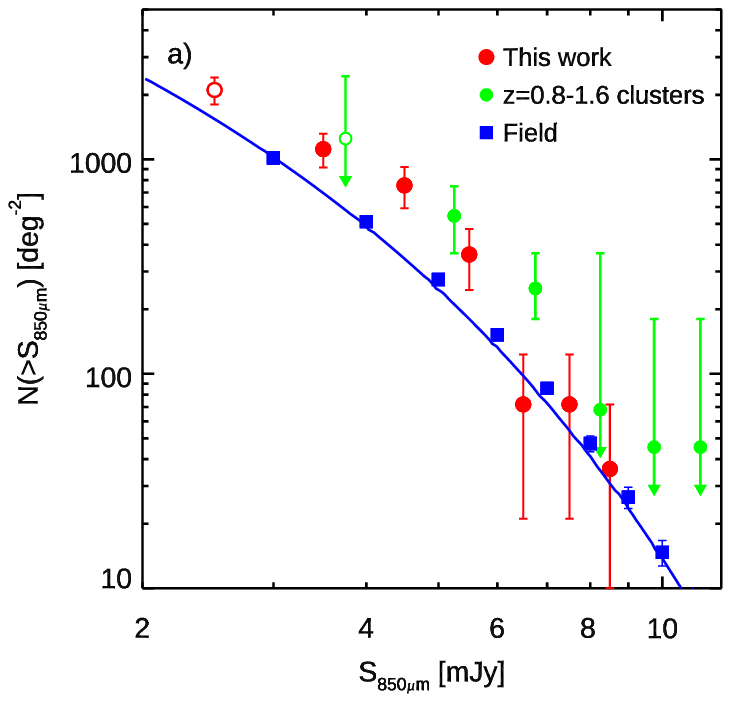, width=3.5in, angle=0} \hspace*{-0.5in}
\psfig{file=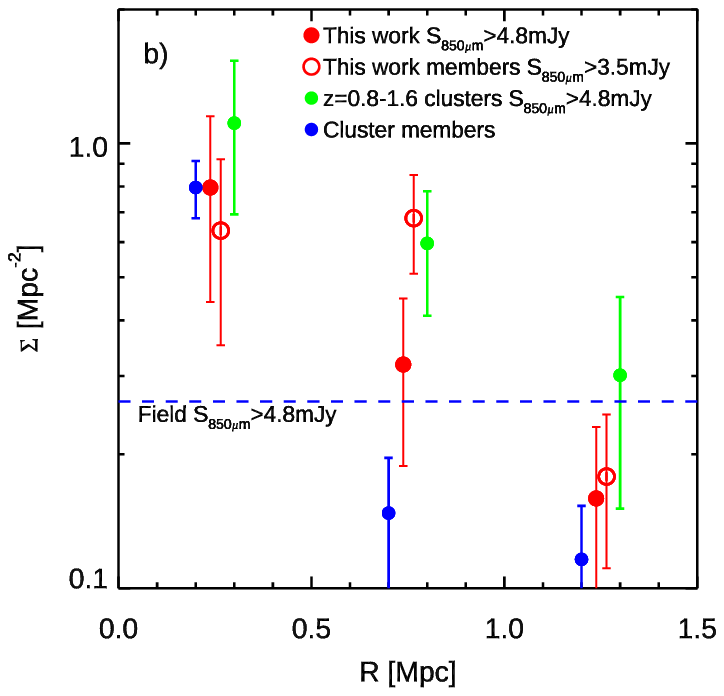, width=3.5in, angle=0}}
\caption{\small {\it  a)} Mean cumulative surface density  of SMG sources in the central $\sim$\,1\,Mpc (2\,arcmin radius) of the eight clusters in this study compared to the SCUBA-2 sources in the fields of the  $z$\,$=$\,0.8--1.6 cluster sample from   \citet{Cooke19} and the field SCUBA-2 counts from the S2CLS survey \citep{Geach17}.
Both cluster samples show moderate excesses above the field counts at 850\,$\mu$m flux densities of $S_{\rm 850\mu m}$\,$\sim$\,3--6\,mJy.
{\it b)} The mean radial density distribution  of submillimetre sources around the clusters in this study.
Two samples are shown, one is simply flux-limited at $S_{\rm 850\mu m}$\,$>$\,4.8\,mJy for comparison to
the values plotted for the lower-redshift clusters from \citet{Cooke19} above the same flux limit.   The second
shows {\sc Main} sample SCUBA-2 sources brighter than $S_{\rm 850\mu m}$\,$\gs$\,3.5\,mJy (corrected for residual field contamination as described in \S3.4) with IRAC counterparts that have colours consistent with being cluster members.   
The radial number density of all IRAC colour-selected cluster members from \S3.2  is also shown for comparison (arbitrarily normalised).  There is a clear overdensity of 850-$\mu$m sources in the central regions of the $z$\,$=$\,1.6--2.0  clusters, although this is  less significant in the raw counts than that seen in the somewhat more massive clusters at $z$\,$=$\,0.8--1.6  from \citet{Cooke19}.  Application of an IRAC colour selection to the SCUBA-2 counterparts indicates a significant overdensity, 4\,$\pm$\,1, of submillimetre sources within the central 1\,Mpc radius of the clusters.
}
\end{figure*}

The observed properties of the {\sc Main} 850\,$\mu$m sample are presented in Table~3 with the lower significance {\sc Supplementary} sample given in Table~4. The listed information is: a short identifier including the cluster name and a catalogue number for the source, peak coordinates, 850\,$\mu$m  signal-to-noise and deboosted 850\,$\mu$m flux density.

The full sample presented in this work comprises 74 850-$\mu$m sources:  56 in the {\sc Main} selection with  SNR\,$>$\,3.5 and a further 18 in the {\sc Supplementary} selection with SNR\,$=$\,3.0--3.5 and $\theta$\,$<$\,2 arcmin.
The {\sc Main} sample is effectively complete for sources with deboosted $S_{\rm 850\mu m}$\,$\gs$\,3.0\,mJy (raw, peak fluxes of $S^{\rm raw}_{\rm 850\mu m}$\,$\gs$\,4\,mJy) within $\theta$\,$\leq$\,2 arcmin, with the {\sc Supplementary} sample having deboosted fluxes of $S_{\rm 850\mu m}$\,$\sim$\,2.5\,mJy. Restricted to just the central $\theta$\,$<$\,2 arcmin of the clusters, there are a total of 48 850-$\mu$m sources: 30 from the {\sc Main} sample and  18 from the lower-significance {\sc Supplementary} sample, which contribute $\sim$\,25 per cent of the total integrated 850-$\mu$m flux density.  The median deboosted  850\,$\mu$m flux density of this sample is $S_{\rm 850\mu m}$\,$=$\,3.8\,$\pm$\,1.0\,mJy, this compares to a median of $S_{\rm 850\mu m}$\,$=$\,3.5\,$\pm$\,1.0\,mJy for the sources detected in the fields of  $z$\,$=$\,0.8--1.6 massive clusters by \citet{Cooke19}.

The mean cumulative number density of 850-$\mu$m sources in the central $\theta$\,$\leq$\,2\,arcmin regions of the eight clusters (Figure~2a)  showed a modest excess above the counts of SCUBA-2 sources in the general field from the S2CLS survey of \citet{Geach17}.  This excess is similar to that reported in the number of SCUBA-2 sources in the central regions of the $z$\,$=$\,0.8--1.6 clusters from \citet{Cooke19}.    

\subsubsection{Comparisons with previous 850$\mu$m observations}

Two of the clusters in this study,  JKCS041 and ClJ1449, were reanalyses of  SCUBA-2 observations taken and analysed by \citet{Smith20} and \citet{Smith19} respectively, although those works presented catalogues over a larger field of view than that analysed here.  In addition there are ALMA 870-$\mu$m observations of ClJ1449 in  \cite{Coogan18} that are discussed in the next section.

The source catalogues from \citet{Smith19}  and \citet{Smith20} were compared to the {\sc Main}+{\sc Supplementary} sample  from the previous section (using an 8$''$ matching radius).  This recovered ten matches  to the eleven sources from this work  in  JKCS041 and eight matches to the ten sources found in ClJ1449, with a median positional offset of 1.8$''$\,$\pm$\,0.3$''$ between the two studies.    The sources that were missing matches from \citet{Smith19}  or \citet{Smith20} were all from the {\sc Supplementary} sample:  JKCS041.009, SNR$_{850}$\,$=$\,3.3,   $S_{\rm 850\mu m}$\,$=$\,2.1$\pm$1.4\,mJy; ClJ1449.008, $_{850}$\,$=$\,3.3,  $S_{\rm 850\mu m}$\,$=$\,2.2$\pm$1.9\,mJy;  ClJ1449.009, SNR$_{850}$\,$=$\,3.2,  $S_{\rm 850\mu m}$\,$=$\,2.0$\pm$1.4\,mJy.   The omission of these sources reflected differences in the data reduction and source detection, but suggested that these only become significant for the lowest SNR sources in the {\sc Supplementary} catalogue.  For homogeneity with the other clusters, the source catalogues for JKCS041  and ClJ1449 derived in this work were used in the subsequent analysis.

\section{Analysis}

%
%
\begin{figure*}
\centerline{\psfig{file=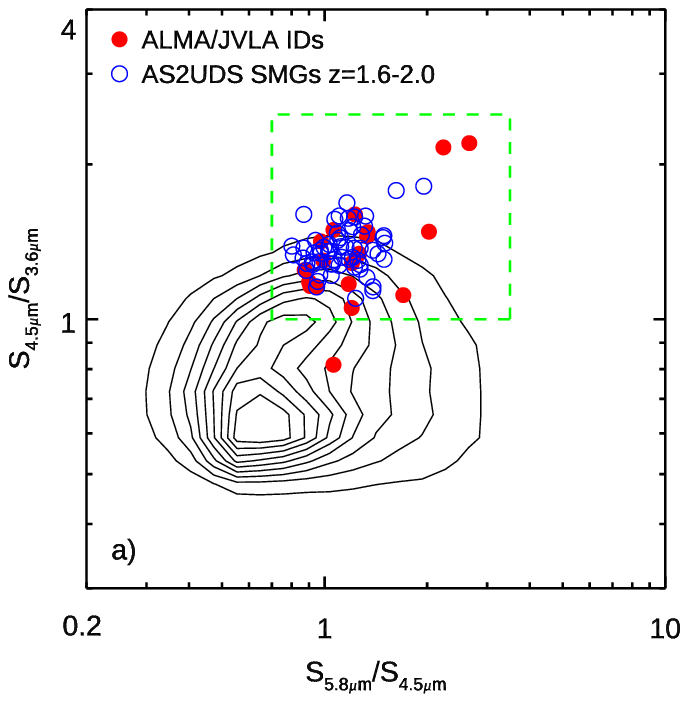, width=3.5in, angle=0} \hspace*{-0.5in}
\psfig{file=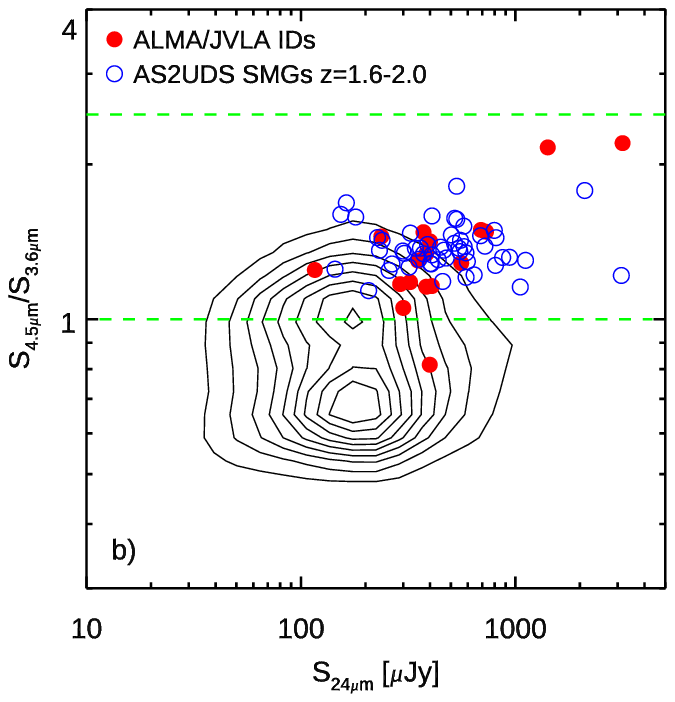, width=3.5in, angle=0}}
\caption{\small {\it  a)} IRAC 5.8/4.5\,$\mu$m versus 4.5/3.6\,$\mu$m colour--colour plot for the ALMA/JVLA interferometrically identified counterparts to the SCUBA-2 sources in the cluster sample and  a comparison sample of  ALMA-identified 850-$\mu$m selected galaxies at $z$\,$=$\,1.6--2.0 from the AS2UDS survey \citep{Dudzeviciute20}. The distribution of these two samples was used to define a colour selection, shown by the dashed lines, that  contains galaxies with colours similar to those of dusty star-forming galaxies at the redshifts of the target clusters, $z$\,$=$\,1.6--2.0.  The one JVLA identification outside of the selection box is the
  {\sc Supplementary} source counterpart LH146.015.0, which is likely to be at significantly lower redshift. The contours show the number density of the general IRAC-detected galaxy population in the cluster fields (the contours are in 0.1\,dex steps down from the peak density).   The colour selection significantly reduced the contamination from unrelated foreground galaxies,  by $\sim$\,80 per cent, in the search for counterparts to the SCUBA-2 sources.  The median photometric errors are similar to the plotted point sizes.
  {\it  b)}   MIPS 24\,$\mu$m flux density versus IRAC 4.5/3.6\,$\mu$m colour for the ALMA/JVLA interferometrically identified counterparts in the clusters and  the AS2UDS sources at $z$\,$=$\,1.6--2.0  from \citep{Dudzeviciute20}.   70 per cent of the interferometric sources have MIPS counterparts, most of which have 24-$\mu$m flux densities brighter than 100\,$\mu$Jy.  The contours now show the number density of all 24-$\mu$m detected IRAC sources in the cluster fields.
}
\end{figure*}

To assess which of the submillimetre sources detected in the eight  fields are likely to be members of the clusters required identification of the stellar counterparts to the submillimetre emission, so that the estimated redshift of the counterpart can be compared with that of the corresponding cluster (Table~1).   The modest spatial resolution of the JCMT at 850\,$\mu$m, $\sim$\,14$''$ FWHM, combined with the high dust content and typically high redshifts of submillimetre galaxies \citep[e.g.,][]{Dudzeviciute20},  complicates this identification process. Nevertheless, certain characteristics of the typical spectral-energy distributions (SEDs) of submillimetre galaxies can be employed to statistically identify possible counterparts \citep[see,][]{An18,An19}.  In addition, this process is also aided in this study because the target clusters are at lower redshifts than the bulk of the submillimetre population, $z$\,$\sim$\,2--3 \citep[e.g.,][]{Chapman05,Dudzeviciute20}, meaning that any counterparts that are cluster members will be typically brighter in the near-/mid-infrared than the background submillimetre field population.

The characteristics of submillimetre galaxies that are frequently used to identify their stellar counterparts are their relative brightness in the sub-/millimetre and radio wavebands and their typically red near-/mid-infrared colours \citep[e.g.,][]{Smail99,Frayer04,Yun08,Chen16}.     The  clusters studied in this work all have homogeneous multi-band coverage from Spitzer IRAC and MIPS \citep{Fazio04,Rieke04}, along with more heterogeneous sub-/millimetre and radio interferometric observations.  Hence, the latter were used primarily to aid in defining regions of Spitzer IRAC/MIPS flux/colour space where submillimetre-emitting cluster members were likely to be found, that could then be used to determine statistical identifications and membership of the submillimetre sources.

\subsection{Interferometric identifications}

The  ALMA archive was searched to identify any public sub-/millimetre observations of the eight clusters.  No public ALMA observations were found in JKCS041, LH146 (unsurprisingly as it is at $+$57 Declination), IDCSJ1426 or IDCSJ1433.   In the remaining four clusters a mix of band 3, 4, 6 and 7 observations were found. Analyses of the available ALMA data products in SpARCSJ0225 produced no identifications for the SCUBA-2 sources in that field. But  in SpARCSJ0224,  band 7 continuum counterparts were uncovered for three SCUBA-2 sources: SpARCSJ0224.013, SpARCSJ0224.014 and two counterparts for SpARCSJ0224.004, as well as band 3 CO(2--1) detections of all three systems, which confirmed that they were all cluster members. In ClJ1449 the search uncovered two SCUBA-2 identifications, one of these sources had already been published by \citet{Coogan18}, the new identification was for ClJ1449.009  from a band 3 continuum counterpart.  In XLSSC122  a single SCUBA-2 counterpart was identified, this source had previously been reported  by \citet{vanMarrewijk23}.   Further details are given in the notes in Tables 5 and 6.

A  search was also undertaken for  deep radio catalogues of the cluster fields.   This indicated suitably sensitive catalogues from the JVLA had been published covering  LH146 from the  Lockman Hole catalogue in \citet{Biggs06} and \citet{Ibar09} and that  JKCS041 and SpARCSJ0225 were covered by the VIDEO/XMM-LSS catalogue from \citet{Heywood20}.    As a result only  IDCSJ1426 and IDCSJ1433 lacked some interferometric coverage, although the  ALMA data is generally very sparse  in the clusters with available observations.  In total there were 20 interferometrically identified counterparts to SCUBA-2 sources over the six clusters with some observations.

%
%
\begin{figure*}
\centerline{\psfig{file=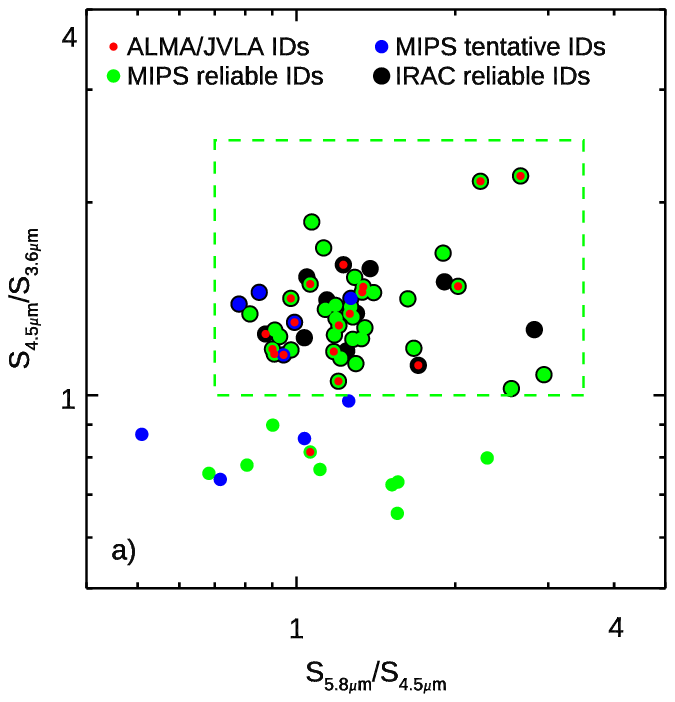, width=2.8in, angle=0} \hspace*{-0.6in}
\psfig{file=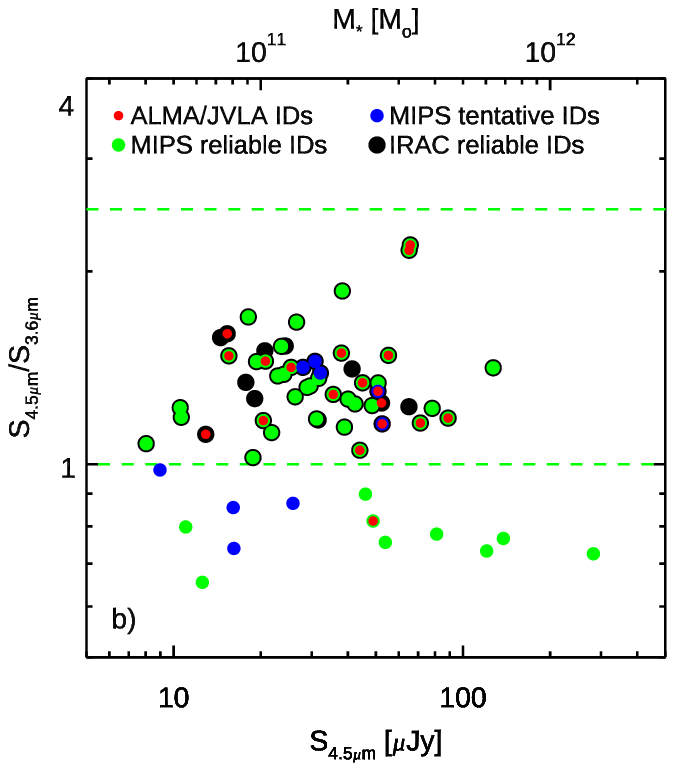, width=2.8in, angle=0} \hspace*{-0.6in}
\psfig{file=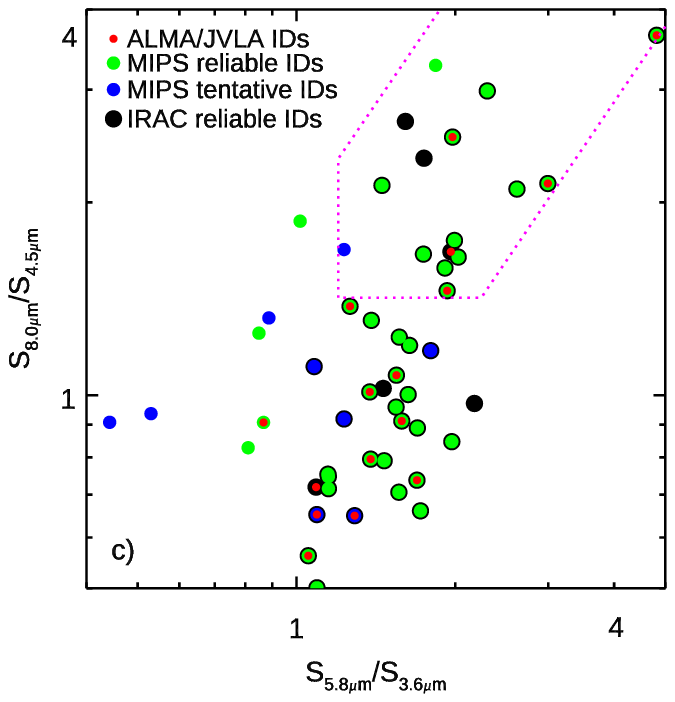, width=2.8in, angle=0}}
\caption{\small {\it  a)} IRAC 5.8/4.5\,$\mu$m versus 4.5/3.6\,$\mu$m colour--colour plot for the reliably identified submillimetre counterparts in the eight clusters.   
  {\it  b)}   IRAC 4.5\,$\mu$m flux density versus IRAC 4.5/3.6\,$\mu$m colour for the reliably identified submillimetre counterparts in the cluster fields.
 {\it  c)} IRAC 5.8/3.6\,$\mu$m versus 8.0/4.5\,$\mu$m colour--colour plot for the reliably identified submillimetre counterparts.  The magenta dotted line denotes the boundary of the AGN selection region from \citet{Donley12}.
  These panels show the effect of the application of the IRAC colour selection in removing interferometric- or MIPS-identified counterparts with   colours that were inconsistent with being cluster members (only sources marked as {\it ``IRAC reliable IDs''}  are probable  members).
}
\end{figure*}

The counterparts detected in the ALMA sub/millimetre covering the SCUBA-2 sources were assumed to be the correct
identification of the 850-$\mu$m source as these bands were tracing the same dust continuum  emission.  However, the radio emission is a more indirect tracer of the submillimetre emission and so a probabilistic analysis was employed to assess whether there were any likely radio counterparts to the submillimetre sources in the  clusters \citep[following e.g.,][]{Lilly99,Ivison02,An19,Hyun23}.  This involved a search within a radius of 6.5\,arcsec \citep{An18} of each SCUBA-2 source position and the assessment of the likelihood that any radio sources found within this radius were chance matches based on their radio fluxes and radial offset, following \cite{Downes86} and \cite{Dunlop89}.

The details of any reliably identified radio counterparts (defined as a likelihood of a random match of $<$\,5 per cent)  are given in the notes in Tables 5 and 6.  The radio  catalogue covering LH146 from the  \citet{Biggs06} and \citet{Ibar09} provided reliable radio counterparts for LH146.001 (which has no IRAC counterpart, see below), LH146.002, LH146.003, LH146.004, LH146.006, LH146.007, LH146.009, LH146.011, LH146.015 and LH146.017. Similarly the VIDEO/XMM-LSS catalogue from \citet{Heywood20}  yielded identifications for  JKCS041.004 and JKCS041.009, but no new identifications in SpARCSJ0225.   

As noted earlier, to ensure uniformity and completeness in the analysis, given the disparate and sparse interferometric coverage, those data were  used primarily to guide the identification and selection of likely submillimetre-detected cluster members from the Spitzer  IRAC and MIPS imaging that uniformly  covers all eight clusters.

%
%
\begin{figure*}
\centerline{\psfig{file=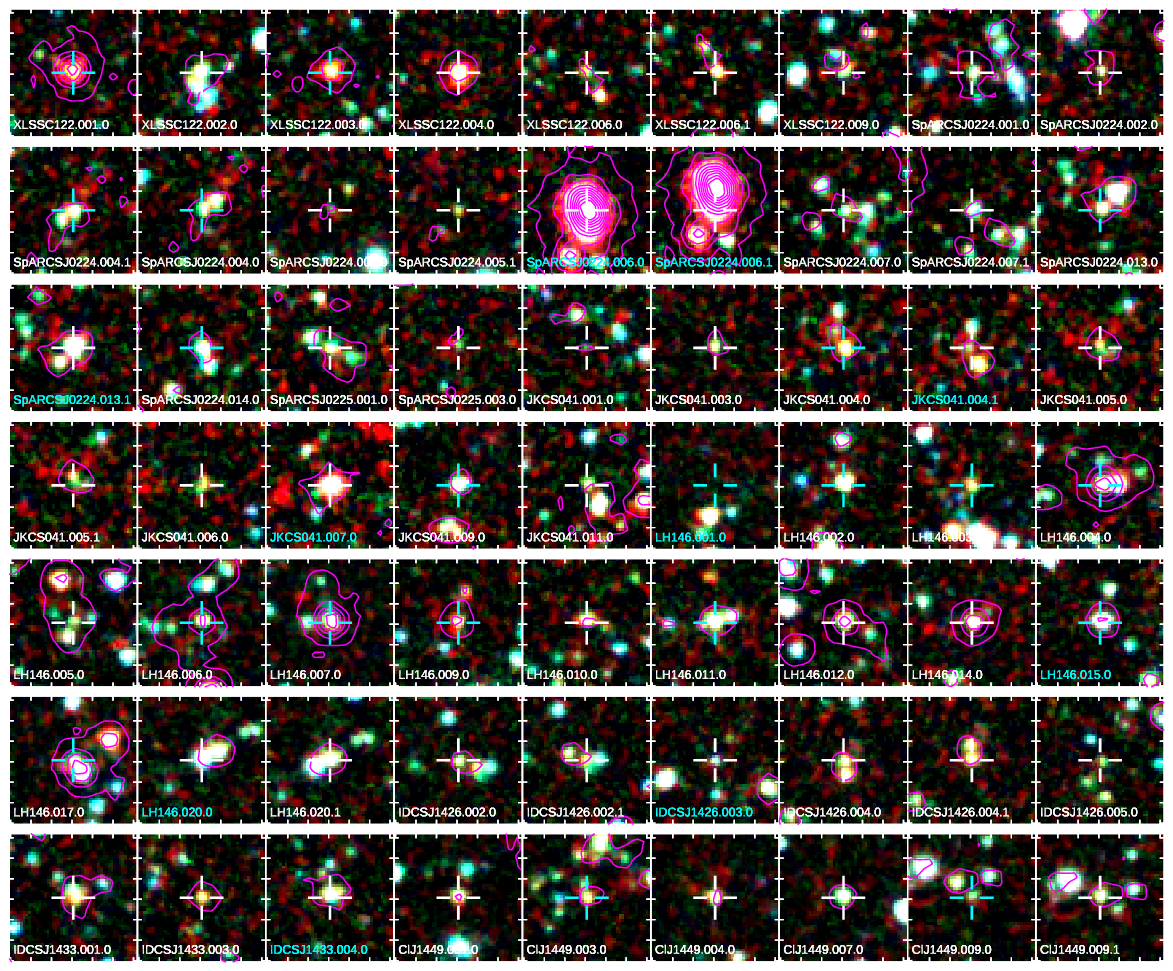, width=7.0in, angle=0}}
\caption{\small 30$''$\,$\times$\,30$''$ images of the reliably-identified IRAC counterparts to the 850-$\mu$m sources in the cluster fields.  The images comprise   IRAC 3.6\,$\mu$m (blue), 4.5\,$\mu$m (green) and 5.8+8.0\,$\mu$m (red).   The  MIPS 24\,$\mu$m emission is contoured in magenta (contours start at 2$\sigma$ in 3-$\sigma$ steps).  The crosshair marks the counterpart (the outer radius of the crosshair shows the 6.5\,arcsec search radius) and these are plotted in cyan where the identification was from interferometric observations with ALMA or JVLA. Identifications labelled in white are potential cluster members based on their IRAC colours, those labelled in cyan are likely non-members.
}
\end{figure*}

\subsection{Spitzer identifications}

The Spitzer satellite carried a powerful complement of instruments for the study of high-redshift, dust obscured galaxies \citep[e.g.,][]{Ivison02}.  These provided thermal-infrared (restframe optical/near-infrared) imaging from the IRAC camera and mid-infrared imaging from MIPS, both  at higher angular resolution ($\sim$\,2$''$ and $\sim$\,6$''$ FWHM respectively) than the SCUBA-2 850-$\mu$m maps.   The Spitzer Enhanced Imaging Products (SEIP\footnote{\url{https://irsa.ipac.caltech.edu/data/SPITZER/Enhanced/SEIP}}) provides uniformly reduced images and catalogues for sources detected with  IRAC, including matched MIPS photometry (Table~2).   SEIP-produced data products have  been used in a previous study of high-redshift clusters by \citet{Rettura18}.    The SEIP data cover all eight clusters in this study (Figure~1) and  catalogues for regions within 10 arcmin radius of the clusters were retrieved from the Spitzer Heritage Archive.  These comprised band-merged catalogues of $>$3\,$\sigma$ detections in the four IRAC channels constructed using {\sc SExtractor} (photometry in a 3.8$''$ diameter aperture, with a point-source aperture correction) along with position-matched PSF-fit photometry from the MIPS 24\,$\mu$m channel measured with {\sc mopex/apex}.  To ensure more reliable detections and photometry, the catalogues were cut to a 5\,$\sigma$ limit at 4.5\,$\mu$m (Table~2), with a median  limit of $S^{\rm 5\sigma}_{\rm 4.5\mu m}$\,$\sim$\,3\,$\mu$Jy and a median 3\,$\sigma$ limit at 24\,$\mu$m of $S^{\rm 3\sigma}_{\rm 24\mu m}$\,$\sim$\,110\,$\mu$Jy. The lowest 4.5/850\,$\mu$m and 24/850\,$\mu$m flux ratios for ALMA-identified submillimetre galaxies
at $z$\,$=$\,1.6--2.0 in \citet{Dudzeviciute20} are $S_{\rm 4.5\mu m}/S_{\rm 850\mu m}$\,$\sim$\,0.002 and $S_{\rm 24\mu m}/S_{\rm 850\mu m}$\,$\sim$\,0.05. For a submillimetre source with $S_{\rm 850\mu m}$\,$\gs$\,3\,mJy  these correspond to limits of $S_{\rm 4.5\mu m}$\,$\geq$\,6\,$\mu$Jy  and $S_{\rm 24\mu m}$\,$\geq$\,150\,$\mu$Jy  indicating that the SEIP catalogue depths (Table~2) should be sufficient to detect the majority of cluster member counterparts to the SCUBA-2 sources in these fields.  Sources in the  exterior regions of each  field were used as a control in the analysis of the corresponding cluster.

The Spitzer catalogues were  matched  to the peak positions of the SCUBA-2 sources in the full  {\sc Main}+{\sc Supplementary} sample with a matching radius of 6.5\,arcsec \citep{An18}.   This yielded 169 IRAC/MIPS sources within this radius of the 74 SCUBA-2 sources across the eight fields.  These 169 sources were then assessed to determine whether any were likely to be potential counterparts to the submillimetre sources.  Seven SCUBA-2 sources returned no IRAC/MIPS matches within 6.5\,arcsec to the 5\,$\sigma$ limit of the Spitzer catalogues, suggesting that they were likely to be high redshift and hence not cluster members (one of these has an interferometric identification: LH146.001.0).

To assess which of these IRAC/MIPS sources were potential counterparts to the SCUBA-2 sources and which were likely to be just chance alignments, the corrected-Poisson probabilistic analysis from \citet{Downes86} was employed (see also \citealt{Dunlop89}).  This started with those sources detected with MIPS at 24\,$\mu$m, as this band traces (warm) dust emission -- more closely linked to the cool dust emission seen by SCUBA-2 -- at the relevant redshifts, while the IRAC channels are predominantly measuring stellar emission.    The  probability calculation  was used to estimate $P_{\rm MIPS}$ for each source \citep{Downes86,Dunlop89}.  As there was no detectable  variation in the surface density of MIPS-detected sources as a function of projected angular radius from the cluster centres in the eight clusters, a uniform surface density was adopted in the  calculation with MIPS source counts as a function of flux derived from the surrounding control region in each cluster.\footnote{At this stage of the analysis the intention was to reliably identify as many SCUBA-2 counterparts as possible, hence an IRAC colour cut was not applied.  However, if a colour selection had been applied to the MIPS catalogue prior to this search then the MIPS-detected sources with  IRAC colours consistent with cluster membership does show a weak central concentration with a profile of $\Sigma_{\rm MIPS}$\,$\sim$\,2.5\,$\theta^{-0.6}$ (with $\Sigma_{\rm MIPS}$ in units of galaxies per arcmin$^2$ and $\theta$ in arcmin).  Using this radial density distribution in the probability calculation does not remove any reliable cluster counterparts as the  application of the IRAC colour cut also reduces the number density of MIPS sources by $\sim$\,70 per cent.}  Following \citet{Ivison07} a probability range of $P_{\rm MIPS}$\,$=$\,0--5 per cent was chosen to identify {\it ``reliable''} counterparts and $P_{\rm MIPS}$\,$=$\,5--10 per cent for {\it ``tentative''} counterparts.  This search yielded 55 MIPS counterparts to 47 SCUBA-2 sources across the eight clusters, 46 of which were classed as reliable (eight of which were pairs of possible counterparts to the same submillimetre sources) and a further nine as tentative (four of which are pairs of counterparts).    For galaxies at $z$\,$=$\,1.6--2.0 the 24\,$\mu$m MIPS filter, 20--26\,$\mu$m FWHM, covers a mix of 7.7 and 8.6\,$\mu$m PAH emission features and 9.8\,$\mu$m silicate absorption \citep[e.g.,][]{Menendez09}, making the 24\,$\mu$m flux density an uncertain tracer of star formation rate \citep[e.g.,][]{Papovich07} and so the following analysis relied on restframe $\sim$\,300\,$\mu$m luminosities from SCUBA-2 to estimate star formation rates.

The last round of identifications of potential counterparts to the SCUBA-2 sources used the  IRAC colours to both attempt to identify associations (following \citealt{Chen16,An19}) and also to determine potential cluster membership from the characteristic variation of colours with redshift.  To determine the IRAC-colour space populated by dusty star-forming galaxies at $z$\,$=$\,1.6--2.0, corresponding to the range of the cluster sample, two ``training'' samples were used.  One training sample was the interferometric identifications for sources from \S3.1 (noting that not all of these are necessarily cluster members, although the CO detections and spectroscopic redshifts confirm that several lie in their respective clusters, Tables~5 and 6).  These were supplemented  by the ALMA-identified SCUBA-2 counterparts from \citet{Dudzeviciute20} with photometric redshifts of $z$\,$=$\,1.6--2.0 derived from 22-band imaging. The  $S_{\rm 4.5\mu m}/S_{\rm 3.6\mu m}$ and $S_{\rm 5.8\mu m}/S_{\rm 4.5\mu m}$  colours of these interferometrically-identified submillimetre counterparts are shown in Figure~3. This combination of filters was used as the IRAC 4.5\,$\mu$m channel roughly covers the restframe 1.6\,$\mu$m H$^-$ opacity minimum in stellar atmospheres, and hence the corresponding SED peak, for galaxies at $z$\,$\sim$\,1.6--2.0.  On the basis of the colours of the two training samples  the following  cuts were selected to isolate sources at $z$\,$=$\,1.6--2.0: $S_{\rm 5.8\mu m}/S_{\rm 4.5\mu m}$\,$=$\,0.7--3.5 and $S_{\rm 4.5\mu m}/S_{\rm 3.6\mu m}$\,$=$\,1.0--2.5.  Figure~3 also shows the number density distribution of the general field population in the regions around the  clusters illustrating that the submillimetre galaxy counterparts have redder  colours than the majority of the (lower redshift) IRAC-detected field population in these regions \citep{Yun08,An18}. Hence the use of the colour cut   reduced the foreground contamination by $\sim$\,80 per cent in the search for potential counterparts to the SCUBA-2 sources.

The same IRAC colour-selection was also used to map the distribution of potential cluster members  with $S_{\rm 4.5\mu m}$\,$>$\,5$\mu$Jy as a function of projected angular separation from the adopted cluster centres.  The radial distribution of these galaxies showed  overdensities around the average cluster on scales out to $\sim$\,2--3 arcmin ($\sim$\,1--1.5 Mpc) as illustrated in Figure~2b.  The field-corrected density profile for these colour-selected cluster members  was fit by $\Sigma_{\rm IRAC}$\,$\sim$\,16\,$\theta^{-1.25}$ (with $\Sigma_{\rm IRAC}$ in units of galaxies per arcmin$^2$ and $\theta$ in arcmin) and this radial variation in density was corrected for in the probability calculation below (note that this correction assumes that the SCUBA-2 counterparts are a small fraction of the total IRAC-detected population at all radii).  The mean density of colour-selected sources brighter than $S_{\rm 4.5\mu m}$\,$>$\,5$\mu$Jy  was 2.76 arcmin$^{-2}$ across the fields (or roughly one per ten 6.5-arcsec radius error circles). 

The  colour selection was applied to the 169 IRAC sources found within 6.5\,arcsec radius of the 74 SCUBA-2 sources, which yielded 85 lying in the colour selection box.  A further nine IRAC sources had 3.6- and 4.5-$\mu$m detections, but 5.8-$\mu$m upper limits, where the $S_{\rm 4.5\mu m}/S_{\rm 3.6\mu m}$ colour  and $S_{\rm 5.8\mu m}/S_{\rm 4.5\mu m}$ limit would have been consistent with the  selection, however all of these limits were relatively blue with $S_{\rm 5.8\mu m}/S_{\rm 4.5\mu m}$\,$\ls$\,0.7--1.3 and so these were conservatively excluded from the analysis.    For the 85 IRAC sources with the appropriate colours, the likelihood of a source with their observed  4.5-$\mu$m flux density and radial offset from the corresponding SCUBA-2 position was calculated \citep{Downes86,Dunlop89}, classifying those with $P^{\rm mem}_{\rm IRAC}$\,$=$\,0--5 per cent as {\it ``reliable''} counterparts and $P^{\rm mem}_{\rm IRAC}$\,$=$\,5--10 per cent as {\it ``tentative''} counterparts.    This produced 40 reliable IRAC counterparts and a further 18 tentative counterparts to a total of 46 SCUBA-2 sources.

\subsection{Final combined identifications}

The final step to identify the galaxy counterparts to the SCUBA-2 sources was to combine the various identifications to provide firstly a list of reliably identified counterparts and to then assess which of these were potential cluster members. The following order was used to determine which identifications would be adopted:  firstly the 20 sources with interferometric identifications were taken as correct, to these were added the 44 sources with reliable MIPS counterparts -- together this yielded 52 unique identifications.  Then two sources with tentative MIPS identifications and reliable/tentative colour-selected IRAC counterparts were included (all the remaining tentative MIPS identifications  had blue IRAC colours and were thus likely to be foreground galaxies), and then finally nine sources with reliable IRAC colour identifications that had not otherwise been  selected were included.   This resulted in 63 reliable counterparts  to 52 SCUBA-2 sources, including 12 submillimetre sources that have two counterparts.    The IRAC colours and fluxes of these reliably-identified counterparts are plotted in Figure 4.  These counterparts were given identifiers to their corresponding SCUBA-2 source with the priority in the numbering  (.0 being the most reliable) increasing from:  interferometric, reliable MIPS identification, tentative MIPS, reliable IRAC, tentative IRAC  and then decreasing 4.5$\mu$Jy flux density.  Figure~4c also shows the IRAC 5.8/3.6\,$\mu$m versus 8.0/4.5\,$\mu$m  colour--colour distribution of the galaxies and the region of this colour space from \citet{Donley12} where potential AGN host galaxies are expected to fall.   16 of the 63 reliably identified counterparts (25\,$\pm$\,6 per cent) were flagged as potential AGN hosts (15 of the 16 potential AGN which were classed as possible members from their IRAC colours, giving a rate of 28\,$\pm$\,6 per cent).  These estimates are just consistent with the upper limit on the AGN fraction  in ALMA-identified submillimetre galaxies in the $z$\,$\ls$\,3 field   of $\ls$\,28 per cent from \citet{Stach19}, potentially allowing for some modest AGN excess in the clusters \citep{Alberts16}.     Figure~5 shows a three-colour IRAC representation of the 63 reliable submillimetre counterparts in the cluster fields with any associated MIPS 24\,$\mu$m emission indicated by contours.

Owing to the different identification criteria used, these 63 reliable counterparts are expected to comprise a mix of cluster and field and are not ``complete'' in a formal sense.  To isolate the probable cluster members from this list the IRAC colour selection was then applied to the interferometric/MIPS identifications to remove any remaining probable foreground or background sources as shown in Figure~4 (noting that any counterparts that had used IRAC colour  as part of their selection already complied with this requirement).    This removed two interferometric identifications: LH146.001.0 (which had no IRAC counterpart) and LH146.015.0, as well as a further eight counterparts with reliable MIPS identifications but blue IRAC colours (Figure~4).  This reduced the total sample of reliable counterparts  which have IRAC colours consistent with $z$\,$=$\,1.6--2.0  to  53 galaxies matched to 45 SCUBA-2 sources (including eight submillimetre sources with pairs of counterparts) and these probable member galaxies are identified in Figure~4 as {\it ``IRAC reliable IDs''}.   While this process reduced the contamination from higher redshift (and the small number of lower redshift) submillimetre sources along the line of sight to the clusters, it is not expected to  completely remove all non-cluster sources, so an additional  correction was applied to account for this residual contamination as described in \S3.4.

The final list of reliably identified submillimetre  counterparts are reported in Tables~5 and 6.  The tables  include the position of the counterpart (either from the interferometric or the IRAC counterpart), the IRAC 3.6, 4.5 and 5.8\,$\mu$m flux densities, the MIPS 24\,$\mu$m flux density where detected, the counterpart's offset from the SCUBA-2 position and the corrected Poisson probabilities (as per cent) for the counterpart identification with MIPS or IRAC colour selection (these are in bold font where the identification is reliable).   The footnotes to the tables give more information about any interferometric identifications.  Sources with IDs in bold are possible cluster members lying within the central 1\,Mpc radius, while those with IDs in italics have redshifts or IRAC colours that are inconsistent with being cluster members.   Sources identified as potential AGN using the IRAC photometric classification of \citet{Donley12} shown in Figure~4c are flagged in Table~5.   The number of reliably identified submillimetre members within the central 1-Mpc radius of the clusters comprised: XLSSC122, 7/6  (the first value is the number of counterparts, the second the number of distinct SCUBA-2 sources they correspond to);  SpARCSJ0224, 7/5; SpARCSJ0225, 1/1; JKCS041, 5/5; LH146, 7/7; IDCSJ1426, 2/1; IDCSJ1433, 0/0; ClJ1449, 6/5.

These reliably identified submillimetre counterparts have accurate positions from either the interferometric observations or their IRAC counterparts and these positions were used to determine which galaxies had  Hubble Space Telescope (HST) imaging from the HST archive.   From the 63 reliably identified counterparts, 23 had useable multi-band HST imaging, typically from WFC3.   These consisted of two sources in XLSSC122, three in JKCS041, two in IDCSJ1426 and four in ClJ1449 all with WFC3 F105W and F140W imaging; six in SpARCSJ0224 with WFC3 F105W, F140W and F160W imaging; and six counterparts in LH146 with WFPC2 imaging in F606W and F814W.    Figure~6 shows the HST imaging for these sources using the available filters.

%
%
\begin{figure*}
\centerline{\psfig{file=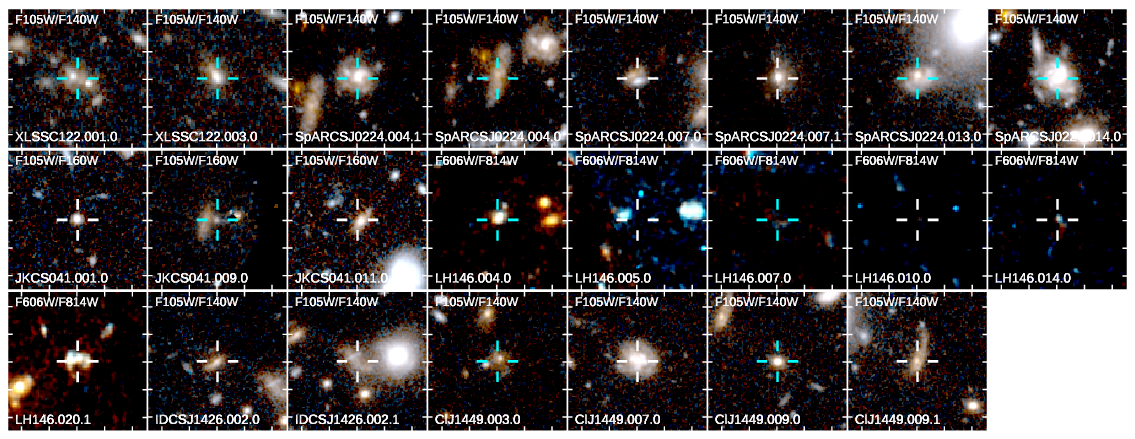, height=6.5in, angle=90}}
\caption{\small 10$''$\,$\times$\,10$''$ log-scaled images of the archival HST observations for the reliably identified  850-$\mu$m counterparts that are potential cluster members in the eight  fields.  The filter combinations used in the images are labelled, the majority of the fields have WFC3 imaging in the F105W and F140W filters.    The crosshair marks the counterpart and these are plotted in cyan where the identification was from interferometric observations with ALMA or JVLA.  XLSSC122.001.0, XLSSC122.003 and ClJ1449.003.0 were classed as potential AGN hosts based on their IRAC colours (see Figure~4c).  However, there are no bright point sources visible suggesting that if present, the AGN must be moderately obscured.
}
\end{figure*}

\subsection{Field contamination}

As the assessment of cluster membership for the submillimetre counterparts relied on a coarse IRAC colour selection, it was necessary to assess the likely contamination from any residual, unrelated (``field'') submillimetre galaxies. This could be achieved by exploiting the analysis of the large S2CLS sample of submillimetre sources in the well studied UKIDSS UDS field \citep{Geach17,Stach19,Dudzeviciute20}. By replicating the same selection of counterparts as used in the cluster fields and then assessing the number density of selected sources, it was possible to determine the likely level of contamination in the cluster fields arising from unrelated field sources, as well as the reliability of the identified counterparts using the ALMA follow up of the complete SCUBA-2 sample in this field from \citet{Stach19}.

To ensure that the  analysis was as close as possible to that applied to the cluster fields, the SEIP catalogue of Spitzer IRAC/MIPS sources was retrieved for the $\sim$\,0.9 degree$^2$ UDS field.   Uniform Spitzer coverage ($S^{5\sigma}_{\rm 4.5mu m}$\,$=$\,0.7\,$\mu$Jy) was available for a total area of 0.48 degree$^2$  containing 570 SCUBA-2 850\,$\mu$m sources from \citet{Geach17}.    A search was undertaken  in 6.5\,arcsec radius areas around these for either MIPS or IRAC sources with the same IRAC colour selection as that used in the clusters.   Across the 1720-arcmin$^2$ area 69 SCUBA-2 sources brighter than the $S_{\rm 850\mu m}$\,$=$\,3.5\,mJy catalogue limit were found with  one or more Spitzer counterparts consistent with the cluster member selection. This compared to 35 brighter than $S_{\rm 850\mu m}$\,$\sim$\,3.5\,mJy  across 308 arcmin$^2$ in the eight clusters. Scaled to the full cluster survey area, the  analysis of the UDS sample would predict 12.3\,$\pm$\,1.5 sources, or 34\,$\pm$\,4 per cent contamination.    However, when restricted to the central 1\,Mpc of the clusters ($\sim$\,100\,arcmin$^2$ total area) there were 20 $S_{\rm 850\mu m}$\,$\geq$\,3.5\,mJy SCUBA-2 sources  in the cluster fields with reliable Spitzer counterparts with cluster member colours, where the UDS sample  predicted 4.0\,$\pm$\,0.5, or 20\,$\pm$\,3 per cent contamination.   The integrated star formation rate in the clusters measured in the next section  was corrected for this estimated contamination.

In terms of reliability, in the analysed S2CLS UDS sample 50 of the 69 unique SCUBA-2 sources brighter than $S_{\rm 850\mu m}$\,$=$\,3.5\,mJy had an ALMA-detected counterpart in the complete AS2UDS follow-up survey undertaken by \citet{Stach19} that matched those selected from the SEIP MIPS/IRAC catalogues, corresponding to 72 per cent \citep[c.f.,][]{Hodge13,An18}.

\subsection{Scaling relations for physical properties}

The availability of the $\sim$\,700 ALMA-identified counterparts to field submillimetre galaxies from AS2UDS
\citep{Stach19}  also allowed the leveraging of the more extensive 22-band photometric coverage in that field to provide rough transformations between the observed properties of submillimetre galaxies and key physical quantities such as stellar mass or star formation rate.     Using the {\sc magphys}-based SED analysis of the ALMA-identified submillimetre sample in the UDS from \citet{Dudzeviciute20},  correlations were derived between the observed SCUBA-2 850-$\mu$m flux density and the estimated star formation rate for the submillimetre galaxies at $z$\,$=$\,1.6--2.0:  $\log_{10}(\rm SFR)$\,$=$\,$(0.75 \pm 0.12) \times \log_{10}(S_{\rm 850\mu m})+(1.88 \pm 0.06)$ with a 0.25\,dex dispersion (primarily reflecting the variation in the far-infrared SEDs in the population),  with units of M$_\odot$\,yr$^{-1}$ for SFR and mJy for $S_{\rm 850\mu m}$.\footnote{The median  ratio of 8--1000\,$\mu$m luminosity to star formation rate was  $ L_{\rm IR}/{\rm SFR}$\,$=$\, $(1.30\pm 0.07)\times 10^{10}$\,L$_\odot$\,yr\,M$_\odot^{-1}$  for sources at $z$\,$=$\,1.6--2.0 from \citet{Dudzeviciute20}.} A similar fit by \cite{Cooke19} for the AS2UDS sample in the $z$\,$=$\,0.8--1.6  redshift range gave: $\log_{10}(\rm SFR)$\,$=$\,$(0.87 \pm 0.06) \times \log_{10}(S_{\rm 850\mu m})+(1.85 \pm 0.04)$.  For consistency with the \cite{Cooke19} results, the normalisation was fixed to that from their fit, this gave a best-fit $z$\,$=$\,1.6--2.0  relation of: $\log_{10}(\rm SFR)$\,$=$\,$(0.81 \pm 0.06) \times \log_{10}(S_{\rm 850\mu m})+(1.85 \pm 0.04)$.   The latter fit was used to estimate the probable star formation rates of the submillimetre galaxies in this cluster sample. 

A similar analysis was undertaken to relate the observed 4.5-$\mu$m flux density of the submillimetre galaxies to their stellar masses from the  {\sc magphys} analysis in \cite{Dudzeviciute20},  taking advantage of the fact that  the IRAC 4.5\,$\mu$m channel samples the SEDs of cluster members close to restframe $\sim$\,1.6\,$\mu$m.  This gave a median scaling of $M_\ast$/$S_{\rm 4.5\mu m}$\,$=$\,10$^{9.8}$\,M$_\odot$\,$\mu$Jy$^{-1}$ with a 0.3\,dex dispersion for submillimetre galaxies at $z$\,$=$\,1.6--2.0
(scaled to a median redshift of $z$\,$=$\,1.8) and $M_\ast$/$S_{\rm 3.6\mu m}$\,$=$\,10$^{9.4}$\,M$_\odot$\,$\mu$Jy$^{-1}$ with a 0.2\,dex dispersion at  $z$\,$=$\,0.8--1.6 (corresponding to the redshift range of the clusters in \citealt{Cooke19}). The typical flux limit for the IRAC catalogues, $S_{\rm 4.5\mu m}$\,$\sim$\,3\,$\mu$Jy, then corresponded to a mass limit of $\sim$\,2\,$\times$\,10$^{10}$\,M$_\odot$.

For consistency with the analysis undertaken in \cite{Cooke19}, the Herschel SPIRE observations of the clusters in this work were not included in the analysis, even though those data may have improved the constraints on the far-infrared luminosities of the sources.   The modest resolution of the SPIRE maps, 18--36$''$ FWHM, and the lack of robust interferometric identifications for the majority of the 850-$\mu$m counterparts \cite[c.f.,][]{Swinbank14,Dudzeviciute20} meant that the  complication of deblending the emission from potentially several contributing sources across all the cluster fields was judged to be unwarranted (c.f., \citealt{Smith19}). Nevertheless, a simple consistency check was undertaken using the HELP/HerMES {\sc xid+} deblended SPIRE 250\,$\mu$m photometry \citep{Roseboom10,Hurley17} based on MIPS 24\,$\mu$m priors.   The HELP database provided  matches to 21 reliably identified cluster members in XLSSC122, SpARCSJ0224, JKCS041, IDCSJ1426 and LH146 (the other clusters were not available).   A scaling relation was derived between observed SPIRE 250\,$\mu$m flux density and far-infrared luminosity, $L_{\rm IR}$, and thence to star formation rate for the ALMA-identified submillimetre galaxies at $z$\,$=$\,1.6--2.0 in \cite{Dudzeviciute20}, which yielded:  SFR/$S_{\rm 250\mu m}$\,$=$\,7.0\,$\pm$\,3.0\,M$_\odot$\,yr$^{-1}$\,mJy$^{-1}$.  When applied to the deblended {\sc xid+} 250-$\mu$m flux densities the ratio of the predicted star formation rates to those derived from the 850-$\mu$m observations was SFR$_{\rm 850\mu m}$/SFR$_{\rm 250\mu m}$\,=\,1.1\,$\pm$\,0.3, indicating that the two estimates were in reasonable agreement.

\section{Results \& Discussion}

\subsection{850$\mu$m overdensities}

Figure~1 shows three-colour IRAC images of the eight clusters with the SCUBA-2 850\,$\mu$m signal-to-noise maps overlaid as contours and the {\sc Main} and {\sc Supplementary} catalogue sources identified.  Potential cluster members at $z$\,$=$\,1.6--2.0 with SEDs that peak at restframe 1.6\,$\mu$m would be brightest in the 4.5-$\mu$m IRAC filter that is shown as the ``green'' channel.  There are clear concentrations of  galaxies with  colours consistent with cluster membership in the central regions of several of the fields. However, focusing on the central 1\,Mpc radius of the clusters, there is also a considerable dispersion in the numbers of detected SCUBA-2 sources: ranging from one  in SpARCSJ0225 to ten in LH146 and a median of 7.0\,$\pm$\,1.9 {\sc Main}+{\sc Supplementary} sources per cluster core. 

Figure~2 quantifies the significance of the raw 850-$\mu$m overdensities in these cluster fields in two ways.  Figure~2a shows the mean cumulative surface density of sources as a function of 850-$\mu$m flux density in the central 2\,arcmin radius ($\sim$\,1\,Mpc) regions of the  clusters compared to that expected in a blank field (from S2CLS, \citealt{Geach17}).  While Figure~2b shows the variation in the mean surface density of sources brighter than $S_{\rm 850\mu m}$\,$=$\,4.8\,mJy (for consistency with the measurements from \citealt{Cooke19}) as a function of radius in the eight clusters.  Both plots indicate modest overdensities of 850-$\mu$m sources in the central $\sim$\,0.5--1\,Mpc of the clusters at flux densities around  $S_{\rm 850\mu m}$\,$\sim$\,3--6\,mJy.   The  overdensity of 850-$\mu$m selected sources is a factor of 1.5\,$\pm$\,0.3  in the central 1\,Mpc radius  brighter than $S_{\rm 850\mu m}$\,$=$\,4.0\,mJy (Figure~2a) and 3.0\,$\pm$\,1.4  in the central 0.5\,Mpc radius  brighter than $S_{\rm 850\mu m}$\,$=$\,4.8\,mJy  (Figure~2b).    The significances of these overdensities are slightly lower than those reported for the similar sized sample of clusters at $z$\,$=$\,0.8--1.6 in \citet{Cooke19}.    However, after the application of the  colour cut to identify submillimetre counterparts with IRAC colours consistent with $z$\,$=$\,1.6--2.0,   a significant mean overdensity  of a factor of 4\,$\pm$\,1 is seen out to $\sim$\,1\,Mpc radius in the clusters for sources brighter than $S_{\rm 850\mu m}$\,$=$\,3.5\,mJy (Figure~2b).

\subsection{850$\mu$m galaxy properties}

The Spitzer and HST imaging of the eight clusters can provide useful insights into the properties of the likely submillimetre cluster members, including  key characteristics such as their stellar masses,  potential triggers for their strong star formation  and evidence of their local environments.  It should be stressed that these identifications are statistical in nature and while the overdensities of SCUBA-2 sources are robust, it may be that individual source identifications are either incorrect, or if the counterpart is correct,  it is not a member of the cluster.

Figure~5 shows $\sim$\,250-kpc regions from the IRAC imaging centered on the reliably identified counterparts to the SCUBA-2 sources.  The colour scheme is the same as was used in Figure~1 (with the MIPS 24\,$\mu$m emission now shown as contours), so that galaxies that appear green (those with SEDs peaking at observed wavelengths of $\sim$\,4.5\,$\mu$m) are possible cluster members, while galaxies appearing blue or red are likely to be in the foreground or background respectively.    This figure illustrates that 18\,$\pm$\,6 per cent  of the  SCUBA-2 sources have multiple counterparts.   This rate is similar to that reported from ALMA identification of SCUBA-2 sources at comparable flux densities by \cite{Stach18}, suggesting little variation in the proportion of submillimetre-bright galaxies with a second submillimetre-bright source within $\sim$\,10$''$ ($\sim$\,100\,kpc in projection)  between these clusters and the field (see also \citealt{Ivison07,Hodge13,Miettinen15,Simpson20,Shim22,Hyun23}).

The low angular resolution of the IRAC imaging means it is difficult to assess whether these multiple counterparts are  physically associated and so could be interacting with each other, or indeed if other nearby galaxies may be responsible for triggering the active star formation in these galaxies.  However, in addition to the IRAC imaging, around a third of the cluster member counterparts to the 850-$\mu$m sources had archival HST imaging.  This provides much higher spatial resolution information about the sources, FWHM of $\sim$\,0.15$''$ or $\sim$\,1\,kpc,   and this is shown in Figure~6.  The majority of this imaging was taken with WFC3 in the F105W and F140W (or F160W) filters (LH146 was the outlier with only bluer and shallower WFPC2 restframe UV F606W and F814W imaging).

%
%
\begin{figure}
\centerline{\psfig{file=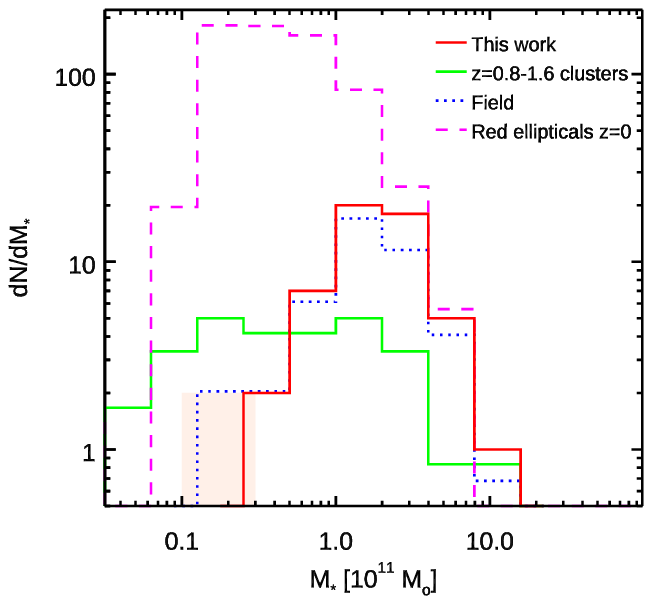, width=3.5in,angle=0}}
\caption{\small The distributions of estimated stellar mass for the IRAC counterparts to the SCUBA-2 cluster sources in this study.  These are plotted along with similar estimates for the IRAC counterparts in the $z$\,$=$\,0.8--1.6 clusters from \citet{Cooke19} and to a ``field'' sample comprising  ALMA-identified SCUBA-2 counterparts at $z$\,$=$\,1.6--2.0 from \citet{Dudzeviciute20}, where the stellar masses were  calculated in the same manner as those in the clusters.   These distributions are also compared to the   $z$\,$=$\,0  morphologically classified red elliptical galaxies from the PM2GC survey \citep{Calvi11,Calvi12}, with an arbitrary normalisation.
The submillimetre galaxies identified in the $z$\,$=$\,1.6--2.0 clusters (some of which already harbour populations of apparently quiescent massive galaxies, e.g., \citealt{Newman14,Andreon14,Nantais16}) are comparable in stellar mass to those at the same redshift in the field and correspond to the most massive galaxies seen in clusters today.  The flux limit of the IRAC catalogues in the cluster fields imposes  a minimum mass limit on the counterparts indicated by the shaded region, this may explain some of the differences seen in the stellar mass distributions of the submillimetre galaxies in the $z$\,$=$\,0.8--1.6  and $z$\,$=$\,1.6--2.0 clusters.
}
\end{figure}

The morphologies of these galaxies can be compared to those for similar WFC3 imaging of ALMA-identified submillimetre field galaxies from the  ALESS \citep{Chen15}, SUPERGOODS \citep{Cowie18} and AS2UDS \citep{Stach19} surveys. Perhaps unsurprisingly the HST imaging of the cluster submillimetre sources typically show more galaxies in their local environment  ($\sim$\,10$''$, $\sim$\,100\,kpc in projection).   As also expected they  appear brighter on average in the observed $H$-band than the field population which extends to much higher redshifts.

Focusing on the WFC3 imaging, it appears that around half of the cluster systems may  have potentially associated companions (either a second source within $\sim$\,1--2$''$ or a component within the galaxy,  although the clear evidence for  interactions between galaxies is not strong (e.g., IDCSJ1426.002.1, ClJ1449.009.1).  This indicates that the majority of the submillimetre galaxies in these clusters are not obvious major mergers based on WFC3 imaging.  \citet{Delahaye17} have similarly suggested there is no clear excess of mergers in the general galaxy population in SpARCSJ0224 or SpARCSJ0225, compared to the rates in the field, while \citet{Coogan18} have suggested that there is enhanced merging contributing to the strongly star-forming population in the core of ClJ1449 (see also \citealt[][]{Watson19}).  However, JWST has shown that care is needed when interpreting restframe UV/optical morphologies of dust-obscured sources at high redshifts \citep[e.g.,][]{Chen22,Cheng23,Smail23}.  Hence the available HST imaging needs to be viewed with caution, especially the optical WFPC2 data, but  even the WFC3 near-infrared imaging provides only restframe $V$-band coverage for cluster members in this work, which are typically expected to suffer average $V$-band extinctions (for stellar populations detectable in the restframe $K$-band) of $A_V$\,$\sim$\,3 \citep{Dudzeviciute20}.   With a spatially inhomogeneous distribution within the galaxies  this extinction can potentially create ``false'' components within galaxies or spurious, apparently disturbed, morphologies.

%
%
\begin{figure*}
\centerline{\psfig{file=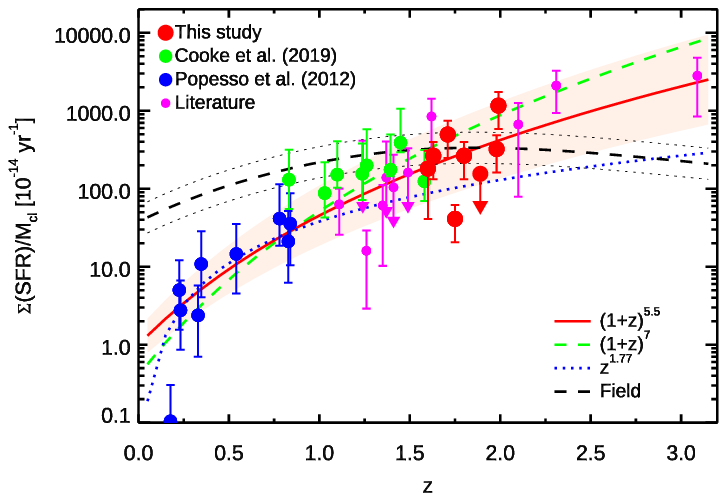, height=4.0in, angle=0}}
\caption{\small The variation in the mass-normalised star formation rate for the $z$\,$=$\,1.6--2.0 clusters from this study (corrected for residual field contamination) compared to clusters from other similar far-infrared/submillimetre studies in the literature  (the \citet{Popesso12} measurements have been corrected to the same luminosity limit and typical cluster mass as the higher redshift observations).  A selection of power-law models are also shown (all normalised at $z$\,$=$\,0.8), as well as  the evolution in the halo mass normalised star formation rate  for the field population from \citet{Madau14} with the  normalisation described in \S4.3, including an offset of $\sim$\,0.2\,dex (and uncertainty of $\pm$\,0.2\,dex shown by the thin dotted lines) to account for baryons not bound in halos \citep{Popesso15b,Faltenbacher10}.  The best-fit trend to the mass-normalised star formation rate in massive clusters  is  $\Sigma_{\rm SFR}/M_{\rm cl} \propto (1+z)^{5.5\pm 0.6}$ (the shaded region shows the uncertainty). This agrees with the results from \citet{Webb13} who found $(1+z)^{5.4\pm 1.9}$ using MIPS 24\,$\mu$m observations of massive clusters out to $z$\,$\sim$\,1.  The best-fit cluster trend intercepts the expected mass-normalised star formation rate for halos in the surrounding field at $z$\,$\sim$\,1.8\,$\pm$\,0.2.} 
\end{figure*}

The star formation rates  for the individual submillimetre-detected cluster members in the central 1-Mpc radius were estimated from their deboosted 850-$\mu$m fluxes using the relation derived in \S3.5 (see also \citealt{Dudzeviciute20}) which indicates a range of SFR\,$=$\,130--350\,M$_\odot$\,yr$^{-1}$ and a median of 220\,$\pm$\,60\,M$_\odot$\,yr$^{-1}$, but noting that this includes five examples of pairs of potential counterparts to individual SCUBA-2 sources,  hence  some of these counterparts will have lower star formation rates.   The submillimetre sources in the equivalent \citet{Cooke19} sample  have an identical median of SFR\,$=$\,210\,$\pm$\,40\,M$_{\odot}$\,yr$^{-1}$ using the $z$\,$=$\,0.8--1.6  conversion, and the faintest sources in both samples have SFR\,$\sim$\,100\,M$_{\odot}$\,yr$^{-1}$ or $L_{\rm IR}$\,$\sim$\,$1\times 10^{12}$\,L$_\odot$.
Converted to star formation rate density the radial number density profile in Figure~2b
indicates a mean star formation rate density of $\sim$\,120\,$\pm$\,25\,M$_{\odot}$\,yr$^{-1}$\,Mpc$^{-2}$ within the
central 1\,Mpc radius  of the clusters, compared to $\sim$\,50\,$\pm$\,20\,M$_{\odot}$\,yr$^{-1}$\,Mpc$^{-2}$
at 1--1.5\,Mpc.  However, normalising these densities using  the  profile of colour-selected cluster members shows a marginally lower level of average activity at $<$\,1\,Mpc compared to 1--1.5\,Mpc.

The stellar masses of the submillimetre cluster members were also estimated from the corresponding scaling relations from \S3.5. The median stellar masses for the reliably-identified members was $M_\ast$\,$=$\,(1.9\,$\pm$\,0.2)\,$\times$\,10$^{11}$\,M$_\odot$
(with a full range of 0.5--10\,$\times$\,10$^{11}$\,M$_\odot$) and this is shown in Figure~7.  Removing the potential AGN hosts from the sample did not change the median mass.

The median mass of the submillimetre-detected $z$\,$=$\,1.6--2.0 cluster members is similar to that estimated for the co-eval submillimetre field population taken from the AS2UDS survey of \cite{Dudzeviciute20}: $M_\ast$\,$=$\,(1.6\,$\pm$\,0.2)\,$\times$\,10$^{11}$\,M$_\odot$. However, the median stellar mass of the cluster population  is higher than the estimate of  $M_\ast$\,$=$\,(0.7\,$\pm$\,0.3)\,$\times$\,10$^{11}$\,M$_\odot$ for the likely cluster IRAC counterparts at $z$\,$=$\,0.8--1.6 from \cite{Cooke19}.  This estimate used  SEIP-derived IRAC photometry and a similar calibration of 3.6\,$\mu$m flux density to stellar mass, where again this wavelength falls close to restframe 1.6\,$\mu$m at the cluster redshifts.  The three distributions are plotted in Figure~7, which also shows the effective mass limit of the IRAC samples in the higher redshift clusters, indicating that this  may explain the apparent differences in the median masses between the two cluster samples.  However,  Figure~7 also compares the masses of the submillimetre cluster galaxies to that derived for  $z$\,$=$\,0 passive, morphologically classified elliptical galaxies from the PM2GC survey lying in groups and clusters \citep{Calvi11,Calvi12}.   This indicates that the dusty, actively star-forming galaxies found in the $z$\,$=$\,1.5--2.0 clusters,  added to any   existing populations of massive, quiescent galaxies \citep[e.g.,][]{Newman14,Nantais16}, that may already exist in the cores of these structures \citep[e.g.,][]{Andreon14}, 
are likely to  correspond to the most  massive galaxy populations found in clusters at the present day. 

Combining the star formation rates and stellar masses, the  median specific star formation rate (sSFR\,$=$\,SFR/$M_{\ast}$) for the cluster sample is 1.06\,$\pm$\,0.14\,Gyr$^{-1}$ with a 16$^{\rm th}$--84${\rm th}$ percentile range of 0.50--1.60\,Gyr$^{-1}$, which indicates that the typical cluster submillimetre galaxy lies on or above the so-called ``main sequence'' at $z$\,$\sim$\,1.8 (sSFR\,$\sim$\,0.4\,Gyr$^{-1}$ at $M_\ast$\,$\sim$\,10$^{11}$\,M$_\odot$, \citealt{Karim11}).

\subsection{Mass-normalised cluster star formation rates}

For each cluster the  integrated star formation rate was calculated by summing the 850-$\mu$m flux densities of the reliably identified submillimetre sources lying within 1-Mpc radius of the cluster centre (where there were two reliable counterparts to a single SCUBA-2 source, the flux was assigned to the more reliable for this calculation).  This was then converted to star formation rates and corrected for residual field contamination, with the final values reported in Table~1.    The median integrated star formation rate  is  $\Sigma_{\rm SFR}$\,$=$\,530\,$\pm$\,80\,M$_\odot$\,yr$^{-1}$ per cluster which is comparable to that measured in the    $z$\,$=$\,0.8--1.6 clusters from \citet{Cooke19}, $\Sigma_{\rm SFR}$\,$=$\,750\,$\pm$\,190\,M$_\odot$\,yr$^{-1}$.

Three clusters from this study have published integrated star formation rates from the literature.
Using a combined Herschel SPIRE+SCUBA-2 analysis of JKCS041, \cite{Smith20} estimated a total SFR\,$=$\,660\,$\pm$\,240\,M$_\odot$\,yr$^{-1}$ for the 850\,$\mu$m-detected sources
within 1\,Mpc, in reasonable agreement with the measurement of SFR\,$=$\,530\,$\pm$\,50\,M$_\odot$\,yr$^{-1}$ derived here (which used the same SCUBA-2 observations).   \citet{Alberts16} used Herschel PACS observations of IDCSJ1426 to estimate SFR\,$=$\,98\,$\pm$\,54\,M$_\odot$\,yr$^{-1}$, which again agrees well with the SFR\,$=$\,160\,$\pm$\,30\,M$_\odot$\,yr$^{-1}$ measured here  using SCUBA-2.  Finally, for ClJ1449 \citet{Strazzullo18} report SFR\,$=$\,700\,$\pm$\,100\,M$_\odot$\,yr$^{-1}$ from ALMA continuum observations of a small region in the cluster core (corresponding to ClJ1449.003), while the SCUBA-2 detected sources in
the inner 1-Mpc from the Herschel/SPIRE and SCUBA-2 study of \cite{Smith19} gave SFR\,$=$\,1300\,$\pm$\,130\,M$_\odot$\,yr$^{-1}$ (or 940\,$\pm$\,90\,M$_\odot$\,yr$^{-1}$ when restricted to deblended components with $L_{\rm IR}$\,$\geq$\,10$^{12}$\,L$_\odot$) compared to   SFR\,$=$\,620\,$\pm$\,70\,M$_\odot$\,yr$^{-1}$ measured here.  While the estimates in the first two clusters agreed well, the differences between the estimated star formation rates in ClJ1449 suggest difference arising from the methodologies or other unidentified uncertainties, e.g., applying the statistical correction for residual field contamination used here to the estimates from \cite{Smith19} would bring them into closer agreement with the measurements in this work.   To reflect these potential uncertainties, the differences between the  various star formation rates in ClJ1449 were used to estimate a conservative uncertainty of 50 per cent (including the systematic uncertainties) that was applied to all the cluster measurements.  

The integrated star formation rates for the eight $z$\,$=$\,1.6--2.0 clusters (Table~1)
were then  normalised by the estimated cluster masses (Table~1) to give  mass-normalised integrated star formation rates, $\Sigma_{\rm SFR}/M_{\rm cl}$, for each system.    These have a mean of $\Sigma_{\rm SFR}/M_{\rm cl}$\,$=$\,(360\,$\pm$\,60)\,$\times$\,10$^{-14}$\,yr$^{-1}$, compared to  $\Sigma_{\rm SFR}/M_{\rm cl}$\,$=$\,(180\,$\pm$\,40)\,$\times$\,10$^{14}$\,\,yr$^{-1}$ for those in \citet{Cooke19}.\footnote{This difference is larger than  expected from the difference in mass of the two samples given the  weak dependence of $\Sigma_{\rm SFR}/M_{\rm cl}$ on halo mass, $M_{200}^{-0.4}$ reported by \citet{Popesso15b}, suggesting that the difference is due to redshift evolution.}  Errors on the cluster values were derived from bootstrap uncertainties on the mean and so reflect the variation due to excluding individual sources from the sum, with a minimum uncertainty  of 50 per cent assumed for the individual $\Sigma_{\rm SFR}/M_{\rm cl}$ measurements.  These estimates are broadly consistent with recent theoretical simulations of the activity in  massive proto-clusters at $z$\,$\sim$\,1.5--2.0 from \citet{Lim21} and \citet{Fukushima23}.

Figure~8 illustrates the variation in mass-normalised star formation rate in the central regions of massive clusters as a  function of redshift including the eight clusters from this work.   Also shown in the plot are  samples taken from the literature with integrated star formation rates within $R_{200}$, which are typically of order $\sim$\,1\,Mpc (or $\sim$\,2\,arcmin at these redshifts). To homogenise these  studies only those that used restframe far-infrared star formation tracers (either Herschel PACS/SPIRE or sub-/millimetre data from SCUBA-2 or ALMA) on individual clusters were used and proto-/clusters were included  only if these were originally identified via either X-ray emission or as overdensities of much less-active galaxies ($L_{\rm IR}$\,$\ll$\,10$^{12}$\,L$_\odot$) in spectroscopic, photometric or narrow-band surveys. Hence any systems identified on the basis of overdensities of far-infrared sources or using active galaxies  as a signpost of an overdensity are excluded.  The integrated star formation rates were derived from the restframe far-infrared detected sources in the  clusters and so excluded any contribution from less active, but potentially more numerous, cluster populations.   

The $z$\,$\gs$\,1 comparison samples shown in Figure~8  all comprise massive clusters ($M_{\rm cl}$\,$\gs$\,10$^{14}$\,M$_\odot$) with halo masses typically estimated from their X-ray luminosities.
All these studies have comparable depths, roughly corresponding to far-infrared luminosities of $L_{\rm IR}$\,$\geq$\,10$^{12}$\,L$_\odot$.  
They consist of X-ray detected clusters at $z$\,$=$\,0.8--1.6  from \cite{Cooke19} and \cite{Santos15},  infrared-selected $z$\,$=$\,1.1--1.6 clusters from \cite{Alberts16} (three of which have X-ray or SZ based halo masses, the other five use weak lensing masses), and X-ray detected clusters from \cite{Smail14} and \cite{Santos14}.
At $z$\,$>$\,2 only three systems are included: 
a proto-cluster at  $z$\,$=$\,2.1  in COSMOS from \cite{Hung16}, the $z$\,$=$\,2.3 system from \cite{Lacaille19}  and the   $z$\,$=$\,3.09 SA22 proto-cluster from \cite{Umehata15}.    For these systems where the choice of cluster centre is increasingly uncertain, two estimates were made \citep{Casey16}, one centered on the most likely centre for the structure and a second that maximised the total star formation in the aperture, the means of these are plotted and the differences were added in quadrature to the  uncertainties.  The estimated halo masses for these $z$\,$>$\,2 systems are based on abundance matching and are therefore quite uncertain, see \citet{Casey16}.

The lowest redshift sample shown in Figure~8  come from the Herschel/PACS observations of high-mass X-ray-detected clusters at $z$\,$=$\,0.2--0.8 by \cite{Popesso12}. These clusters are typically more massive than the higher-redshift systems and the observations are also deeper, probing down to $L_{\rm IR}$\,$\ls$\,10$^{12}$\,L$_\odot$.  The \citet{Popesso12} measurements were therefore corrected to account for these differences, firstly by scaling the $\Sigma_{\rm SFR}/M_{\rm cl}$ in the high-mass cluster sample  to the median mass of the higher-redshift samples assuming the mass dependence of $M_{200}^{-0.4}$ from \cite{Popesso15b}, which increased the integrated star formation rates by a factor of $\sim$\,2.  Then the $\Sigma_{\rm SFR}/M_{\rm cl}$  estimates were corrected to match the luminosity limit of the high-redshift samples, using the luminosity functions in \cite{Popesso15}, which reduced the estimates by $\sim$\,65 per cent, almost  cancelling out the correction applied for the cluster masses.

The mass-normalised star formation rate in Figure~8  shows a  rapid increase in  clusters at higher redshifts.   A fit to the evolution in $\Sigma_{\rm SFR}/M_{\rm cl}$ of the form $\Sigma_{\rm SFR}/M_{\rm cl}  \propto (1+z)^\gamma$  gave a median trend with $\gamma$\,$=$\,5.5\,$\pm$\,0.6 and a dispersion of 0.4\,dex.  Including or excluding the $z$\,$>$\,2 proto-clusters did not change the fit, while a fit to just this sample and those in \citet{Cooke19} gave marginally weaker evolution: $\gamma$\,$\sim$\,3.7\,$\pm$\,1.2.  The clusters plotted in Figure~8 also  appear to show a fairly well defined upper bound in $\Sigma_{\rm SFR}/M_{\rm cl}$ around $\sim$\,0.5\,dex above the median trend and the  scatter around the best-fit trend does not increase strongly with redshift out to $z$\,$\sim$\,2. 

The form of the redshift evolution of $\Sigma_{\rm SFR}/M_{\rm cl}$  derived here is in excellent agreement  with that reported by \cite{Webb13} who estimated  $\gamma$\,$=$\,5.4\,$\pm$\,1.9  in an independent analysis that used MIPS 24\,$\mu$m observations of a sample of 42 massive clusters at $z$\,$=$\,0.3--1.0.   The best-fit trend also agrees with estimates of $\gamma$\,$=$\,5.9\,$\pm$\,0.8 at  $M_{200}$\,$\sim$\,10$^{14}$\,M$_\odot$ from \cite{Popesso15b} and  $\gamma$\,$\sim$\,6 from \cite{Cooke19} and \cite{Smith20}  (see also, \citealt{Bai09,Alberts16}), although all of those datasets are included in the fit here.   However, the  measured $\gamma$\,$=$\,5.5\,$\pm$\,0.6 evolution is in more tension with claims of $\gamma$\,$\sim$\,7 by \cite{Smith19}, see also \cite{Geach06} and \cite{Smail14}, and indeed the weaker evolution reported in \cite{Popesso12}.

The evolution in $\Sigma_{\rm SFR}/M_{\rm cl}$ for the clusters  can also be compared to the average mass-normalised star formation activity of halos in the surrounding field.    The expected evolution of the analogous measure for the field population, ${\rm SFR}/M_{\rm halo}$,  was estimated following \cite{Popesso12, Popesso15b} and \cite{Behroozi13} by taking the cosmic star formation rate density from \citet{Madau14} and dividing it by the mean comoving density of the Universe ($\Omega_{\rm M} \times \rho_{\rm crit}$, where $\rho_{\rm crit}$ is the critical density of the Universe), and then applying a $\sim$\,0.2\,dex correction (with a $\pm$\,0.2\,dex uncertainty) to account for baryons not tied to halos  \citep{Faltenbacher10,Popesso15b}.    Figure~8 shows that the median mass-normalised integrated star formation rate for the clusters  increases to match that estimated for an average halo in the field at $z$\,$\sim$\,1.8\,$\pm$\,0.2. If the evolution of the cluster activity continues beyond this epoch then that will result in the wide spread reversal of the local SFR--density relation seen in massive clusters, with the galaxy populations in clusters at $z$\,$\gs$\,1.8 having enhanced star formation activity compared to the surrounding field \citep[e.g.,][]{Elbaz07,Tran10,Koyama13,Alberts14,Smith19}.  This is consistent with theoretical work by \citet{Hwang19}  who suggested that a reversal occured at $z$\,$\gs$\,1.5  in the star formation activity of galaxy populations within $M_{\rm cl}$\,$\gs$\,10$^{14}$\,M$_\odot$ clusters, driven by a combination of accelerated evolution in high density regions at higher redshifts and increasing environmental quenching at lower redshifts.

\section{Conclusions}

This paper reports the results from a SCUBA-2 850\,$\mu$m survey of eight  clusters of galaxies at $z$\,$=$\,1.6--2.0 with a median mass within their central 1-Mpc radius cores of $M_{\rm cl}$\,$\sim$\,2\,$\times$\,10$^{14}$\,M$_\odot$  (these  are expected to grow into   $M_{\rm cl}$\,$\sim$\,6\,$\times$\,10$^{14}$\,M$_\odot$ systems by the present day).   The survey was designed to extend to higher redshift the previous SCUBA-2 study by \cite{Cooke19} of a similar sized sample of massive clusters at $z$\,$=$\,0.8--1.6.   

The SCUBA-2 observations were a mix of new and archival observations and reached a median depth of $\sigma_{\rm 850\mu m}$\,$=$\,1.0\,$\pm$\,0.1\,mJy.  The eight maps detected 56 sources at significance levels above 3.5\,$\sigma$  (the ``{\sc Main}'' sample) out to 4\,arcmin radius and a further 18 at 3.0--3.5$\sigma$ in the central 2\,arcmin ($\sim$\,1\,Mpc), termed the ``{\sc Supplementary}'' sample.    Within the central 2\,arcmin ($\sim$\,1\,Mpc) radius of the clusters a mean overdensity of a factor of 1.5\,$\pm$\,0.3 was measured compared to the integrated field counts of submillimetre sources brighter than $S_{\rm 850\mu m}$\,$\geq$\,4\,mJy and a factor of 3.0\,$\pm$\,1.4 in the central 1\,Mpc diameter cores for sources with $S_{\rm 850\mu m}$\,$\geq$\,4.8\,mJy.  Applying an IRAC colour selection to attempt to isolate those submillimetre counterparts that are likely to be cluster members increases the significance of the mean overdensity in the central 1\,Mpc radius of the clusters to  4\,$\pm$\,1 for sources brighter than $S_{\rm 850\mu m}$\,$=$\,3.5\,mJy.

Archival sub/millimetre and radio interferometry, as well as Spitzer MIPS and IRAC imaging, were used to identify likely galaxy counterparts to the SCUBA-2 sources.  This yielded a sample of 53 reliably identified counterparts with IRAC colours consistent with $z$\,$=$\,1.6--2.0, that were matched to 45 SCUBA-2 sources.  This included eight submillimetre sources with pairs of counterparts, corresponding to a multiplicity fraction of $\sim$\,18\,per cent, consistent with that found in ALMA field studies at similar flux densities.  These are statistical identifications and so both the individual source identifications and their assignment as cluster members are uncertain.   Nevertheless, the detection of excesses of submillimetre sources in the cluster cores should be robust.

The analysis also showed that both the overdensities of submillimetre sources  in the central regions of these clusters at $z$\,$\sim$\,1.5--2,  and the integrated star formation activity associated with these luminous star-forming galaxies, were comparable to those seen in similar mass clusters at  $z$\,$\sim$\,0.8--1.6 and two orders of magnitude higher than  massive clusters at $z$\,$\sim$\,0.

Normalising the integrated star formation rates by the estimated cluster masses showed that the mass-normalised integrated star formation rate of the clusters evolves as  $\Sigma_{\rm SFR}/M_{\rm cl}  \propto (1+z)^{5.5\pm 0.6}$ in good agreement with  previous estimates of the evolutionary rate in clusters at $z$\,$<$\,0--1.5 from \citet{Webb13} and \citet{Popesso15b}.

Moreover, the  $z$\,$\sim$\,1.5--2 clusters were found to have mass-normalised star formation rates comparable to those for average halos in the surrounding field, with the best-fit cluster $\Sigma_{\rm SFR}/M_{\rm cl}$  trend matching  the estimate for  halos in the field at $z$\,$\sim$\,1.8\,$\pm$\,0.2.  This  indicates a reversal in the star formation rate-density relation for massive clusters beyond this epoch, consistent with theoretical expectations \citep{Hwang19}.

This work has highlighted a number of challenges that need to be overcome in future studies of the environmental influences on star formation in  clusters at $z$\,$\gs$\,1.5.  The majority of this activity is occuring in dust-obscured systems \citep[e.g.,][]{McKinney22} and so such studies have to be undertaken in the far-infrared or submillimetre wavebands.  However, to make further progress on this issue it will be necessary to first construct more robust samples of $z$\,$\gs$\,1.5 clusters, ideally by detecting the X-ray emission from their virialised cores or through their SZ decrements (although this is complicated by the potential dilution of the decrements caused by the presence of bright submillimetre sources), to enable both better localisation of the cluster centres and also estimation of their masses.   

With reliable positions for the cluster cores,  mosaiced observations can be undertaken with interferometers such as SMA, ACA or ALMA in  submillimetre wavebands with better sensitivity than single-dish observations and hence lower shot-noise in the detected star-forming population.  Such interferometric surveys  provide both  the continuum sensitivity needed to map the obscured star-forming galaxies down to relatively low luminosities, as well as the spatial resolution necessary to directly identify counterparts and potentially also yield confirmation of redshifts from the detection of CO emission lines (although the best frequency ranges for this, and to maximise the  primary beam area, are in tension with the desire for higher frequencies to better estimate star formation rates).  Notwithstanding these challenges, such studies continue to be critical as they provide one of the few methods to directly link the properties and evolution of the galaxy populations found in present-day clusters to their progenitors at high redshifts and so understand the formation of some of the most massive and oldest galaxies in the Universe.

\section*{Acknowledgements}

The referee is thanked for their constructive comments that helped clarify the text of this paper.
Mark Swinbank, Jack Birkin, Soh Ikarashi, Lizzie Cooke, Minhee Hyun and James Simpson are thanked for their comments, help and work on the earlier stages of this project.   Thomas Greve, Chian-Chou Chen and Stefano Andreon are thanked for comments on the manuscript.   IRS acknowledges support from STFC (ST/T000244/1 and ST/X001075/1).

This work is dedicated to the memories of Richard Bower, Wayne Holland and Nick Kaiser, all three are deeply missed.

This study made use of data from JCMT project IDs M12AI01, M15AI09, M15AI39, M15AI51, M15BI038, M16AP047, M16AP087, M21BP030 and M22AP039. The James Clerk Maxwell Telescope is operated by the East Asian Observatory on behalf of The National Astronomical Observatory of Japan; Academia Sinica Institute of Astronomy and Astrophysics; the Korea Astronomy and Space Science Institute; the Operation, Maintenance and Upgrading Fund for Astronomical Telescopes and Facility Instruments, budgeted from the Ministry of Finance (MOF) of China and administrated by the Chinese Academy of Sciences (CAS), as well as the National Key R\&D Program of China (No.\ 2017YFA0402700). Additional funding support is provided by the Science and Technology Facilities Council of the United Kingdom and participating universities in the United Kingdom and Canada.
This paper made use of observations from  the following ALMA projects: 2021.1.01257.S, 2018.1.00974.S, 2016.1.01107.S.    ALMA is a partnership of ESO (representing its member states), NSF (USA), and NINS (Japan), together with NRC (Canada), NSC and ASIAA (Taiwan), and KASI (Republic of Korea), in cooperation with the Republic of Chile. The Joint ALMA Observatory is operated by ESO, AUI/NRAO, and NAOJ.
This research also used the facilities of the Canadian Astronomy Data Centre operated by the National Research Council of Canada with the support of the Canadian Space Agency. 
This work is based in part on archival data obtained with the Spitzer Space Telescope and the NASA/IPAC Extragalactic Database (NED), which are operated by the Jet Propulsion Laboratory, California Institute of Technology under a contract with the National Aeronautics and Space Administration. This study has made use of NASA’s Astrophysics Data System Bibliographic Services.
This research has made use of the NASA/IPAC Infrared Science Archive, which is funded by the National Aeronautics and Space Administration and operated by the California Institute of Technology.
The Herschel Extragalactic Legacy Project, (HELP), is a European Commission Research Executive Agency funded project under the SP1-Cooperation, Collaborative project, Small or medium-scale focused research project, FP7-SPACE-2013-1 scheme, Grant Agreement Number 607254.
Extensive use was made of {\sc topcat} \citep{Taylor05} in this study.

Facilities: JCMT, Spitzer, HST, ALMA, JVLA.

\section*{Data availability}

The data used in this paper can be obtained from the   JCMT,  Spitzer, HST and ALMA data archives.
The Spitzer Enhanced Imaging Products (SEIP) DOI is 10.26131/IRSA433.

\bibliography{paper}{}

\begin{thebibliography}{}
\makeatletter
\relax
\def\mn@urlcharsother{\let\do\@makeother \do\$\do\&\do\#\do\^\do\_\do\%\do\~}
\def\mn@doi{\begingroup\mn@urlcharsother \@ifnextchar [ {\mn@doi@}
  {\mn@doi@[]}}
\def\mn@doi@[#1]#2{\def\@tempa{#1}\ifx\@tempa\@empty \href
  {http://dx.doi.org/#2} {doi:#2}\else \href {http://dx.doi.org/#2} {#1}\fi
  \endgroup}
\def\mn@eprint#1#2{\mn@eprint@#1:#2::\@nil}
\def\mn@eprint@arXiv#1{\href {http://arxiv.org/abs/#1} {{\tt arXiv:#1}}}
\def\mn@eprint@dblp#1{\href {http://dblp.uni-trier.de/rec/bibtex/#1.xml}
  {dblp:#1}}
\def\mn@eprint@#1:#2:#3:#4\@nil{\def\@tempa {#1}\def\@tempb {#2}\def\@tempc
  {#3}\ifx \@tempc \@empty \let \@tempc \@tempb \let \@tempb \@tempa \fi \ifx
  \@tempb \@empty \def\@tempb {arXiv}\fi \@ifundefined
  {mn@eprint@\@tempb}{\@tempb:\@tempc}{\expandafter \expandafter \csname
  mn@eprint@\@tempb\endcsname \expandafter{\@tempc}}}

\bibitem[\protect\citeauthoryear{{Alberts} \& {Noble}}{{Alberts} \&
  {Noble}}{2022}]{Alberts22}
{Alberts} S.,  {Noble} A.,  2022, \mn@doi [Universe] {10.3390/universe8110554},
  \href {https://ui.adsabs.harvard.edu/abs/2022Univ....8..554A} {8, 554}

\bibitem[\protect\citeauthoryear{{Alberts} et~al.,}{{Alberts}
  et~al.}{2014}]{Alberts14}
{Alberts} S.,  et~al., 2014, \mn@doi [\mnras] {10.1093/mnras/stt1897}, \href
  {https://ui.adsabs.harvard.edu/abs/2014MNRAS.437..437A} {437, 437}

\bibitem[\protect\citeauthoryear{{Alberts} et~al.,}{{Alberts}
  et~al.}{2016}]{Alberts16}
{Alberts} S.,  et~al., 2016, \mn@doi [\apj] {10.3847/0004-637X/825/1/72}, \href
  {https://ui.adsabs.harvard.edu/abs/2016ApJ...825...72A} {825, 72}

\bibitem[\protect\citeauthoryear{{Alberts} et~al.,}{{Alberts}
  et~al.}{2021}]{Alberts21}
{Alberts} S.,  et~al., 2021, \mn@doi [\mnras] {10.1093/mnras/staa3357}, \href
  {https://ui.adsabs.harvard.edu/abs/2021MNRAS.501.1970A} {501, 1970}

\bibitem[\protect\citeauthoryear{{An} et~al.,}{{An} et~al.}{2018}]{An18}
{An} F.~X.,  et~al., 2018, \mn@doi [\apj] {10.3847/1538-4357/aacdaa}, \href
  {https://ui.adsabs.harvard.edu/abs/2018ApJ...862..101A} {862, 101}

\bibitem[\protect\citeauthoryear{{An} et~al.,}{{An} et~al.}{2019}]{An19}
{An} F.~X.,  et~al., 2019, \mn@doi [\apj] {10.3847/1538-4357/ab4d53}, \href
  {https://ui.adsabs.harvard.edu/abs/2019ApJ...886...48A} {886, 48}

\bibitem[\protect\citeauthoryear{{Andreon}, {Maughan}, {Trinchieri}  \&
  {Kurk}}{{Andreon} et~al.}{2009}]{Andreon09}
{Andreon} S.,  {Maughan} B.,  {Trinchieri} G.,   {Kurk} J.,  2009, \mn@doi
  [\aap] {10.1051/0004-6361/200912299}, \href
  {https://ui.adsabs.harvard.edu/abs/2009A&A...507..147A} {507, 147}

\bibitem[\protect\citeauthoryear{{Andreon}, {Newman}, {Trinchieri}, {Raichoor},
  {Ellis}  \& {Treu}}{{Andreon} et~al.}{2014}]{Andreon14}
{Andreon} S.,  {Newman} A.~B.,  {Trinchieri} G.,  {Raichoor} A.,  {Ellis}
  R.~S.,   {Treu} T.,  2014, \mn@doi [\aap] {10.1051/0004-6361/201323077},
  \href {https://ui.adsabs.harvard.edu/abs/2014A&A...565A.120A} {565, A120}

\bibitem[\protect\citeauthoryear{{Andreon} et~al.,}{{Andreon}
  et~al.}{2021}]{Andreon21}
{Andreon} S.,  et~al., 2021, \mn@doi [\mnras] {10.1093/mnras/stab1639}, \href
  {https://ui.adsabs.harvard.edu/abs/2021MNRAS.505.5896A} {505, 5896}

\bibitem[\protect\citeauthoryear{{Andreon} et~al.,}{{Andreon}
  et~al.}{2023}]{Andreon23}
{Andreon} S.,  et~al., 2023, \mn@doi [\mnras] {10.1093/mnras/stad1270}, \href
  {https://ui.adsabs.harvard.edu/abs/2023MNRAS.522.4301A} {522, 4301}

\bibitem[\protect\citeauthoryear{{Babyk} \& {Vavilova}}{{Babyk} \&
  {Vavilova}}{2014}]{Babyk14}
{Babyk} I.,  {Vavilova} I.,  2014, \mn@doi [\apss] {10.1007/s10509-014-2057-x},
  \href {https://ui.adsabs.harvard.edu/abs/2014Ap&SS.353..613B} {353, 613}

\bibitem[\protect\citeauthoryear{{Bai}, {Rieke}, {Rieke}, {Christlein}  \&
  {Zabludoff}}{{Bai} et~al.}{2009}]{Bai09}
{Bai} L.,  {Rieke} G.~H.,  {Rieke} M.~J.,  {Christlein} D.,   {Zabludoff}
  A.~I.,  2009, \mn@doi [\apj] {10.1088/0004-637X/693/2/1840}, \href
  {https://ui.adsabs.harvard.edu/abs/2009ApJ...693.1840B} {693, 1840}

\bibitem[\protect\citeauthoryear{{Behroozi}, {Wechsler}  \&
  {Conroy}}{{Behroozi} et~al.}{2013}]{Behroozi13}
{Behroozi} P.~S.,  {Wechsler} R.~H.,   {Conroy} C.,  2013, \mn@doi [\apj]
  {10.1088/0004-637X/770/1/57}, \href
  {https://ui.adsabs.harvard.edu/abs/2013ApJ...770...57B} {770, 57}

\bibitem[\protect\citeauthoryear{{Best}}{{Best}}{2002}]{Best02}
{Best} P.~N.,  2002, \mn@doi [\mnras] {10.1046/j.1365-8711.2002.05896.x}, \href
  {https://ui.adsabs.harvard.edu/abs/2002MNRAS.336.1293B} {336, 1293}

\bibitem[\protect\citeauthoryear{{Biggs} \& {Ivison}}{{Biggs} \&
  {Ivison}}{2006}]{Biggs06}
{Biggs} A.~D.,  {Ivison} R.~J.,  2006, \mn@doi [\mnras]
  {10.1111/j.1365-2966.2006.10730.x}, \href
  {https://ui.adsabs.harvard.edu/abs/2006MNRAS.371..963B} {371, 963}

\bibitem[\protect\citeauthoryear{{Bower}, {Ellis}, {Rose}  \&
  {Sharples}}{{Bower} et~al.}{1990}]{Bower90}
{Bower} R.~G.,  {Ellis} R.~S.,  {Rose} J.~A.,   {Sharples} R.~M.,  1990,
  \mn@doi [\aj] {10.1086/115347}, \href
  {https://ui.adsabs.harvard.edu/abs/1990AJ.....99..530B} {99, 530}

\bibitem[\protect\citeauthoryear{{Brodwin} et~al.,}{{Brodwin}
  et~al.}{2012}]{Brodwin12}
{Brodwin} M.,  et~al., 2012, \mn@doi [\apj] {10.1088/0004-637X/753/2/162},
  \href {https://ui.adsabs.harvard.edu/abs/2012ApJ...753..162B} {753, 162}

\bibitem[\protect\citeauthoryear{{Brodwin}, {McDonald}, {Gonzalez}, {Stanford},
  {Eisenhardt}, {Stern}  \& {Zeimann}}{{Brodwin} et~al.}{2016}]{Brodwin16}
{Brodwin} M.,  {McDonald} M.,  {Gonzalez} A.~H.,  {Stanford} S.~A.,
  {Eisenhardt} P.~R.,  {Stern} D.,   {Zeimann} G.~R.,  2016, \mn@doi [\apj]
  {10.3847/0004-637X/817/2/122}, \href
  {https://ui.adsabs.harvard.edu/abs/2016ApJ...817..122B} {817, 122}

\bibitem[\protect\citeauthoryear{{Calvi}, {Poggianti}  \& {Vulcani}}{{Calvi}
  et~al.}{2011}]{Calvi11}
{Calvi} R.,  {Poggianti} B.~M.,   {Vulcani} B.,  2011, \mn@doi [\mnras]
  {10.1111/j.1365-2966.2011.19088.x}, \href
  {https://ui.adsabs.harvard.edu/abs/2011MNRAS.416..727C} {416, 727}

\bibitem[\protect\citeauthoryear{{Calvi}, {Poggianti}, {Fasano}  \&
  {Vulcani}}{{Calvi} et~al.}{2012}]{Calvi12}
{Calvi} R.,  {Poggianti} B.~M.,  {Fasano} G.,   {Vulcani} B.,  2012, \mn@doi
  [\mnras] {10.1111/j.1745-3933.2011.01168.x10.1086/122140}, \href
  {https://ui.adsabs.harvard.edu/abs/2012MNRAS.419L..14C} {419, L14}

\bibitem[\protect\citeauthoryear{{Casey}}{{Casey}}{2016}]{Casey16}
{Casey} C.~M.,  2016, \mn@doi [\apj] {10.3847/0004-637X/824/1/36}, \href
  {https://ui.adsabs.harvard.edu/abs/2016ApJ...824...36C} {824, 36}

\bibitem[\protect\citeauthoryear{{Casey} et~al.,}{{Casey}
  et~al.}{2015}]{Casey15}
{Casey} C.~M.,  et~al., 2015, \mn@doi [\apjl] {10.1088/2041-8205/808/2/L33},
  \href {https://ui.adsabs.harvard.edu/abs/2015ApJ...808L..33C} {808, L33}

\bibitem[\protect\citeauthoryear{{Chabrier}}{{Chabrier}}{2003}]{Chabrier03}
{Chabrier} G.,  2003, \mn@doi [\pasp] {10.1086/376392}, \href
  {https://ui.adsabs.harvard.edu/abs/2003PASP..115..763C} {115, 763}

\bibitem[\protect\citeauthoryear{{Chapin}, {Berry}, {Gibb}, {Jenness}, {Scott},
  {Tilanus}, {Economou}  \& {Holland}}{{Chapin} et~al.}{2013}]{Chapin13}
{Chapin} E.~L.,  {Berry} D.~S.,  {Gibb} A.~G.,  {Jenness} T.,  {Scott} D.,
  {Tilanus} R. P.~J.,  {Economou} F.,   {Holland} W.~S.,  2013, \mn@doi
  [\mnras] {10.1093/mnras/stt052}, \href
  {https://ui.adsabs.harvard.edu/abs/2013MNRAS.430.2545C} {430, 2545}

\bibitem[\protect\citeauthoryear{{Chapman}, {Blain}, {Smail}  \&
  {Ivison}}{{Chapman} et~al.}{2005}]{Chapman05}
{Chapman} S.~C.,  {Blain} A.~W.,  {Smail} I.,   {Ivison} R.~J.,  2005, \mn@doi
  [\apj] {10.1086/428082}, \href
  {https://ui.adsabs.harvard.edu/abs/2005ApJ...622..772C} {622, 772}

\bibitem[\protect\citeauthoryear{{Chen} et~al.,}{{Chen} et~al.}{2015}]{Chen15}
{Chen} C.-C.,  et~al., 2015, \mn@doi [\apj] {10.1088/0004-637X/799/2/194},
  \href {https://ui.adsabs.harvard.edu/abs/2015ApJ...799..194C} {799, 194}

\bibitem[\protect\citeauthoryear{{Chen} et~al.,}{{Chen} et~al.}{2016}]{Chen16}
{Chen} C.-C.,  et~al., 2016, \mn@doi [\apj] {10.3847/0004-637X/820/2/82}, \href
  {https://ui.adsabs.harvard.edu/abs/2016ApJ...820...82C} {820, 82}

\bibitem[\protect\citeauthoryear{{Chen} et~al.,}{{Chen} et~al.}{2022}]{Chen22}
{Chen} C.-C.,  et~al., 2022, \mn@doi [\apjl] {10.3847/2041-8213/ac98c6}, \href
  {https://ui.adsabs.harvard.edu/abs/2022ApJ...939L...7C} {939, L7}

\bibitem[\protect\citeauthoryear{{Cheng} et~al.,}{{Cheng}
  et~al.}{2019}]{Cheng19}
{Cheng} T.,  et~al., 2019, \mn@doi [\mnras] {10.1093/mnras/stz2640}, \href
  {https://ui.adsabs.harvard.edu/abs/2019MNRAS.490.3840C} {490, 3840}

\bibitem[\protect\citeauthoryear{{Cheng} et~al.,}{{Cheng}
  et~al.}{2023}]{Cheng23}
{Cheng} C.,  et~al., 2023, \mn@doi [\apjl] {10.3847/2041-8213/aca9d0}, \href
  {https://ui.adsabs.harvard.edu/abs/2023ApJ...942L..19C} {942, L19}

\bibitem[\protect\citeauthoryear{{Coia} et~al.,}{{Coia} et~al.}{2005}]{Coia05}
{Coia} D.,  et~al., 2005, \mn@doi [\aap] {10.1051/0004-6361:20041782}, \href
  {https://ui.adsabs.harvard.edu/abs/2005A&A...431..433C} {431, 433}

\bibitem[\protect\citeauthoryear{{Cole} \& {Kaiser}}{{Cole} \&
  {Kaiser}}{1989}]{Cole89}
{Cole} S.,  {Kaiser} N.,  1989, \mn@doi [\mnras] {10.1093/mnras/237.4.1127},
  \href {https://ui.adsabs.harvard.edu/abs/1989MNRAS.237.1127C} {237, 1127}

\bibitem[\protect\citeauthoryear{{Coogan} et~al.,}{{Coogan}
  et~al.}{2018}]{Coogan18}
{Coogan} R.~T.,  et~al., 2018, \mn@doi [\mnras] {10.1093/mnras/sty1446}, \href
  {https://ui.adsabs.harvard.edu/abs/2018MNRAS.479..703C} {479, 703}

\bibitem[\protect\citeauthoryear{{Cooke}, {Smail}, {Stach}, {Swinbank},
  {Bower}, {Chen}, {Koyama}  \& {Thomson}}{{Cooke} et~al.}{2019}]{Cooke19}
{Cooke} E.~A.,  {Smail} I.,  {Stach} S.~M.,  {Swinbank} A.~M.,  {Bower} R.~G.,
  {Chen} C.-C.,  {Koyama} Y.,   {Thomson} A.~P.,  2019, \mn@doi [\mnras]
  {10.1093/mnras/stz955}, \href
  {https://ui.adsabs.harvard.edu/abs/2019MNRAS.486.3047C} {486, 3047}

\bibitem[\protect\citeauthoryear{{Coppin} et~al.,}{{Coppin}
  et~al.}{2006}]{Coppin06}
{Coppin} K.,  et~al., 2006, \mn@doi [\mnras]
  {10.1111/j.1365-2966.2006.10961.x}, \href
  {https://ui.adsabs.harvard.edu/abs/2006MNRAS.372.1621C} {372, 1621}

\bibitem[\protect\citeauthoryear{{Cowie}, {Gonz{\'a}lez-L{\'o}pez}, {Barger},
  {Bauer}, {Hsu}  \& {Wang}}{{Cowie} et~al.}{2018}]{Cowie18}
{Cowie} L.~L.,  {Gonz{\'a}lez-L{\'o}pez} J.,  {Barger} A.~J.,  {Bauer} F.~E.,
  {Hsu} L.~Y.,   {Wang} W.~H.,  2018, \mn@doi [\apj]
  {10.3847/1538-4357/aadc63}, \href
  {https://ui.adsabs.harvard.edu/abs/2018ApJ...865..106C} {865, 106}

\bibitem[\protect\citeauthoryear{{Delahaye} et~al.,}{{Delahaye}
  et~al.}{2017}]{Delahaye17}
{Delahaye} A.~G.,  et~al., 2017, \mn@doi [\apj] {10.3847/1538-4357/aa756a},
  \href {https://ui.adsabs.harvard.edu/abs/2017ApJ...843..126D} {843, 126}

\bibitem[\protect\citeauthoryear{{Dempsey} et~al.,}{{Dempsey}
  et~al.}{2013}]{Dempsey13}
{Dempsey} J.~T.,  et~al., 2013, \mn@doi [\mnras] {10.1093/mnras/stt090}, \href
  {https://ui.adsabs.harvard.edu/abs/2013MNRAS.430.2534D} {430, 2534}

\bibitem[\protect\citeauthoryear{{Donley} et~al.,}{{Donley}
  et~al.}{2012}]{Donley12}
{Donley} J.~L.,  et~al., 2012, \mn@doi [\apj] {10.1088/0004-637X/748/2/142},
  \href {https://ui.adsabs.harvard.edu/abs/2012ApJ...748..142D} {748, 142}

\bibitem[\protect\citeauthoryear{{Downes}, {Peacock}, {Savage}  \&
  {Carrie}}{{Downes} et~al.}{1986}]{Downes86}
{Downes} A.~J.~B.,  {Peacock} J.~A.,  {Savage} A.,   {Carrie} D.~R.,  1986,
  \mn@doi [\mnras] {10.1093/mnras/218.1.31}, \href
  {https://ui.adsabs.harvard.edu/abs/1986MNRAS.218...31D} {218, 31}

\bibitem[\protect\citeauthoryear{{Dressler}}{{Dressler}}{1980}]{Dressler80}
{Dressler} A.,  1980, \mn@doi [\apj] {10.1086/157753}, \href
  {https://ui.adsabs.harvard.edu/abs/1980ApJ...236..351D} {236, 351}

\bibitem[\protect\citeauthoryear{{Dudzevi{\v{c}}i{\={u}}t{\.{e}}}
  et~al.,}{{Dudzevi{\v{c}}i{\={u}}t{\.{e}}} et~al.}{2020}]{Dudzeviciute20}
{Dudzevi{\v{c}}i{\={u}}t{\.{e}}} U.,  et~al., 2020, \mn@doi [\mnras]
  {10.1093/mnras/staa769}, \href
  {https://ui.adsabs.harvard.edu/abs/2020MNRAS.494.3828D} {494, 3828}

\bibitem[\protect\citeauthoryear{{Dunlop}, {Peacock}, {Savage}, {Lilly},
  {Heasley}  \& {Simon}}{{Dunlop} et~al.}{1989}]{Dunlop89}
{Dunlop} J.~S.,  {Peacock} J.~A.,  {Savage} A.,  {Lilly} S.~J.,  {Heasley}
  J.~N.,   {Simon} A.~J.~B.,  1989, \mn@doi [\mnras]
  {10.1093/mnras/238.4.1171}, \href
  {https://ui.adsabs.harvard.edu/abs/1989MNRAS.238.1171D} {238, 1171}

\bibitem[\protect\citeauthoryear{{Elbaz} et~al.,}{{Elbaz}
  et~al.}{2007}]{Elbaz07}
{Elbaz} D.,  et~al., 2007, \mn@doi [\aap] {10.1051/0004-6361:20077525}, \href
  {https://ui.adsabs.harvard.edu/abs/2007A&A...468...33E} {468, 33}

\bibitem[\protect\citeauthoryear{{Faltenbacher}, {Finoguenov}  \&
  {Drory}}{{Faltenbacher} et~al.}{2010}]{Faltenbacher10}
{Faltenbacher} A.,  {Finoguenov} A.,   {Drory} N.,  2010, \mn@doi [\apj]
  {10.1088/0004-637X/712/1/484}, \href
  {https://ui.adsabs.harvard.edu/abs/2010ApJ...712..484F} {712, 484}

\bibitem[\protect\citeauthoryear{{Fazio} et~al.,}{{Fazio}
  et~al.}{2004}]{Fazio04}
{Fazio} G.~G.,  et~al., 2004, \mn@doi [\apjs] {10.1086/422843}, \href
  {https://ui.adsabs.harvard.edu/abs/2004ApJS..154...10F} {154, 10}

\bibitem[\protect\citeauthoryear{{Finn} et~al.,}{{Finn} et~al.}{2005}]{Finn05}
{Finn} R.~A.,  et~al., 2005, \mn@doi [\apj] {10.1086/431642}, \href
  {https://ui.adsabs.harvard.edu/abs/2005ApJ...630..206F} {630, 206}

\bibitem[\protect\citeauthoryear{{Frayer}, {Reddy}, {Armus}, {Blain},
  {Scoville}  \& {Smail}}{{Frayer} et~al.}{2004}]{Frayer04}
{Frayer} D.~T.,  {Reddy} N.~A.,  {Armus} L.,  {Blain} A.~W.,  {Scoville} N.~Z.,
    {Smail} I.,  2004, \mn@doi [\aj] {10.1086/380943}, \href
  {https://ui.adsabs.harvard.edu/abs/2004AJ....127..728F} {127, 728}

\bibitem[\protect\citeauthoryear{{Fukushima}, {Nagamine}  \&
  {Shimizu}}{{Fukushima} et~al.}{2023}]{Fukushima23}
{Fukushima} K.,  {Nagamine} K.,   {Shimizu} I.,  2023, \mn@doi [\mnras]
  {10.1093/mnras/stad2526}, \href
  {https://ui.adsabs.harvard.edu/abs/2023MNRAS.525.3760F} {525, 3760}

\bibitem[\protect\citeauthoryear{{Geach} et~al.,}{{Geach}
  et~al.}{2006}]{Geach06}
{Geach} J.~E.,  et~al., 2006, \mn@doi [\apj] {10.1086/506469}, \href
  {https://ui.adsabs.harvard.edu/abs/2006ApJ...649..661G} {649, 661}

\bibitem[\protect\citeauthoryear{{Geach} et~al.,}{{Geach}
  et~al.}{2017}]{Geach17}
{Geach} J.~E.,  et~al., 2017, \mn@doi [\mnras] {10.1093/mnras/stw2721}, \href
  {https://ui.adsabs.harvard.edu/abs/2017MNRAS.465.1789G} {465, 1789}

\bibitem[\protect\citeauthoryear{{Gobat} et~al.,}{{Gobat}
  et~al.}{2011}]{Gobat11}
{Gobat} R.,  et~al., 2011, \mn@doi [\aap] {10.1051/0004-6361/201016084}, \href
  {https://ui.adsabs.harvard.edu/abs/2011A&A...526A.133G} {526, A133}

\bibitem[\protect\citeauthoryear{{Gobat} et~al.,}{{Gobat}
  et~al.}{2013}]{Gobat13}
{Gobat} R.,  et~al., 2013, \mn@doi [\apj] {10.1088/0004-637X/776/1/9}, \href
  {https://ui.adsabs.harvard.edu/abs/2013ApJ...776....9G} {776, 9}

\bibitem[\protect\citeauthoryear{{Gobat} et~al.,}{{Gobat}
  et~al.}{2019}]{Gobat19}
{Gobat} R.,  et~al., 2019, \mn@doi [\aap] {10.1051/0004-6361/201935862}, \href
  {https://ui.adsabs.harvard.edu/abs/2019A&A...629A.104G} {629, A104}

\bibitem[\protect\citeauthoryear{{Greenslade} et~al.,}{{Greenslade}
  et~al.}{2018}]{Greenslade18}
{Greenslade} J.,  et~al., 2018, \mn@doi [\mnras] {10.1093/mnras/sty023}, \href
  {https://ui.adsabs.harvard.edu/abs/2018MNRAS.476.3336G} {476, 3336}

\bibitem[\protect\citeauthoryear{{Henry}, {Aoki}, {Finoguenov}, {Fotopoulou},
  {Hasinger}, {salvato}, {Suh}  \& {Tanaka}}{{Henry} et~al.}{2014}]{Henry14}
{Henry} J.~P.,  {Aoki} K.,  {Finoguenov} A.,  {Fotopoulou} S.,  {Hasinger} G.,
  {salvato} M.,  {Suh} H.,   {Tanaka} M.,  2014, \mn@doi [\apj]
  {10.1088/0004-637X/780/1/58}, \href
  {https://ui.adsabs.harvard.edu/abs/2014ApJ...780...58H} {780, 58}

\bibitem[\protect\citeauthoryear{{Heywood}, {Hale}, {Jarvis}, {Makhathini},
  {Peters}, {Sebokolodi}  \& {Smirnov}}{{Heywood} et~al.}{2020}]{Heywood20}
{Heywood} I.,  {Hale} C.~L.,  {Jarvis} M.~J.,  {Makhathini} S.,  {Peters}
  J.~A.,  {Sebokolodi} M.~L.~L.,   {Smirnov} O.~M.,  2020, \mn@doi [\mnras]
  {10.1093/mnras/staa1770}, \href
  {https://ui.adsabs.harvard.edu/abs/2020MNRAS.496.3469H} {496, 3469}

\bibitem[\protect\citeauthoryear{{Hodge} et~al.,}{{Hodge}
  et~al.}{2013}]{Hodge13}
{Hodge} J.~A.,  et~al., 2013, \mn@doi [\apj] {10.1088/0004-637X/768/1/91},
  \href {https://ui.adsabs.harvard.edu/abs/2013ApJ...768...91H} {768, 91}

\bibitem[\protect\citeauthoryear{{Holland} et~al.,}{{Holland}
  et~al.}{2013}]{Holland13}
{Holland} W.~S.,  et~al., 2013, \mn@doi [\mnras] {10.1093/mnras/sts612}, \href
  {https://ui.adsabs.harvard.edu/abs/2013MNRAS.430.2513H} {430, 2513}

\bibitem[\protect\citeauthoryear{{Hung} et~al.,}{{Hung} et~al.}{2016}]{Hung16}
{Hung} C.-L.,  et~al., 2016, \mn@doi [\apj] {10.3847/0004-637X/826/2/130},
  \href {https://ui.adsabs.harvard.edu/abs/2016ApJ...826..130H} {826, 130}

\bibitem[\protect\citeauthoryear{{Hurley} et~al.,}{{Hurley}
  et~al.}{2017}]{Hurley17}
{Hurley} P.~D.,  et~al., 2017, \mn@doi [\mnras] {10.1093/mnras/stw2375}, \href
  {https://ui.adsabs.harvard.edu/abs/2017MNRAS.464..885H} {464, 885}

\bibitem[\protect\citeauthoryear{{Hwang}, {Shin}  \& {Song}}{{Hwang}
  et~al.}{2019}]{Hwang19}
{Hwang} H.~S.,  {Shin} J.,   {Song} H.,  2019, \mn@doi [\mnras]
  {10.1093/mnras/stz2136}, \href
  {https://ui.adsabs.harvard.edu/abs/2019MNRAS.489..339H} {489, 339}

\bibitem[\protect\citeauthoryear{{Hyun} et~al.,}{{Hyun} et~al.}{2023}]{Hyun23}
{Hyun} M.,  et~al., 2023, \mn@doi [\apjs] {10.3847/1538-4365/ac9bf4}, \href
  {https://ui.adsabs.harvard.edu/abs/2023ApJS..264...19H} {264, 19}

\bibitem[\protect\citeauthoryear{{Ibar}, {Ivison}, {Biggs}, {Lal}, {Best}  \&
  {Green}}{{Ibar} et~al.}{2009}]{Ibar09}
{Ibar} E.,  {Ivison} R.~J.,  {Biggs} A.~D.,  {Lal} D.~V.,  {Best} P.~N.,
  {Green} D.~A.,  2009, \mn@doi [\mnras] {10.1111/j.1365-2966.2009.14866.x},
  \href {https://ui.adsabs.harvard.edu/abs/2009MNRAS.397..281I} {397, 281}

\bibitem[\protect\citeauthoryear{{Ivison} et~al.,}{{Ivison}
  et~al.}{2002}]{Ivison02}
{Ivison} R.~J.,  et~al., 2002, \mn@doi [\mnras]
  {10.1046/j.1365-8711.2002.05900.x}, \href
  {https://ui.adsabs.harvard.edu/abs/2002MNRAS.337....1I} {337, 1}

\bibitem[\protect\citeauthoryear{{Ivison} et~al.,}{{Ivison}
  et~al.}{2007}]{Ivison07}
{Ivison} R.~J.,  et~al., 2007, \mn@doi [\mnras]
  {10.1111/j.1365-2966.2007.12044.x}, \href
  {https://ui.adsabs.harvard.edu/abs/2007MNRAS.380..199I} {380, 199}

\bibitem[\protect\citeauthoryear{{Jenness}, {Berry}, {Cavanagh}, {Currie},
  {Draper}  \& {Economou}}{{Jenness} et~al.}{2009}]{Jenness09}
{Jenness} T.,  {Berry} D.~S.,  {Cavanagh} B.,  {Currie} M.~J.,  {Draper} P.~W.,
    {Economou} F.,  2009, in {Bohlender} D.~A.,  {Durand} D.,   {Dowler} P.,
  eds,  Astronomical Society of the Pacific Conference Series Vol. 411,
  Astronomical Data Analysis Software and Systems XVIII. p.~418

\bibitem[\protect\citeauthoryear{{Karim} et~al.,}{{Karim}
  et~al.}{2011}]{Karim11}
{Karim} A.,  et~al., 2011, \mn@doi [\apj] {10.1088/0004-637X/730/2/61}, \href
  {https://ui.adsabs.harvard.edu/abs/2011ApJ...730...61K} {730, 61}

\bibitem[\protect\citeauthoryear{{Kato} et~al.,}{{Kato} et~al.}{2016}]{Kato16}
{Kato} Y.,  et~al., 2016, \mn@doi [\mnras] {10.1093/mnras/stw1237}, \href
  {https://ui.adsabs.harvard.edu/abs/2016MNRAS.460.3861K} {460, 3861}

\bibitem[\protect\citeauthoryear{{Kodama}, {Balogh}, {Smail}, {Bower}  \&
  {Nakata}}{{Kodama} et~al.}{2004}]{Kodama04}
{Kodama} T.,  {Balogh} M.~L.,  {Smail} I.,  {Bower} R.~G.,   {Nakata} F.,
  2004, \mn@doi [\mnras] {10.1111/j.1365-2966.2004.08271.x}, \href
  {https://ui.adsabs.harvard.edu/abs/2004MNRAS.354.1103K} {354, 1103}

\bibitem[\protect\citeauthoryear{{Koyama} et~al.,}{{Koyama}
  et~al.}{2013}]{Koyama13}
{Koyama} Y.,  et~al., 2013, \mn@doi [\mnras] {10.1093/mnras/stt1035}, \href
  {https://ui.adsabs.harvard.edu/abs/2013MNRAS.434..423K} {434, 423}

\bibitem[\protect\citeauthoryear{{Lacaille} et~al.,}{{Lacaille}
  et~al.}{2019}]{Lacaille19}
{Lacaille} K.~M.,  et~al., 2019, \mn@doi [\mnras] {10.1093/mnras/stz1742},
  \href {https://ui.adsabs.harvard.edu/abs/2019MNRAS.488.1790L} {488, 1790}

\bibitem[\protect\citeauthoryear{{Lilly}, {Eales}, {Gear}, {Hammer}, {Le
  F{\`e}vre}, {Crampton}, {Bond}  \& {Dunne}}{{Lilly} et~al.}{1999}]{Lilly99}
{Lilly} S.~J.,  {Eales} S.~A.,  {Gear} W. K.~P.,  {Hammer} F.,  {Le F{\`e}vre}
  O.,  {Crampton} D.,  {Bond} J.~R.,   {Dunne} L.,  1999, \mn@doi [\apj]
  {10.1086/307310}, \href
  {https://ui.adsabs.harvard.edu/abs/1999ApJ...518..641L} {518, 641}

\bibitem[\protect\citeauthoryear{{Lim}, {Scott}, {Babul}, {Barnes}, {Kay},
  {McCarthy}, {Rennehan}  \& {Vogelsberger}}{{Lim} et~al.}{2021}]{Lim21}
{Lim} S.,  {Scott} D.,  {Babul} A.,  {Barnes} D.~J.,  {Kay} S.~T.,  {McCarthy}
  I.~G.,  {Rennehan} D.,   {Vogelsberger} M.,  2021, \mn@doi [\mnras]
  {10.1093/mnras/staa3693}, \href
  {https://ui.adsabs.harvard.edu/abs/2021MNRAS.501.1803L} {501, 1803}

\bibitem[\protect\citeauthoryear{{Ma} et~al.,}{{Ma} et~al.}{2015}]{Ma15}
{Ma} C.~J.,  et~al., 2015, \mn@doi [\apj] {10.1088/0004-637X/806/2/257}, \href
  {https://ui.adsabs.harvard.edu/abs/2015ApJ...806..257M} {806, 257}

\bibitem[\protect\citeauthoryear{{MacKenzie} et~al.,}{{MacKenzie}
  et~al.}{2017}]{MacKenzie17}
{MacKenzie} T.~P.,  et~al., 2017, \mn@doi [\mnras] {10.1093/mnras/stx512},
  \href {https://ui.adsabs.harvard.edu/abs/2017MNRAS.468.4006M} {468, 4006}

\bibitem[\protect\citeauthoryear{{Madau} \& {Dickinson}}{{Madau} \&
  {Dickinson}}{2014}]{Madau14}
{Madau} P.,  {Dickinson} M.,  2014, \mn@doi [\araa]
  {10.1146/annurev-astro-081811-125615}, \href
  {https://ui.adsabs.harvard.edu/abs/2014ARA&A..52..415M} {52, 415}

\bibitem[\protect\citeauthoryear{{Mairs} et~al.,}{{Mairs}
  et~al.}{2021}]{Mairs21}
{Mairs} S.,  et~al., 2021, \mn@doi [\aj] {10.3847/1538-3881/ac18bf}, \href
  {https://ui.adsabs.harvard.edu/abs/2021AJ....162..191M} {162, 191}

\bibitem[\protect\citeauthoryear{{Mantz} et~al.,}{{Mantz}
  et~al.}{2014}]{Mantz14}
{Mantz} A.~B.,  et~al., 2014, \mn@doi [\apj] {10.1088/0004-637X/794/2/157},
  \href {https://ui.adsabs.harvard.edu/abs/2014ApJ...794..157M} {794, 157}

\bibitem[\protect\citeauthoryear{{Marcillac}, {Rigby}, {Rieke}  \&
  {Kelly}}{{Marcillac} et~al.}{2007}]{Marcillac07}
{Marcillac} D.,  {Rigby} J.~R.,  {Rieke} G.~H.,   {Kelly} D.~M.,  2007, \mn@doi
  [\apj] {10.1086/509107}, \href
  {https://ui.adsabs.harvard.edu/abs/2007ApJ...654..825M} {654, 825}

\bibitem[\protect\citeauthoryear{{Martinache} et~al.,}{{Martinache}
  et~al.}{2018}]{Martinache18}
{Martinache} C.,  et~al., 2018, \mn@doi [\aap] {10.1051/0004-6361/201833198},
  \href {https://ui.adsabs.harvard.edu/abs/2018A&A...620A.198M} {620, A198}

\bibitem[\protect\citeauthoryear{{McKinney}, {Ramakrishnan}, {Lee}, {Pope},
  {Alberts}, {Chiang}  \& {Popescu}}{{McKinney} et~al.}{2022}]{McKinney22}
{McKinney} J.,  {Ramakrishnan} V.,  {Lee} K.-S.,  {Pope} A.,  {Alberts} S.,
  {Chiang} Y.-K.,   {Popescu} R.,  2022, \mn@doi [\apj]
  {10.3847/1538-4357/ac5110}, \href
  {https://ui.adsabs.harvard.edu/abs/2022ApJ...928...88M} {928, 88}

\bibitem[\protect\citeauthoryear{{Mei} et~al.,}{{Mei} et~al.}{2015}]{Mei15}
{Mei} S.,  et~al., 2015, \mn@doi [\apj] {10.1088/0004-637X/804/2/117}, \href
  {https://ui.adsabs.harvard.edu/abs/2015ApJ...804..117M} {804, 117}

\bibitem[\protect\citeauthoryear{{Men{\'e}ndez-Delmestre}
  et~al.,}{{Men{\'e}ndez-Delmestre} et~al.}{2009}]{Menendez09}
{Men{\'e}ndez-Delmestre} K.,  et~al., 2009, \mn@doi [\apj]
  {10.1088/0004-637X/699/1/667}, \href
  {https://ui.adsabs.harvard.edu/abs/2009ApJ...699..667M} {699, 667}

\bibitem[\protect\citeauthoryear{{Miettinen} et~al.,}{{Miettinen}
  et~al.}{2015}]{Miettinen15}
{Miettinen} O.,  et~al., 2015, \mn@doi [\aap] {10.1051/0004-6361/201425032},
  \href {https://ui.adsabs.harvard.edu/abs/2015A&A...577A..29M} {577, A29}

\bibitem[\protect\citeauthoryear{{Nantais} et~al.,}{{Nantais}
  et~al.}{2016}]{Nantais16}
{Nantais} J.~B.,  et~al., 2016, \mn@doi [\aap] {10.1051/0004-6361/201628663},
  \href {https://ui.adsabs.harvard.edu/abs/2016A&A...592A.161N} {592, A161}

\bibitem[\protect\citeauthoryear{{Nelan}, {Smith}, {Hudson}, {Wegner}, {Lucey},
  {Moore}, {Quinney}  \& {Suntzeff}}{{Nelan} et~al.}{2005}]{Nelan05}
{Nelan} J.~E.,  {Smith} R.~J.,  {Hudson} M.~J.,  {Wegner} G.~A.,  {Lucey}
  J.~R.,  {Moore} S. A.~W.,  {Quinney} S.~J.,   {Suntzeff} N.~B.,  2005,
  \mn@doi [\apj] {10.1086/431962}, \href
  {https://ui.adsabs.harvard.edu/abs/2005ApJ...632..137N} {632, 137}

\bibitem[\protect\citeauthoryear{{Newman}, {Ellis}, {Andreon}, {Treu},
  {Raichoor}  \& {Trinchieri}}{{Newman} et~al.}{2014}]{Newman14}
{Newman} A.~B.,  {Ellis} R.~S.,  {Andreon} S.,  {Treu} T.,  {Raichoor} A.,
  {Trinchieri} G.,  2014, \mn@doi [\apj] {10.1088/0004-637X/788/1/51}, \href
  {https://ui.adsabs.harvard.edu/abs/2014ApJ...788...51N} {788, 51}

\bibitem[\protect\citeauthoryear{{Noble} et~al.,}{{Noble}
  et~al.}{2017}]{Noble17}
{Noble} A.~G.,  et~al., 2017, \mn@doi [\apjl] {10.3847/2041-8213/aa77f3}, \href
  {https://ui.adsabs.harvard.edu/abs/2017ApJ...842L..21N} {842, L21}

\bibitem[\protect\citeauthoryear{{Nowotka}, {Chen}, {Battaia}, {Fumagalli},
  {Cai}, {Lusso}, {Prochaska}  \& {Yang}}{{Nowotka} et~al.}{2022}]{Nowotka22}
{Nowotka} M.,  {Chen} C.-C.,  {Battaia} F.~A.,  {Fumagalli} M.,  {Cai} Z.,
  {Lusso} E.,  {Prochaska} J.~X.,   {Yang} Y.,  2022, \mn@doi [\aap]
  {10.1051/0004-6361/202040133}, \href
  {https://ui.adsabs.harvard.edu/abs/2022A&A...658A..77N} {658, A77}

\bibitem[\protect\citeauthoryear{{Papovich} et~al.,}{{Papovich}
  et~al.}{2007}]{Papovich07}
{Papovich} C.,  et~al., 2007, \mn@doi [\apj] {10.1086/521090}, \href
  {https://ui.adsabs.harvard.edu/abs/2007ApJ...668...45P} {668, 45}

\bibitem[\protect\citeauthoryear{{Poggianti} et~al.,}{{Poggianti}
  et~al.}{2001}]{Poggianti01}
{Poggianti} B.~M.,  et~al., 2001, \mn@doi [\apj] {10.1086/323767}, \href
  {https://ui.adsabs.harvard.edu/abs/2001ApJ...563..118P} {563, 118}

\bibitem[\protect\citeauthoryear{{Polletta}, {Dole}, {Martinache}, {Lehnert},
  {Frye}  \& {Kneissl}}{{Polletta} et~al.}{2022}]{Polletta22}
{Polletta} M.,  {Dole} H.,  {Martinache} C.,  {Lehnert} M.~D.,  {Frye} B.~L.,
  {Kneissl} R.,  2022, \mn@doi [\aap] {10.1051/0004-6361/202142255}, \href
  {https://ui.adsabs.harvard.edu/abs/2022A&A...662A..85P} {662, A85}

\bibitem[\protect\citeauthoryear{{Popesso} et~al.,}{{Popesso}
  et~al.}{2012}]{Popesso12}
{Popesso} P.,  et~al., 2012, \mn@doi [\aap] {10.1051/0004-6361/201117973},
  \href {https://ui.adsabs.harvard.edu/abs/2012A&A...537A..58P} {537, A58}

\bibitem[\protect\citeauthoryear{{Popesso} et~al.,}{{Popesso}
  et~al.}{2015a}]{Popesso15}
{Popesso} P.,  et~al., 2015a, \mn@doi [\aap] {10.1051/0004-6361/201424711},
  \href {https://ui.adsabs.harvard.edu/abs/2015A&A...574A.105P} {574, A105}

\bibitem[\protect\citeauthoryear{{Popesso} et~al.,}{{Popesso}
  et~al.}{2015b}]{Popesso15b}
{Popesso} P.,  et~al., 2015b, \mn@doi [\aap] {10.1051/0004-6361/201424715},
  \href {https://ui.adsabs.harvard.edu/abs/2015A&A...579A.132P} {579, A132}

\bibitem[\protect\citeauthoryear{{Rettura}, {Chary}, {Krick}  \&
  {Ettori}}{{Rettura} et~al.}{2018}]{Rettura18}
{Rettura} A.,  {Chary} R.,  {Krick} J.,   {Ettori} S.,  2018, \mn@doi [\apj]
  {10.3847/1538-4357/aad818}, \href
  {https://ui.adsabs.harvard.edu/abs/2018ApJ...867...12R} {867, 12}

\bibitem[\protect\citeauthoryear{{Rieke} et~al.,}{{Rieke}
  et~al.}{2004}]{Rieke04}
{Rieke} G.~H.,  et~al., 2004, \mn@doi [\apjs] {10.1086/422717}, \href
  {https://ui.adsabs.harvard.edu/abs/2004ApJS..154...25R} {154, 25}

\bibitem[\protect\citeauthoryear{{Rigby} et~al.,}{{Rigby}
  et~al.}{2014}]{Rigby14}
{Rigby} E.~E.,  et~al., 2014, \mn@doi [\mnras] {10.1093/mnras/stt2019}, \href
  {https://ui.adsabs.harvard.edu/abs/2014MNRAS.437.1882R} {437, 1882}

\bibitem[\protect\citeauthoryear{{Roseboom} et~al.,}{{Roseboom}
  et~al.}{2010}]{Roseboom10}
{Roseboom} I.~G.,  et~al., 2010, \mn@doi [\mnras]
  {10.1111/j.1365-2966.2010.17634.x}, \href
  {https://ui.adsabs.harvard.edu/abs/2010MNRAS.409...48R} {409, 48}

\bibitem[\protect\citeauthoryear{{Rotermund} et~al.,}{{Rotermund}
  et~al.}{2021}]{Rotermund21}
{Rotermund} K.~M.,  et~al., 2021, \mn@doi [\mnras] {10.1093/mnras/stab103},
  \href {https://ui.adsabs.harvard.edu/abs/2021MNRAS.502.1797R} {502, 1797}

\bibitem[\protect\citeauthoryear{{Santos} et~al.,}{{Santos}
  et~al.}{2014}]{Santos14}
{Santos} J.~S.,  et~al., 2014, \mn@doi [\mnras] {10.1093/mnras/stt2376}, \href
  {https://ui.adsabs.harvard.edu/abs/2014MNRAS.438.2565S} {438, 2565}

\bibitem[\protect\citeauthoryear{{Santos} et~al.,}{{Santos}
  et~al.}{2015}]{Santos15}
{Santos} J.~S.,  et~al., 2015, \mn@doi [\mnras] {10.1093/mnrasl/slu180}, \href
  {https://ui.adsabs.harvard.edu/abs/2015MNRAS.447L..65S} {447, L65}

\bibitem[\protect\citeauthoryear{{Shim} et~al.,}{{Shim} et~al.}{2022}]{Shim22}
{Shim} H.,  et~al., 2022, \mn@doi [\mnras] {10.1093/mnras/stac1105}, \href
  {https://ui.adsabs.harvard.edu/abs/2022MNRAS.514.2915S} {514, 2915}

\bibitem[\protect\citeauthoryear{{Simpson} et~al.,}{{Simpson}
  et~al.}{2019}]{Simpson19}
{Simpson} J.~M.,  et~al., 2019, \mn@doi [\apj] {10.3847/1538-4357/ab23ff},
  \href {https://ui.adsabs.harvard.edu/abs/2019ApJ...880...43S} {880, 43}

\bibitem[\protect\citeauthoryear{{Simpson} et~al.,}{{Simpson}
  et~al.}{2020}]{Simpson20}
{Simpson} J.~M.,  et~al., 2020, \mn@doi [\mnras] {10.1093/mnras/staa1345},
  \href {https://ui.adsabs.harvard.edu/abs/2020MNRAS.495.3409S} {495, 3409}

\bibitem[\protect\citeauthoryear{{Smail}, {Ivison}, {Kneib}, {Cowie}, {Blain},
  {Barger}, {Owen}  \& {Morrison}}{{Smail} et~al.}{1999}]{Smail99}
{Smail} I.,  {Ivison} R.~J.,  {Kneib} J.~P.,  {Cowie} L.~L.,  {Blain} A.~W.,
  {Barger} A.~J.,  {Owen} F.~N.,   {Morrison} G.,  1999, \mn@doi [\mnras]
  {10.1046/j.1365-8711.1999.02819.x}, \href
  {https://ui.adsabs.harvard.edu/abs/1999MNRAS.308.1061S} {308, 1061}

\bibitem[\protect\citeauthoryear{{Smail} et~al.,}{{Smail}
  et~al.}{2014}]{Smail14}
{Smail} I.,  et~al., 2014, \mn@doi [\apj] {10.1088/0004-637X/782/1/19}, \href
  {https://ui.adsabs.harvard.edu/abs/2014ApJ...782...19S} {782, 19}

\bibitem[\protect\citeauthoryear{{Smail} et~al.,}{{Smail}
  et~al.}{2023}]{Smail23}
{Smail} I.,  et~al., 2023, \mn@doi [\apj] {10.3847/1538-4357/acf931}, \href
  {https://ui.adsabs.harvard.edu/abs/2023ApJ...958...36S} {958, 36}

\bibitem[\protect\citeauthoryear{{Smith}}{{Smith}}{2020}]{Smith20}
{Smith} C.,  2020, PhD thesis, Cardiff University

\bibitem[\protect\citeauthoryear{{Smith}, {Gear}, {Smith}, {Papageorgiou}  \&
  {Eales}}{{Smith} et~al.}{2019}]{Smith19}
{Smith} C.~M.~A.,  {Gear} W.~K.,  {Smith} M.~W.~L.,  {Papageorgiou} A.,
  {Eales} S.~A.,  2019, \mn@doi [\mnras] {10.1093/mnras/stz1090}, \href
  {https://ui.adsabs.harvard.edu/abs/2019MNRAS.486.4304S} {486, 4304}

\bibitem[\protect\citeauthoryear{{Spitzer} \& {Baade}}{{Spitzer} \&
  {Baade}}{1951}]{Spitzer51}
{Spitzer} Lyman J.,  {Baade} W.,  1951, \mn@doi [\apj] {10.1086/145406}, \href
  {https://ui.adsabs.harvard.edu/abs/1951ApJ...113..413S} {113, 413}

\bibitem[\protect\citeauthoryear{{Stach}, {Swinbank}, {Smail}, {Hilton},
  {Simpson}  \& {Cooke}}{{Stach} et~al.}{2017}]{Stach17}
{Stach} S.~M.,  {Swinbank} A.~M.,  {Smail} I.,  {Hilton} M.,  {Simpson} J.~M.,
   {Cooke} E.~A.,  2017, \mn@doi [\apj] {10.3847/1538-4357/aa93f6}, \href
  {https://ui.adsabs.harvard.edu/abs/2017ApJ...849..154S} {849, 154}

\bibitem[\protect\citeauthoryear{{Stach} et~al.,}{{Stach}
  et~al.}{2018}]{Stach18}
{Stach} S.~M.,  et~al., 2018, \mn@doi [\apj] {10.3847/1538-4357/aac5e5}, \href
  {https://ui.adsabs.harvard.edu/abs/2018ApJ...860..161S} {860, 161}

\bibitem[\protect\citeauthoryear{{Stach} et~al.,}{{Stach}
  et~al.}{2019}]{Stach19}
{Stach} S.~M.,  et~al., 2019, \mn@doi [\mnras] {10.1093/mnras/stz1536}, \href
  {https://ui.adsabs.harvard.edu/abs/2019MNRAS.487.4648S} {487, 4648}

\bibitem[\protect\citeauthoryear{{Stevens} et~al.,}{{Stevens}
  et~al.}{2003}]{Stevens03}
{Stevens} J.~A.,  et~al., 2003, \mn@doi [\nat]
  {10.48550/arXiv.astro-ph/0309495}, \href
  {https://ui.adsabs.harvard.edu/abs/2003Natur.425..264S} {425, 264}

\bibitem[\protect\citeauthoryear{{Strazzullo} et~al.,}{{Strazzullo}
  et~al.}{2018}]{Strazzullo18}
{Strazzullo} V.,  et~al., 2018, \mn@doi [\apj] {10.3847/1538-4357/aacd10},
  \href {https://ui.adsabs.harvard.edu/abs/2018ApJ...862...64S} {862, 64}

\bibitem[\protect\citeauthoryear{{Swinbank} et~al.,}{{Swinbank}
  et~al.}{2014}]{Swinbank14}
{Swinbank} A.~M.,  et~al., 2014, \mn@doi [\mnras] {10.1093/mnras/stt2273},
  \href {https://ui.adsabs.harvard.edu/abs/2014MNRAS.438.1267S} {438, 1267}

\bibitem[\protect\citeauthoryear{{Tadaki} et~al.,}{{Tadaki}
  et~al.}{2012}]{Tadaki12}
{Tadaki} K.-i.,  et~al., 2012, \mn@doi [\mnras]
  {10.1111/j.1365-2966.2012.21063.x}, \href
  {https://ui.adsabs.harvard.edu/abs/2012MNRAS.423.2617T} {423, 2617}

\bibitem[\protect\citeauthoryear{{Taylor}}{{Taylor}}{2005}]{Taylor05}
{Taylor} M.~B.,  2005, in {Shopbell} P.,  {Britton} M.,   {Ebert} R.,  eds,
  Astronomical Society of the Pacific Conference Series Vol. 347, Astronomical
  Data Analysis Software and Systems XIV. p.~29

\bibitem[\protect\citeauthoryear{{Tran} et~al.,}{{Tran} et~al.}{2010}]{Tran10}
{Tran} K.-V.~H.,  et~al., 2010, \mn@doi [\apjl] {10.1088/2041-8205/719/2/L126},
  \href {https://ui.adsabs.harvard.edu/abs/2010ApJ...719L.126T} {719, L126}

\bibitem[\protect\citeauthoryear{{Umehata} et~al.,}{{Umehata}
  et~al.}{2015}]{Umehata15}
{Umehata} H.,  et~al., 2015, \mn@doi [\apjl] {10.1088/2041-8205/815/1/L8},
  \href {https://ui.adsabs.harvard.edu/abs/2015ApJ...815L...8U} {815, L8}

\bibitem[\protect\citeauthoryear{{Wagner}, {Courteau}, {Brodwin}, {Stanford},
  {Snyder}  \& {Stern}}{{Wagner} et~al.}{2017}]{Wagner17}
{Wagner} C.~R.,  {Courteau} S.,  {Brodwin} M.,  {Stanford} S.~A.,  {Snyder}
  G.~F.,   {Stern} D.,  2017, \mn@doi [\apj] {10.3847/1538-4357/834/1/53},
  \href {https://ui.adsabs.harvard.edu/abs/2017ApJ...834...53W} {834, 53}

\bibitem[\protect\citeauthoryear{{Warren-Smith} \& {Wallace}}{{Warren-Smith} \&
  {Wallace}}{1993}]{WarrenSmith93}
{Warren-Smith} R.~F.,  {Wallace} P.~T.,  1993, in {Hanisch} R.~J.,
  {Brissenden} R.~J.~V.,   {Barnes} J.,  eds,  Astronomical Society of the
  Pacific Conference Series Vol. 52, Astronomical Data Analysis Software and
  Systems II. p.~229

\bibitem[\protect\citeauthoryear{{Watson} et~al.,}{{Watson}
  et~al.}{2019}]{Watson19}
{Watson} C.,  et~al., 2019, \mn@doi [\apj] {10.3847/1538-4357/ab06ef}, \href
  {https://ui.adsabs.harvard.edu/abs/2019ApJ...874...63W} {874, 63}

\bibitem[\protect\citeauthoryear{{Webb}, {Yee}, {Ivison}, {Hoekstra},
  {Gladders}, {Barrientos}  \& {Hsieh}}{{Webb} et~al.}{2005}]{Webb05}
{Webb} T.~M.~A.,  {Yee} H.~K.~C.,  {Ivison} R.~J.,  {Hoekstra} H.,  {Gladders}
  M.~D.,  {Barrientos} L.~F.,   {Hsieh} B.~C.,  2005, \mn@doi [\apj]
  {10.1086/432524}, \href
  {https://ui.adsabs.harvard.edu/abs/2005ApJ...631..187W} {631, 187}

\bibitem[\protect\citeauthoryear{{Webb} et~al.,}{{Webb} et~al.}{2013}]{Webb13}
{Webb} T.~M.~A.,  et~al., 2013, \mn@doi [\aj] {10.1088/0004-6256/146/4/84},
  \href {https://ui.adsabs.harvard.edu/abs/2013AJ....146...84W} {146, 84}

\bibitem[\protect\citeauthoryear{{Willis} et~al.,}{{Willis}
  et~al.}{2013}]{Willis13}
{Willis} J.~P.,  et~al., 2013, \mn@doi [\mnras] {10.1093/mnras/sts540}, \href
  {https://ui.adsabs.harvard.edu/abs/2013MNRAS.430..134W} {430, 134}

\bibitem[\protect\citeauthoryear{{Wu} et~al.,}{{Wu} et~al.}{2018}]{Wu18}
{Wu} J.~F.,  et~al., 2018, \mn@doi [\apj] {10.3847/1538-4357/aaa0dc}, \href
  {https://ui.adsabs.harvard.edu/abs/2018ApJ...853..195W} {853, 195}

\bibitem[\protect\citeauthoryear{{Yun} et~al.,}{{Yun} et~al.}{2008}]{Yun08}
{Yun} M.~S.,  et~al., 2008, \mn@doi [\mnras]
  {10.1111/j.1365-2966.2008.13565.x}, \href
  {https://ui.adsabs.harvard.edu/abs/2008MNRAS.389..333Y} {389, 333}

\bibitem[\protect\citeauthoryear{{Zeballos} et~al.,}{{Zeballos}
  et~al.}{2018}]{Zeballos18}
{Zeballos} M.,  et~al., 2018, \mn@doi [\mnras] {10.1093/mnras/sty1714}, \href
  {https://ui.adsabs.harvard.edu/abs/2018MNRAS.479.4577Z} {479, 4577}

\bibitem[\protect\citeauthoryear{{Zeimann} et~al.,}{{Zeimann}
  et~al.}{2012}]{Zeimann12}
{Zeimann} G.~R.,  et~al., 2012, \mn@doi [\apj] {10.1088/0004-637X/756/2/115},
  \href {https://ui.adsabs.harvard.edu/abs/2012ApJ...756..115Z} {756, 115}

\bibitem[\protect\citeauthoryear{{Zhang} et~al.,}{{Zhang}
  et~al.}{2022}]{Zhang22}
{Zhang} Y.,  et~al., 2022, \mn@doi [\mnras] {10.1093/mnras/stac824}, \href
  {https://ui.adsabs.harvard.edu/abs/2022MNRAS.512.4893Z} {512, 4893}

\bibitem[\protect\citeauthoryear{{van Marrewijk} et~al.,}{{van Marrewijk}
  et~al.}{2023}]{vanMarrewijk23}
{van Marrewijk} J.,  et~al., 2023, \mn@doi [arXiv e-prints]
  {10.48550/arXiv.2310.06120}, \href
  {https://ui.adsabs.harvard.edu/abs/2023arXiv231006120V} {p. arXiv:2310.06120}

\makeatother
\end{thebibliography}

%
%
\begin{table*}
\caption{{\sc Main}  Sample}
\begin{tabular}{lccrc}
\hline \noalign {\smallskip}
ID  & R.A.\ & Dec.\ & SNR$_{850}$ &  $S_{\rm 850\mu m}$ \\
  &  \multicolumn{2}{c}{(J2000)} & & (mJy)  \\
\hline \noalign {\smallskip}
XLSSC122.001 & 02:17:43.03 & $-$03:45:32.0 & 4.95 & 5.0\,$\pm$\,1.0 \\ 
XLSSC122.002 & 02:17:41.16 & $-$03:48:00.0 & 4.69 & 5.9\,$\pm$\,1.3 \\ 
    XLSSC122.003 & 02:17:41.16 & $-$03:45:32.0 & 4.49 & 4.3\,$\pm$\,1.0 \\ 
    XLSSC122.004 & 02:17:37.69 & $-$03:46:52.0 & 4.19 & 5.1\,$\pm$\,1.2 \\ 
    XLSSC122.005 & 02:17:43.57 & $-$03:49:20.0 & 4.04 & 5.2\,$\pm$\,1.3 \\ 
    XLSSC122.006 & 02:17:41.16 & $-$03:47:36.0 & 3.66 & 3.9\,$\pm$\,1.1 \\ 
    XLSSC122.007 & 02:17:52.92 & $-$03:46:08.0 & 3.65 & 4.5\,$\pm$\,1.2 \\ 
    XLSSC122.008 & 02:17:37.95 & $-$03:46:08.0 & 3.62 & 3.4\,$\pm$\,0.9 \\ 
 \noalign {\smallskip}
 SpARCSJ0224.001 & 02:24:16.45 & $-$03:24:02.8 & 6.59 & 8.4\,$\pm$\,1.3 \\ 
 SpARCSJ0224.002 & 02:24:19.65 & $-$03:22:34.8 & 4.66 & 4.9\,$\pm$\,1.1 \\ 
 SpARCSJ0224.003 & 02:24:19.12 & $-$03:24:18.8 & 4.52 & 4.7\,$\pm$\,1.0 \\ 
 SpARCSJ0224.004 & 02:24:29.80 & $-$03:23:38.8 & 4.38 & 4.2\,$\pm$\,1.0 \\ 
 SpARCSJ0224.005 & 02:24:28.20 & $-$03:26:30.8 & 4.16 & 5.0\,$\pm$\,1.2 \\ 
 SpARCSJ0224.006 & 02:24:34.61 & $-$03:22:42.8 & 3.85 & 4.5\,$\pm$\,1.2 \\ 
 SpARCSJ0224.007 & 02:24:33.01 & $-$03:22:54.8 & 3.86 & 4.1\,$\pm$\,1.1 \\ 
 SpARCSJ0224.008 & 02:24:17.25 & $-$03:25:06.8 & 3.56 & 3.7\,$\pm$\,1.0 \\ 
 SpARCSJ0224.009 & 02:24:19.65 & $-$03:24:50.8 & 3.51 & 3.2\,$\pm$\,0.9 \\ 
 \noalign {\smallskip}
 SpARCSJ0225.001 & 02:25:43.95 & $-$03:56:45.1 & 4.50 & 4.3\,$\pm$\,1.0 \\ 
 SpARCSJ0225.002 & 02:25:34.06 & $-$03:56:05.1 & 4.08 & 4.5\,$\pm$\,1.1 \\ 
 SpARCSJ0225.003 & 02:25:38.60 & $-$03:57:45.1 & 3.71 & 4.0\,$\pm$\,1.1 \\ 
\noalign {\smallskip}
JKCS041.001 & 02:26:42.39 & $-$04:42:16.0 & 5.66 & 5.4\,$\pm$\,1.0 \\ 
     JKCS041.002 & 02:26:46.68 & $-$04:42:16.0 & 4.57 & 4.2\,$\pm$\,0.9 \\ 
     JKCS041.003 & 02:26:42.93 & $-$04:39:56.0 & 4.30 & 4.2\,$\pm$\,1.0 \\ 
     JKCS041.004 & 02:26:32.76 & $-$04:43:04.0 & 4.04 & 4.9\,$\pm$\,1.2 \\ 
     JKCS041.005 & 02:26:53.10 & $-$04:41:32.0 & 3.91 & 4.0\,$\pm$\,1.0 \\ 
     JKCS041.006 & 02:26:49.62 & $-$04:42:48.0 & 3.67 & 3.6\,$\pm$\,1.0 \\ 
     JKCS041.007 & 02:26:42.39 & $-$04:43:56.0 & 3.51 & 3.2\,$\pm$\,0.9 \\ 
\noalign {\smallskip}
LH146.001 & 10:53:22.09 & $+$57:23:12.0 & 5.46 & 5.4\,$\pm$\,1.0 \\ 
       LH146.002 & 10:53:25.56 & $+$57:22:52.0 & 5.52 & 5.7\,$\pm$\,1.0 \\ 
       LH146.003 & 10:53:43.89 & $+$57:25:39.9 & 5.37 & 6.8\,$\pm$\,1.3 \\ 
       LH146.004 & 10:53:14.18 & $+$57:24:12.0 & 5.20 & 5.2\,$\pm$\,1.0 \\ 
       LH146.005 & 10:53:15.16 & $+$57:24:40.0 & 4.64 & 4.4\,$\pm$\,0.9 \\ 
       LH146.006 & 10:53:44.86 & $+$57:23:27.9 & 4.29 & 5.3\,$\pm$\,1.2 \\ 
       LH146.007 & 10:53:17.14 & $+$57:27:24.0 & 4.24 & 4.8\,$\pm$\,1.1 \\ 
       LH146.008 & 10:53:40.92 & $+$57:26:23.9 & 4.00 & 4.6\,$\pm$\,1.2 \\ 
       LH146.009 & 10:53:19.13 & $+$57:21:12.0 & 3.81 & 4.8\,$\pm$\,1.2 \\ 
       LH146.010 & 10:53:16.65 & $+$57:25:12.0 & 3.63 & 3.0\,$\pm$\,0.8 \\ 
       LH146.011 & 10:53:48.33 & $+$57:24:03.8 & 3.58 & 3.9\,$\pm$\,1.1 \\ 
       LH146.012 & 10:53:44.87 & $+$57:24:27.9 & 3.57 & 3.7\,$\pm$\,1.0 \\ 
  \noalign {\smallskip}
  IDCSJ1426.001 & 14:26:38.24 & $+$35:09:17.0 & 4.69 & 4.3\,$\pm$\,0.9 \\ 
   IDCSJ1426.002 & 14:26:36.29 & $+$35:07:13.0 & 4.27 & 3.9\,$\pm$\,0.9 \\ 
   IDCSJ1426.003 & 14:26:28.46 & $+$35:10:17.0 & 3.65 & 3.3\,$\pm$\,0.9 \\ 
   IDCSJ1426.004 & 14:26:45.09 & $+$35:06:57.0 & 3.61 & 3.8\,$\pm$\,1.1 \\ 
   IDCSJ1426.005 & 14:26:37.92 & $+$35:06:17.0 & 3.58 & 3.5\,$\pm$\,1.0 \\ 
  \noalign {\smallskip}
  IDCSJ1433.001 & 14:33:04.32 & $+$33:09:34.2 & 13.09 & 15.7\,$\pm$\,1.2 \\ 
   IDCSJ1433.002 & 14:33:16.73 & $+$33:05:58.2 & 6.80 & 8.0\,$\pm$\,1.2 \\ 
   IDCSJ1433.003 & 14:33:06.55 & $+$33:08:38.2 & 5.68 & 5.1\,$\pm$\,0.9 \\ 
   IDCSJ1433.004 & 14:33:03.68 & $+$33:06:14.2 & 5.06 & 3.8\,$\pm$\,0.8 \\ 
   IDCSJ1433.005 & 14:33:04.96 & $+$33:07:26.2 & 4.75 & 3.5\,$\pm$\,0.7 \\ 
   IDCSJ1433.006 & 14:32:47.77 & $+$33:05:50.1 & 4.17 & 4.5\,$\pm$\,1.1 \\ 
   IDCSJ1433.007 & 14:33:13.55 & $+$33:07:30.2 & 4.06 & 3.6\,$\pm$\,0.9 \\ 
  \noalign {\smallskip}
  ClJ1449.001 & 14:49:12.92 & $+$08:58:13.0 & 6.42 & 7.1\,$\pm$\,1.1 \\ 
     ClJ1449.002 & 14:49:07.52 & $+$08:53:53.0 & 5.43 & 6.6\,$\pm$\,1.2 \\ 
     ClJ1449.003 & 14:49:14.27 & $+$08:56:13.0 & 5.37 & 5.0\,$\pm$\,0.9 \\ 
     ClJ1449.004 & 14:49:08.06 & $+$08:57:25.0 & 4.76 & 4.9\,$\pm$\,1.0 \\ 
  ClJ1449.005 & 14:49:24.53 & $+$08:55:45.0 & 4.73 & 5.7\,$\pm$\,1.2 \\
  \hline \noalign {\smallskip}
\end{tabular}
\end{table*}


%
%
 \begin{table*}
\caption{{\sc Supplementary}  Sample}
\begin{tabular}{lccrc}
\hline \noalign {\smallskip}
ID  & R.A.\ & Dec.\ & SNR$_{850}$ &  $S_{\rm 850\mu m}$  \\
  &  \multicolumn{2}{c}{(J2000)} & & (mJy)  \\
\hline \noalign {\smallskip}
   XLSSC122.009 & 02:17:39.56 & $-$03:47:48.0 & 3.44 & 3.7\,$\pm$\,1.1 \\ 
    XLSSC122.011 & 02:17:44.10 & $-$03:47:52.0 & 3.07 & 2.9\,$\pm$\,0.9 \\ 
  \noalign {\smallskip}
  SpARCSJ0224.011 & 02:24:31.67 & $-$03:23:58.8 & 3.38 & 3.0\,$\pm$\,0.9 \\ 
 SpARCSJ0224.012 & 02:24:29.80 & $-$03:24:14.8 & 3.30 & 2.7\,$\pm$\,0.8 \\ 
 SpARCSJ0224.013 & 02:24:27.13 & $-$03:24:02.8 & 3.26 & 2.5\,$\pm$\,0.8 \\ 
 SpARCSJ0224.014 & 02:24:26.60 & $-$03:23:34.8 & 3.17 & 2.3\,$\pm$\,0.7 \\ 
 SpARCSJ0224.016 & 02:24:19.65 & $-$03:23:50.8 & 3.09 & 2.3\,$\pm$\,0.8 \\ 
  \noalign {\smallskip}
  JKCS041.009 & 02:26:38.92 & $-$04:41:08.0 & 3.32 & 2.6\,$\pm$\,0.8 \\ 
     JKCS041.010 & 02:26:39.18 & $-$04:43:12.0 & 3.31 & 2.9\,$\pm$\,0.9 \\ 
     JKCS041.011 & 02:26:47.75 & $-$04:41:12.0 & 3.10 & 2.3\,$\pm$\,0.7 \\ 
  \noalign {\smallskip}
  LH146.014 & 10:53:26.06 & $+$57:25:36.0 & 3.42 & 2.8\,$\pm$\,0.8 \\ 
       LH146.015 & 10:53:27.04 & $+$57:23:12.0 & 3.39 & 2.6\,$\pm$\,0.8 \\ 
       LH146.017 & 10:53:32.49 & $+$57:24:48.0 & 3.37 & 2.6\,$\pm$\,0.8 \\ 
       LH146.020 & 10:53:17.64 & $+$57:24:00.0 & 3.12 & 2.2\,$\pm$\,0.7 \\ 
       LH146.022 & 10:53:13.68 & $+$57:23:32.0 & 3.05 & 2.3\,$\pm$\,0.8 \\ 
  \noalign {\smallskip}
  IDCSJ1426.007 & 14:26:25.85 & $+$35:09:09.0 & 3.16 & 2.4\,$\pm$\,0.8 \\ 
  \noalign {\smallskip}
  ClJ1449.007 & 14:49:17.78 & $+$08:56:53.0 & 3.31 & 2.6\,$\pm$\,0.8 \\ 
     ClJ1449.009 & 14:49:16.43 & $+$08:56:13.0 & 3.23 & 2.4\,$\pm$\,0.7 \\ 
 \hline \noalign {\smallskip}
\end{tabular}
\end{table*}

%
%
\begin{table*}
  \caption{{\sc Main}  Identifications}
\begin{tabular}{lcccccccc}
\hline \noalign {\smallskip}
ID  & R.A.~~~~~~~~~~~~Dec.\ & $S_{\rm 3.6\mu m}$ &  $S_{\rm 4.5\mu m}$ & $S_{\rm 5.8\mu m}$ & $S_{\rm 24\mu m}$ & $\Delta \theta_{\rm S2}$ & $P_{\rm MIPS}$ & $P^{\rm mem}_{\rm IRAC}$ \\
  &  (J2000) & ($\mu$Jy) & ($\mu$Jy) &($\mu$Jy) & ($\mu$Jy) & ($''$) & (\%) & (\%) \\
\hline \noalign {\smallskip}
{\bf XLSSC122.001.0}$^{1\ast}$ & 02:17:42.78 $-$03:45:31.1 &  29.9\,$\pm$\,0.5 &  65.7\,$\pm$\,1.1 & 174.8\,$\pm$\,4.2 &  3162\,$\pm$\,61 & 3.82 &  {\bf 0.09} &  {\bf 1.04} \\ 
{\bf XLSSC122.002.0}$^2$ & 02:17:41.18 $-$03:47:59.8 &  64.0\,$\pm$\,0.7 &  78.3\,$\pm$\,1.2 & 100.1\,$\pm$\,4.3 &   654\,$\pm$\,56 & 0.31 &  {\bf 0.02} &  {\bf 0.01} \\ 
{\bf XLSSC122.003.0}$^\ast$ & 02:17:41.24 $-$03:45:31.9 &  30.2\,$\pm$\,0.6 &  65.1\,$\pm$\,1.1 & 145.4\,$\pm$\,4.6 &  1416\,$\pm$\,66 & 1.25 &  {\bf 0.05} &  {\bf 0.15} \\ 
{\bf XLSSC122.004.0}$^\ast$ & 02:17:37.69 $-$03:46:50.9 &  89.9\,$\pm$\,0.8 & 127.2\,$\pm$\,1.2 & 206.8\,$\pm$\,5.5 &  1390\,$\pm$\,66 & 1.07 &  {\bf 0.04} &  {\bf 0.03} \\ 
{\bf XLSSC122.006.0} & 02:17:41.19 $-$03:47:35.9 &  16.0\,$\pm$\,0.6 &  26.6\,$\pm$\,1.1 &  50.5\,$\pm$\,4.8 &   237\,$\pm$\,55 & 0.51 &  {\bf 0.20} &  {\bf 0.14} \\ 
{\bf XLSSC122.006.1}$^\ast$ & 02:17:41.01 $-$03:47:41.9 &  29.4\,$\pm$\,0.6 &  41.4\,$\pm$\,1.1 &  47.2\,$\pm$\,4.7 &     ... & 6.37 & ... &  {\bf 3.71} \\ 
\noalign {\smallskip}
SpARCSJ0224.001.0 & 02:24:16.25 $-$03:24:04.4 &  34.1\,$\pm$\,0.6 &  38.9\,$\pm$\,1.0 &  47.2\,$\pm$\,3.8 &   514\,$\pm$\,62 & 3.36 &  {\bf 1.08} &  {\bf 1.52} \\ 
{\bf SpARCSJ0224.002.0}$^\ast$ & 02:24:19.38 $-$03:22:34.0 &  19.5\,$\pm$\,0.5 &  21.8\,$\pm$\,1.0 &  28.3\,$\pm$\,3.8 &   508\,$\pm$\,66 & 4.17 &  {\bf 1.51} &  {\bf 3.97} \\ 
{\bf SpARCSJ0224.004.0}$^3$ & 02:24:29.98 $-$03:23:40.8 &  33.5\,$\pm$\,0.6 &  45.0\,$\pm$\,1.1 &  56.8\,$\pm$\,4.2 &   384\,$\pm$\,65 & 3.33 &  {\bf 0.20} &  {\bf 1.54} \\ 
SpARCSJ0224.005.0 & 02:24:28.38 $-$03:26:34.8 &   9.0\,$\pm$\,0.6 &  10.6\,$\pm$\,0.8 &  17.7\,$\pm$\,4.3 &   200\,$\pm$\,65 & 4.76 &  {\bf 4.89} &  6.46 \\ 
SpARCSJ0224.005.1$^\ast$ & 02:24:28.09 $-$03:26:29.2 &  13.2\,$\pm$\,0.5 &  17.8\,$\pm$\,0.8 &  23.1\,$\pm$\,4.3 &     ... & 2.29 & ... &  {\bf 2.03} \\ 
{\it SpARCSJ0224.006.0} & 02:24:34.76 $-$03:22:40.3 & 389.3\,$\pm$\,1.2 & 282.3\,$\pm$\,1.4 & 428.3\,$\pm$\,5.0 &  4608\,$\pm$\,55 & 3.32 &  {\bf 0.07} & ... \\ 
{\it SpARCSJ0224.006.1} & 02:24:34.75 $-$03:22:45.2 & 164.8\,$\pm$\,0.9 & 120.7\,$\pm$\,1.2 & 187.9\,$\pm$\,4.5 &  2176\,$\pm$\,55 & 3.19 &  {\bf 0.13} & ... \\ 
{\bf SpARCSJ0224.007.0} & 02:24:32.78 $-$03:22:57.1 &  31.7\,$\pm$\,0.5 &  40.1\,$\pm$\,1.1 &  36.5\,$\pm$\,4.5 &   292\,$\pm$\,55 & 4.07 &  {\bf 3.17} &  {\bf 2.05} \\ 
{\bf SpARCSJ0224.007.1} & 02:24:33.17 $-$03:22:51.0 &  52.9\,$\pm$\,0.6 &  65.0\,$\pm$\,1.0 &  67.3\,$\pm$\,3.7 &     ... & 4.49 & ... &  {\bf 1.08} \\ 
\noalign {\smallskip}
{\bf SpARCSJ0225.001.0} & 02:25:43.85 $-$03:56:39.7 &  38.0\,$\pm$\,0.8 &  50.9\,$\pm$\,1.1 &  41.5\,$\pm$\,5.1 &   700\,$\pm$\,50 & 5.55 &  {\bf 3.00} &  {\bf 2.55} \\ 
SpARCSJ0225.003.0$^\ast$ & 02:25:38.40 $-$03:57:46.9 &  10.7\,$\pm$\,0.6 &  18.1\,$\pm$\,1.0 &  20.4\,$\pm$\,5.1 &   222\,$\pm$\,55 & 3.50 &  {\bf 3.28} &  {\bf 3.87} \\ 
\noalign {\smallskip}
{\bf JKCS041.001.0} & 02:26:42.59 $-$04:42:13.9 &   7.5\,$\pm$\,0.6 &   8.0\,$\pm$\,1.1 &  23.7\,$\pm$\,4.1 &   206\,$\pm$\,56 & 3.58 &  {\bf 3.84} &  8.14 \\ 
{\bf JKCS041.003.0} & 02:26:42.80 $-$04:39:56.5 &  22.4\,$\pm$\,0.6 &  29.6\,$\pm$\,1.0 &  37.9\,$\pm$\,4.1 &   402\,$\pm$\,58 & 1.96 &  {\bf 0.70} &  {\bf 1.10} \\ 
JKCS041.004.0$^{4\ast}$ & 02:26:32.53 $-$04:43:06.9 &  37.4\,$\pm$\,0.7 &  55.3\,$\pm$\,1.3 & 111.9\,$\pm$\,5.3 &   730\,$\pm$\,56 & 4.54 & {\bf  1.24} & {\bf  1.38} \\ 
{\it JKCS041.004.1} & 02:26:32.64 $-$04:43:02.7 &  13.8\,$\pm$\,0.6 &  11.0\,$\pm$\,1.1 &  25.3\,$\pm$\,4.5 &   202\,$\pm$\,56 & 2.27 &  {\bf 2.23} & ... \\ 
JKCS041.005.0$^\ast$ & 02:26:53.06 $-$04:41:29.9 &  18.4\,$\pm$\,0.6 &  18.8\,$\pm$\,1.1 &  48.0\,$\pm$\,4.5 &   636\,$\pm$\,56 & 2.18 &  {\bf 0.47} &  {\bf 3.31} \\ 
JKCS041.005.1 & 02:26:53.04 $-$04:41:31.5 &  13.7\,$\pm$\,0.5 &  20.7\,$\pm$\,1.1 &  39.4\,$\pm$\,4.5 &     ... & 0.94 & ... &  {\bf 0.50} \\ 
{\bf JKCS041.006.0} & 02:26:49.44 $-$04:42:50.0 &  15.1\,$\pm$\,0.6 &  19.1\,$\pm$\,1.1 &  53.9\,$\pm$\,4.5 &     ... & 3.34 & ... &  {\bf 3.75} \\ 
{\it JKCS041.007.0} & 02:26:42.25 $-$04:43:52.9 & 104.2\,$\pm$\,0.8 &  81.1\,$\pm$\,1.2 &  65.3\,$\pm$\,4.5 &   596\,$\pm$\,65 & 3.79 &  {\bf 1.17} & ... \\ 
\noalign {\smallskip}
{\it LH146.001.0}$^5$ & 10:53:22.29 ~+57:23:10.5 &   ...  &   ...  &   ...  &     ... & 2.18 & ... & ... \\ 
{\bf LH146.002.0}$^6$ & 10:53:25.62 ~+57:22:48.3 &  42.0\,$\pm$\,0.3 &  52.3\,$\pm$\,0.4 &  45.7\,$\pm$\,1.2 &   116\,$\pm$\,6 & 3.77 & ... &  {\bf 3.55} \\ 
LH146.003.0$^{7\ast}$ & 10:53:43.58 ~+57:25:43.7 &   9.6\,$\pm$\,0.2 &  15.3\,$\pm$\,0.4 &  18.8\,$\pm$\,1.2 &     ... & 4.60 & ... & ... \\ 
{\bf LH146.004.0}$^8$ & 10:53:14.39 ~+57:24:10.5 &  27.7\,$\pm$\,0.2 &  35.6\,$\pm$\,0.5 &  42.9\,$\pm$\,1.1 &   559\,$\pm$\,10 & 2.26 &  {\bf 0.69} &  {\bf 3.08} \\ 
{\bf LH146.005.0} & 10:53:15.24 ~+57:24:38.7 &   8.6\,$\pm$\,0.2 &  10.5\,$\pm$\,0.4 &  14.0\,$\pm$\,1.1 &   182\,$\pm$\,10 & 1.41 &  {\bf 1.58} &  5.56 \\ 
LH146.006.0$^9$ & 10:53:45.18 ~+57:23:29.3 &  18.0\,$\pm$\,0.2 &  25.5\,$\pm$\,0.3 &  24.9\,$\pm$\,1.4 &   400\,$\pm$\,6 & 3.00 &  {\bf 2.00} &  6.18 \\ 
LH146.007.0$^{10}$ & 10:53:17.45 ~+57:27:22.8 &  25.5\,$\pm$\,0.2 &  38.0\,$\pm$\,0.3 &  40.3\,$\pm$\,1.2 &   691\,$\pm$\,11 & 2.81 &  {\bf 0.71} &  {\bf 3.11} \\ 
LH146.009.0$^{11\ast}$ & 10:53:19.24 ~+57:21:08.6 &  10.5\,$\pm$\,0.2 &  15.5\,$\pm$\,0.3 &  20.8\,$\pm$\,1.2 &   373\,$\pm$\,6 & 3.57 &  {\bf 2.63} & ... \\ 
{\bf LH146.010.0} & 10:53:16.68 ~+57:25:15.0 &  17.4\,$\pm$\,0.2 &  24.0\,$\pm$\,0.4 &  28.5\,$\pm$\,1.2 &   196\,$\pm$\,6 & 2.99 &  {\bf 4.67} &  7.71 \\ 
LH146.011.0$^{12}$ & 10:53:48.52 ~+57:23:57.9 &  39.1\,$\pm$\,0.3 &  50.8\,$\pm$\,0.4 &  50.4\,$\pm$\,1.3 &   352\,$\pm$\,6 & 6.09 &  6.48 &  6.47 \\ 
LH146.012.0 & 10:53:45.17 ~+57:24:29.2 &  34.1\,$\pm$\,0.2 &  42.4\,$\pm$\,0.4 &  50.0\,$\pm$\,1.3 &   371\,$\pm$\,10 & 2.76 &  {\bf 1.77} &  {\bf 2.70} \\ 
\noalign {\smallskip}
{\bf IDCSJ1426.002.0} & 14:26:36.45 ~+35:07:14.7 &  15.4\,$\pm$\,0.2 &  23.6\,$\pm$\,0.5 &  30.4\,$\pm$\,1.7 &   277\,$\pm$\,24 & 2.66 &  {\bf 2.23} &  {\bf 3.31} \\ 
{\bf IDCSJ1426.002.1} & 14:26:36.07 ~+35:07:12.7 &  15.9\,$\pm$\,0.2 &  24.3\,$\pm$\,0.5 &  25.4\,$\pm$\,1.7 &     ... & 2.65 & ... &  {\bf 3.20} \\ 
{\it IDCSJ1426.003.0} & 14:26:28.41 ~+35:10:16.1 &  19.2\,$\pm$\,0.2 &  12.6\,$\pm$\,0.4 &  19.5\,$\pm$\,1.7 &   110\,$\pm$\,24 & 1.06 &  {\bf 1.43} & ... \\ 
IDCSJ1426.004.0$^\ast$ & 14:26:45.09 ~+35:06:56.8 &  16.7\,$\pm$\,0.2 &  22.9\,$\pm$\,0.4 &  29.0\,$\pm$\,1.8 &   282\,$\pm$\,23 & 0.19 &  {\bf 0.03} &  {\bf 0.04} \\ 
IDCSJ1426.004.1 & 14:26:45.05 ~+35:06:52.7 &  19.8\,$\pm$\,0.2 &  28.0\,$\pm$\,0.4 &  35.5\,$\pm$\,1.8 &   127\,$\pm$\,23 & 4.25 &  8.43 &  {\bf 4.84} \\ 
IDCSJ1426.005.0 & 14:26:37.73 ~+35:06:17.4 &   9.2\,$\pm$\,0.2 &  14.5\,$\pm$\,0.4 &  20.0\,$\pm$\,1.8 &     ... & 2.35 & ... &  {\bf 3.98} \\ 
\noalign {\smallskip}
IDCSJ1433.001.0$^\ast$ & 14:33:04.17 ~+33:09:32.7 &  20.5\,$\pm$\,0.2 &  38.3\,$\pm$\,0.4 &  40.9\,$\pm$\,1.6 &   331\,$\pm$\,28 & 2.33 &  {\bf 1.39} &  {\bf 1.19} \\ 
IDCSJ1433.003.0$^{\ast}$ & 14:33:06.38 ~+33:08:38.1 &  13.4\,$\pm$\,0.2 &  19.3\,$\pm$\,0.5 &  27.1\,$\pm$\,1.8 &   305\,$\pm$\,27 & 2.10 &  {\bf 1.38} &  {\bf 2.57} \\ 
{\it IDCSJ1433.004.0} & 14:33:03.45 ~+33:06:14.5 &  51.2\,$\pm$\,0.3 &  46.0\,$\pm$\,0.5 &  41.5\,$\pm$\,1.7 &   327\,$\pm$\,26 & 2.88 &  {\bf 1.99} & ... \\
\noalign {\smallskip}
{\bf ClJ1449.001.0} & 14:49:13.03 ~+08:58:14.5 &  23.3\,$\pm$\,0.3 &  31.7\,$\pm$\,0.5 &  36.0\,$\pm$\,2.3 &   139\,$\pm$\,11 & 2.23 &  {\bf 3.29} &  {\bf 1.20} \\ 
{\bf ClJ1449.003.0}$^{13,\ast}$ & 14:49:14.32 ~+08:56:12.7 &  14.3\,$\pm$\,0.3 &  20.8\,$\pm$\,0.4 &  27.7\,$\pm$\,2.3 &   237\,$\pm$\,29 & 0.81 &  {\bf 0.37} &  {\bf 2.46} \\ 
{\bf ClJ1449.004.0} & 14:49:08.26 ~+08:57:26.2 &  21.9\,$\pm$\,0.3 &  28.9\,$\pm$\,0.5 &  34.4\,$\pm$\,2.3 &   209\,$\pm$\,14 & 3.19 &  {\bf 3.60} &  {\bf 2.48} \\ 
 \hline \noalign {\smallskip}
\end{tabular}

\parbox{\textwidth}{
Probable cluster members lying within 1\,Mpc radius are identified by bold IDs, while likely non-members are identified by italicised IDs.  Sources that have IRAC/MIPS probabalistic counterparts with $P$\,$\leq$\,0.05 and thus are classified as ``reliable''
identifications have the $P$ values shown in bold. $^\ast$ Photometrically identified AGN following \citet{Donley12}.  Footnotes identify sources with interferometric identifications from ALMA or VLA. Footnotes:
$^1$ ALMA band 3 \& 4 continuum source from \cite{vanMarrewijk23} $z$\,$=$\,1.19, foreground;
$^2$ ALMA band 4 continuum source from \cite{vanMarrewijk23} $z$\,$=$\,1.96, member;
$^3$ J0224$-$424 ALMA band 3, line at 87.489\,GHz, CO(2--1) $z$\,$=$\,1.635, member;
$^4$ J022632.53$-$044306.7 \citep{Heywood20}  $S_{\rm 1.4GHz}$\,$=$\,368\,$\pm$\,6\,$\mu$Jy $P_{\rm 1.4GHz}$\,$=$\,0.28\%;
$^5$ LH1.4GHzJ105322.3+572310 \citep[counterparts prefixed LH1.4GHz come from][]{Biggs06} $S_{\rm 1.4GHz}$\,$=$\,28\,$\pm$\,9\,$\mu$Jy $P_{\rm 1.4GHz}$\,$=$\,1.1\% No IRAC ID;
$^6$ LH1.4GHzJ105325.6+572248 $S_{\rm 1.4GHz}$\,$=$\,80\,$\pm$\,11\,$\mu$Jy $P_{\rm 1.4GHz}$\,$=$\,1.2\%;
$^7$ LH1.4GHzJ105343.6+572545  $S_{\rm 1.4GHz}$\,$=$\,55\,$\pm$\,14\,$\mu$Jy $P_{\rm 1.4GHz}$\,$=$\,2.6\%
$^8$ LH1.4GHzJ105314.3+572410 $S_{\rm 1.4GHz}$\,$=$\,104\,$\pm$\,11\,$\mu$Jy $P_{\rm 1.4GHz}$\,$=$\,0.37\%;
$^9$ LH1.4GHzJ105345.2+572329  $S_{\rm 1.4GHz}$\,$=$\,44\,$\pm$\,10\,$\mu$Jy $P_{\rm 1.4GHz}$\,$=$\,1.7\%;
$^{10}$ LH1.4GHzJ105317.4+572722 $S_{\rm 1.4GHz}$\,$=$\,132\,$\pm$\,13\,$\mu$Jy $P_{\rm 1.4GHz}$\,$=$\,0.60\%;
$^{11}$ LH1.4GHzJ105319.2+572108  $S_{\rm 1.4GHz}$\,$=$\,108\,$\pm$\,10\,$\mu$Jy $P_{\rm 1.4GHz}$\,$=$\,0.59\%;
$^{12}$ LH1.4GHzJ105348.5+572357 6 $S_{\rm 1.4GHz}$\,$=$\,77\,$\pm$\,11\,$\mu$Jy $P_{\rm 1.4GHz}$\,$=$\,2.3\%;
$^{13}$ A5 ALMA band 7 continuum source from \citet{Coogan18} $S_{\rm 870\mu m}$\,$=$\,6.0\,$\pm$\,0.2\,mJy.}
\end{table*}


%
%
\begin{table*}
\caption{{\sc Supplementary}  Identifications}
\begin{tabular}{lcccccccc}
\hline \noalign {\smallskip}
ID  & R.A.~~~~~~~~~~~~Dec.\ & S$_{\rm 3.6\mu m}$ &  S$_{\rm 4.5\mu m}$ & S$_{\rm 5.8\mu m}$ & S$_{\rm 24\mu m}$ & $\Delta \theta_{\rm S2}$ & $P_{\rm MIPS}$ & $P^{\rm mem}_{\rm IRAC}$ \\
  &  (J2000) & ($\mu$Jy) & ($\mu$Jy) &($\mu$Jy) & ($\mu$Jy) & ($''$) & (\%) & (\%) \\
\hline \noalign {\smallskip}
{\bf XLSSC122.009.0} & 02:17:39.38 $-$03:47:51.5 &  26.9\,$\pm$\,0.6 &  31.6\,$\pm$\,1.1 &  39.3\,$\pm$\,4.7 &     ... & 4.42 & ... &  {\bf 3.14} \\ 
\noalign {\smallskip}
{\bf SpARCSJ0224.013.0}$^{14}$ & 02:24:27.16 $-$03:24:01.3 &  41.8\,$\pm$\,0.6 &  44.0\,$\pm$\,1.1 &  52.8\,$\pm$\,3.7 &   300\,$\pm$\,60 & 1.62 &  {\bf 1.00} &  {\bf 0.63} \\ 
{\it SpARCSJ0224.013.1} & 02:24:26.90 $-$03:23:57.3 & 180.0\,$\pm$\,0.9 & 137.9\,$\pm$\,1.3 & 152.8\,$\pm$\,4.4 &   523\,$\pm$\,58 & 6.47 &  {\bf 2.63} & ... \\ 
{\bf SpARCSJ0224.014.0}$^{15}$ & 02:24:26.33 $-$03:23:30.5 &  75.2\,$\pm$\,0.7 &  88.8\,$\pm$\,1.2 &  80.0\,$\pm$\,4.5 &   323\,$\pm$\,58 & 5.90 &  {\bf 4.29} & ... \\ 
\noalign {\smallskip}
{\bf JKCS041.009.0}$^{16}$ & 02:26:38.74 $-$04:41:05.3 &  61.4\,$\pm$\,0.7 &  71.2\,$\pm$\,1.1 &  64.7\,$\pm$\,4.3 &   408\,$\pm$\,63 & 3.80 &  {\bf 1.84} &  {\bf 0.79} \\ 
{\bf JKCS041.011.0} & 02:26:47.74 $-$04:41:09.2 &  23.2\,$\pm$\,0.6 &  32.2\,$\pm$\,1.2 &  25.0\,$\pm$\,4.3 &   100\,$\pm$\,50 & 2.77 &  7.00 &  {\bf 2.01} \\ 
\noalign {\smallskip}
{\bf LH146.014.0} & 10:53:26.04 ~+57:25:35.8 &  20.6\,$\pm$\,0.2 &  26.3\,$\pm$\,0.4 &  35.5\,$\pm$\,1.4 &   444\,$\pm$\,10 & 0.27 &  {\bf 0.02} &  {\bf 0.12} \\ 
{\it LH146.015.0}$^{17}$& 10:53:27.18 ~+57:23:13.3 &  60.0\,$\pm$\,0.3 &  48.9\,$\pm$\,0.4 &  51.9\,$\pm$\,1.3 &   399\,$\pm$\,6 & 1.69 &  {\bf 0.70} & ... \\ 
{\bf LH146.017.0}$^{18}$ & 10:53:32.86 ~+57:24:49.2 &  17.4\,$\pm$\,0.2 &  20.4\,$\pm$\,0.3 &  24.0\,$\pm$\,1.4 &   290\,$\pm$\,10 & 3.17 &  {\bf 3.22} &  9.48 \\ 
{\it LH146.020.0} & 10:53:17.71 ~+57:24:00.6 &  71.3\,$\pm$\,0.4 &  53.9\,$\pm$\,0.4 &  36.8\,$\pm$\,1.1 &    75\,$\pm$\,6 & 0.83 &  {\bf 1.52} & ... \\ 
{\bf LH146.020.1} & 10:53:17.19 ~+57:24:02.6 &  26.5\,$\pm$\,0.2 &  31.2\,$\pm$\,0.4 &  30.4\,$\pm$\,1.1 &   309\,$\pm$\,6 & 4.45 &  {\bf 5.00} & ... \\ 
\noalign {\smallskip}
{\bf ClJ1449.007.0} & 14:49:17.66 ~+08:56:55.0 &  39.3\,$\pm$\,0.3 &  48.5\,$\pm$\,0.5 &  45.1\,$\pm$\,2.3 &   316\,$\pm$\,11 & 2.63 &  {\bf 1.64} &  {\bf 0.95} \\ 
{\it ClJ1449.009.0}$^{19}$ & 14:49:16.43 ~+08:56:08.5 &  11.6\,$\pm$\,0.3 &  12.9\,$\pm$\,0.5 &  22.0\,$\pm$\,3.2 &     ... & 4.52 & ... & ... \\ 
{\bf ClJ1449.009.1} & 14:49:16.70 ~+08:56:11.4 &  21.3\,$\pm$\,0.3 &  30.8\,$\pm$\,0.5 &  26.2\,$\pm$\,3.2 &   175\,$\pm$\,30 & 4.33 &  6.29 &  5.31 \\ 
 \hline \noalign {\smallskip}
\end{tabular}

\parbox{\textwidth}{Potential non-members are identified by italicised IDs.  Sources that have IRAC/MIPS probabalistic counterparts with $P$\,$\leq$\,0.05 and thus are classified as ``reliable''
identifications have the $P$ values shown in bold.  Footnotes identify sources with interferometric identifications from ALMA or VLA:
$^{14}$ J0224$-$151 or 159? ALMA band 3, line at 87.549\,GHz, CO(2--1) $z$\,$=$\,1.633 member (+companion);
$^{15}$ XMM-113/J0224$-$306 ALMA band 3, line at CO 87.555\,GHz, CO(2--1) $z$\,$=$\,1.633 member;
$^{16}$ J022638.75-044105.9 $S_{\rm 1.4GHz}$\,$=$\,86\,$\pm$\,4\,$\mu$Jy   $P_{\rm 1.4GHz}$\,$=$\,0.57\%;
$^{17}$ LH1.4GHzJ105327.1+572313 $S_{\rm 1.4GHz}$\,$=$\,38\,$\pm$\,9\,$\mu$Jy  $P_{\rm 1.4GHz}$\,$=$\,0.79\%;
$^{18}$ LH1.4GHzJ105332.6+572446 $S_{\rm 1.4GHz}$\,$=$\,63\,$\pm$\,9\,$\mu$Jy  $P_{\rm 1.4GHz}$\,$=$\,0.37\%;
$^{19}$ ALMA band 3, weak line at 85.0\,GHz, possible CO(2--1) $z$\,$=$\,1.712 foreground?}
\end{table*}

\label{lastpage}

\end{document}